\RequirePackage{fix-cm}
\documentclass[smallextended]{svjour3}       % onecolumn (second format)
\smartqed  % flush right qed marks, e.g. at end of proof
\usepackage{graphicx}
\usepackage[authoryear]{natbib}

\usepackage[linesnumbered,algoruled,boxed,lined]{algorithm2e}

\usepackage{color}

\newcommand{\topkmd}{\textit{Top-k m\textsuperscript{th}} }
\newcommand{\topkfirstd}{\textit{Top-k 1\textsuperscript{st}} }

\usepackage{comment}
\usepackage{url}
\usepackage{amsmath}
\usepackage{amssymb}

\begin{document}

\title{Matrix Profile Goes MAD: Variable-Length Motif And Discord Discovery in Data Series%\thanks{Grants or other notes
}

\titlerunning{Matrix Profile Goes MAD}        % if too long for running head

\author{Michele Linardi \and Yan Zhu \and Themis Palpanas \and Eamonn Keogh}

\institute{Michele Linardi \at
             University of Paris\\
              \email{michele.linardi@parisdescartes.fr}           %  \\
\and
    Themis Palpanas \at
	University of Paris and French University Institute (IUF)\\
	\email{themis@mi.parisdescartes.fr}           %  \\
	\and
           Yan Zhu and Eamonn Keogh \at University of California at Riverside \\
           \email{yzhu015@ucr.edu, eamonn@cs.ucr.edu}
}

\date{Received: date / Accepted: date}

\maketitle

\begin{abstract}
	In the last fifteen years, data series \emph{motif} and \emph{discord} discovery have emerged as two useful and well-used primitives for data series mining, 
	%. It is used as both a data exploration tool and a subroutine for higher-level algorithms, which includes classification, clustering, and rule discovery. 
	%Motif discovery has 
	with applications to many domains, including robotics, entomology, seismology, medicine, and climatology. 
	%Recently, there has been significant progress on the scalability of motif discovery, and it is now possible to search for exact motifs in datasets with lengths in the tens of millions in a reasonable amount of time. 
	%With the computational ease of performing motif search, and its diverse application among domain experts, who are not data mining experts, an important issue is brought to the forefront. 
	%The 
	Nevertheless, the state-of-the-art  motif and discord discovery tools still require the user to provide the relative length. 
	Yet, in several cases, the choice of length is critical and unforgiving. 
	Unfortunately, the obvious brute-force solution, which tests all lengths within a given range, is computationally untenable.   
	In this work, we introduce a new framework, which provides an exact and scalable motif and discord discovery algorithm that efficiently finds all motifs and discords in a given range of lengths. 
	We evaluate our approach with five diverse real datasets, and demonstrate that it is up to 20 times faster than the state-of-the-art. 
 	Our results also show that removing the unrealistic assumption that the user knows the correct length, can often produce more intuitive and actionable results, which could have otherwise been missed.
 	(Paper published in Data Mining and Knowledge Discovery Journal - 2020)
\end{abstract}

\section{Introduction}

Data series\footnote{If the dimension that imposes the ordering of the series is time, then we talk about \emph{time series}. However, a series can also be defined through other measures (e.g., angle in radial profiles in astronomy, mass in mass spectroscopy, position in genome sequences, etc.). We use the terms \emph{time series}, \emph{data series}, and \emph{sequence} interchangeably.} have gathered the attention of the data management community for more than two decades~\citep{DBLP:conf/fodo/AgrawalFS93,DBLP:conf/pods/JagadishMM95,Rafiei1998,Chakrabarti02,DBLP:conf/sigmod/PapadimitriouY06,camerra2010isax,DBLP:conf/kdd/KashyapK11,DBLP:journals/pvldb/WangWPWH13,DBLP:journals/kais/CamerraSPRK14,DBLP:journals/pvldb/DallachiesaPI14,DBLP:journals/vldb/ZoumpatianosIP16,dpisax,DBLP:journals/tkde/JensenPT17,DBLP:conf/ieeehpcs/Palpanas17,coconut,conf/bigdata/peng18,KostasThemisTalkICDE,DBLP:journals/vldb/ZoumpatianosLIP18,DBLP:conf/edbt/GogolouTPB19,lernaeanhydra,lernaeanhydra2,dpisaxjournal,coconutpalm,conf/icde/boniol20,conf/icde/peng20,series2graph,parisplus,evolutionofanindex,conf/sigmod/gogolou20}.
They are now one of the most common types of data, present in virtually every scientific and social domain~\citep{DBLP:journals/sigmod/Palpanas15,percomJournal,DBLP:conf/edbt/MirylenkaCPPM16,eamonnaaai2011tutorial,Palpanas2019,DBLP:journals/dagstuhl-reports/BagnallCPZ19}.

Over the last decade, data series motif discovery has emerged as perhaps the most used primitive for data series data mining, and it has many applications to a wide variety of domains~\citep{Whitney,DBLP:conf/kdd/YankovKMCZ07}, including classification, clustering, and rule discovery. More recently, there has been substantial progress on the scalability of motif discovery, and now massive datasets can be routinely searched on conventional hardware~\citep{Whitney}. 

Another critical improvement in motif discovery, is the reduction in the number of parameters that require specification. The first motif discovery algorithm, PROJECTION~\citep{DBLP:conf/kdd/ChiuKL03}, requires that the users set seven parameters, and it still only produces answers that are approximately correct. Researchers have \textit{``chipped''} away at this over the years~\citep{DBLP:conf/sdm/MueenKZCW09,DBLP:conf/ijcai/SariaDK11}, and the current state-of-the-art algorithms only require the user to set a single parameter, which is the desired length of the motifs. 
Paradoxically, the ease with which we can now perform motif discovery has revealed that even this single burden on the user's experience or intuition may be too great. 

For example, AspenTech, a company that makes software for optimizing the manufacturing process for the oil and gas industry, has begun to use motif discovery in their products both as a stand-alone service and also as part of a precursor search tool. They recently noted that, \textit{``our lighthouse (early adopter) customers love motif discovery, and they feel it adds great value [...] but they are frustrated by the finicky setting of the motif length.''}(Noskov Michael - Director, Data Science at Aspen Technology - Personal communication, February $3^{rd}$ 2015).%~\citep{Aspen}. 
The issue, of being restricted to specifying length as an input parameter, has also been noted in other domains that use motif discovery, such as cardiology~\citep{DBLP:journals/tkdd/SyedSKIG10} and speech therapy~\citep{Wordrecognition}, as well as in related problems, such as data series indexing~\citep{ulissevldb,ulisse}.

The obvious solution to this issue is to make the algorithms search over all lengths in a given range and rank the various length motifs discovered. 
Nevertheless, this strategy poses two challenges. 
First, how we can rank motifs of different lengths? 
Second, and most important, how we can search over this much larger solution space in an efficient way, in order to identify the motifs? 

In this work, we describe the first algorithms in the literature that address both problems. 
The proposed solution requires new techniques that significantly extend the state-of-the-art algorithms, including the introduction of a novel lower bounding method, which makes efficiently searching a large number of potential solutions possible. 

\begin{figure}[tb]
	\centering
	\includegraphics[trim={3cm 13cm 9cm 2cm},scale=0.47]{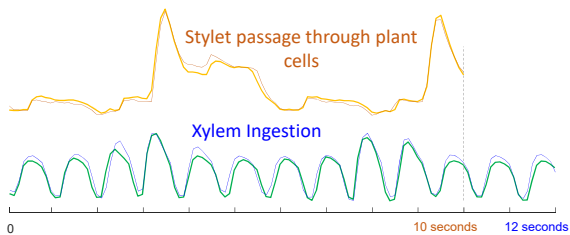}
	\caption{An existence proof of semantically different motifs, of slightly different lengths, extracted from a single dataset.}
	\label{fig:insectVarLeMotifintro}
\end{figure}

Note that even if the user has good knowledge of the data domain, in many circumstances, searching with one single motif length is not enough, because the data can contain motifs of various lengths. 
We show an example in Figure~\ref{fig:insectVarLeMotifintro}, where we report the 10-second and 12-second motifs discovered in the Electrical Penetration Graph (EPG) of an insect called Asian citrus psyllid. The first motif denotes the insect's highly technical probing skill as it searches for a rich leaf vein (stylet passage), whereas the second motif is just a simple repetitive ``sucking'' behavior (xylem ingestion). This example shows the utility of variable length motif discovery. An entomologist using classic motif search, say at the length of 12 seconds, might have plausibly believed that this insect only engaged in xylem ingestion during this time period, and not realized the insect had found it necessary to reposition itself at least twice.

The two motif pairs are radically different, reflecting two different types of insect activities. In order to capture all useful activity information within the data, a fast search of motifs over all lengths is necessary.

Another popular and well-studied data series primitive, the discord~\citep{DBLP:journals/kais/YankovKR08,Keogh2005,YehZUBDDSMK16,DBLP:conf/edbt/Senin0WOGBCF15,Luo2013ParameterFreeSO}, is proposed to discover subsequences that represent outliers. 
Surprisingly, the solutions to this problem that have been proposed in the literature are not as effective and scalable as practice requires. 
The reasons are twofold.
First, they only support fixed-length discord discovery, and as we explained earlier, this rigidity with the subsequence length restricts the search space, and consequently, also the produced solutions and the effectiveness of the algorithm. 
Second, the existing techniques provide poor support for enumerating multiple discords, namely, for the identification of multiple anomalous subsequences. These works have considered only cases with up to \textit{3} anomalous subsequences.

Therefore, we extend our motif discovery framework, and propose the first approach in the literature that deals with the variable-length discord discovery problem.
Our approach leads to a scalable solution, enabling the identification of a large number of anomalous patterns, which can be of different lengths.

In this work\footnote{A preliminary version of this work has appeared elsewhere~\citep{DBLP:conf/sigmod/LinardiZPK18,DBLP:conf/sigmod/LinardiZPK18a}.}, we make the following contributions:
	
\begin{itemize}
\item We define the problems of variable-length motif and discord discovery, which significantly extend the usability of their operations, respectively.
\item We propose a new data series motif and discord framework. 
The Variable Length Motif Discovery algorithm (VALMOD) takes as input a data series $T$, and finds the subsequence pairs with the smallest Euclidean distance of each length in the (user-defined) range [$\ell_{min}$, $\ell_{max}$]. 
VALMOD is based on a novel lower bounding technique, which is specifically designed for the motif discovery problem.
\item
Furthermore, we extend VALMOD to the  discord discovery problem.
We propose a new exact variable-length discord discovery, which aims at finding  the subsequence pairs with the largest Euclidean distances of each length in the (user-defined) range [$\ell_{min}$, $\ell_{max}$]. 
\item
We evaluate our techniques using five diverse real datasets, and demonstrate the scalability of our approach. 
The results show that VALMOD is up to 20x faster than the state-of-the-art techniques.
Furthermore, we present real case studies with datasets from entomology, seismology, and traffic data analysis, which demonstrate the usefulness of our approach.
\end{itemize}

\section{Problem Definition} 
\label{sec:problem}

We begin by defining the data type of interest, data series:

\begin{definition}[Data series]
A data series $T \in \mathbb{R}^n $ is a sequence of real-valued numbers $t_i\in\mathbb{R}$ $[t_1,t_2,...,t_n]$, where $n$ is the length of $T$.	
\end{definition}

We are typically not interested in the global properties of a data series, but in the local regions known as subsequences:

\begin{definition}[Subsequence]
A subsequence $T_{i,\ell} \in \mathbb{R}^\ell$ of a data series $T$ is a continuous subset of the values from $T$ of length $\ell$ starting from position $i$. Formally, $T_{i,\ell} = [t_i, t_{i+1},...,t_{i+\ell-1}]$.	
\end{definition}

\subsection{Motif Discovery}

In this work, a particular local property we are interested in is data series motifs.
A data series motif pair is the pair of the most similar subsequences of a given length, $\ell$, of a data series:
 
\begin{definition}[Data series motif pair]
$T_{a,\ell}$ and $T_{b,\ell}$ is a motif pair iff $dist(T_{a,\ell},T_{b,\ell})  \le dist(T_{i,\ell},T_{j,\ell})$
$\forall i,j \in[1,2,...,n-\ell+1]$, where $a \neq b$ and $i \neq j$, and dist is a function that computes the z-normalized Euclidean distance between the input subsequences \citep{DBLP:conf/kdd/ChiuKL03,DBLP:conf/sdm/MueenKZCW09,Wordrecognition,Whitney,DBLP:conf/kdd/YankovKMCZ07}.
\end{definition}

%Note, that if we remove the motif pair from the dataset, the pair with the second smallest distance will become the new motif pair. 
%In this way, we can produce a ranked list of subsequence pairs, which we call motif pairs of length $\ell$. 
Note, that we can consider more motifs, beyond the top motif pair. 
To that extent, we can simply build a rank of subsequence pairs in $T$ (of length $\ell$), according to their distances in ascending order. 
We call the subsequences pairs of this ranking \emph{motif pairs of length $\ell$.} 

\begin{comment}
We can moreover generalize the definition of data series motif by considering a motifs rank:
\begin{definition}[$k^{th}$-data series motif pair]
A $k^{th}$-data series motif pair of length $m$, is the $k^{th}$ most similar subsequence pair of a data series. Formally, \{$T_{a,m}$,$T_{b,m}$\} is the $k^{th}$-motif pair iff there exist a subsequence pairs set $S$ of size $k-1$, where \{$T_{a,m}$,$T_{x,m}$\}, \{$T_{b,m}$,$T_{x,m}$\}  $\notin S$, for $a \ne b$, $1\le x \le|T|-m+1$ and $\forall$\{$T_{i,m}$,$T_{j,m}$\} $\in S$, \{$T_{w,m}$,$T_{z,m}$\} $\notin S$, dist($T_{i,m}$,$T_{j,m}$)$\le$dist($T_{a,m}$,$T_{b,m}$)$\le$dist($T_{w,m}$,$T_{z,m}$) \citep{DBLP:conf/sdm/MueenKZCW09}.   
\end{definition}
\end{comment}
\begin{comment}
\begin{definition}[Range data series motif]
The range motif of length $m$, is the maximal subsequences set with range $r$. Formally, given a data series $T$, $S$ is a range motif set if $\forall T_{i,m},T_{j,m}\in S$, $dist(T_{i,m},T_{j,m}) \le 2r$ and  $\forall T_{i,m}\in S$, $T_{x,m}\notin S$, such that $1 \le x \le |T|-m+1$, $ dist(T_{i,m},T_{x,m})$ > $2r$ \citep{DBLP:conf/sdm/MueenKZCW09}.   
\end{definition}
\end{comment}

We store the distance between a subsequence of a data series with all the other subsequences from the same data series in an ordered array called a distance profile.

\begin{definition}[Distance profile]
A distance profile $D \in \mathbb{R}^{(n-\ell+1)}$ of a data series $T$ regarding subsequence $T_{i,\ell}$ is a vector that stores $dist(T_{i,\ell},T_{j,\ell})$, $\forall j \in [1,2,...,n-\ell+1]$, where $i \ne j$.
\end{definition}

One of the most efficient ways to locate the exact data series motif is to compute the matrix profile \citep{YehZUBDDSMK16,ZhuZSYFMBK16}, which can be obtained by evaluating the minimum value of every distance profile in the time series.

\begin{definition}[Matrix profile]
A matrix profile $ MP \in \mathbb{R}^{(n-\ell+1)}$ of a data series $T$ is a meta data series that stores the z-normalized Euclidean distance between each subsequence and its nearest neighbor, where $n$ is the length of $T$ and $\ell$ is the given subsequence length. The data series motif can be found by locating the two lowest values in $MP$.
\end{definition}

%Figure \ref{dataSeriesTandMP} illustrates a matrix profile on a small dataset.

%\begin{figure}[h]
% 	\includegraphics[trim={0 6cm 0 4.5cm},scale=0.30]{MP_EX}
% 	\caption{A data series T and its self-join matrix profile P.}
% 	\label{dataSeriesTandMP}	
%\end{figure}

To avoid trivial matches, in which a pattern is matched to itself or a pattern that largely overlaps with itself, the matrix profile incorporates an ``exclusion-zone'' concept, which is a region before and after the location of a given query that should be ignored. The exclusion zone is heuristically set to $\ell$/2.
The recently introduced STOMP algorithm~\citep{ZhuZSYFMBK16} offers a solution to compute the matrix profile $MP$ in $\mathcal{O}(n^2)$ time. This may seem untenable for data series mining, but several factors mitigate this concern. First, note that the time complexity is independent of $\ell$, the length of the subsequences. Secondly, the matrix profile can be computed with an anytime algorithm, and in most domains, in just $\mathcal{O}(nc)$ steps the algorithm converges to what would be the final solution~\citep{YehZUBDDSMK16} ($c$ is a small constant). Finally, the matrix profile can be computed with GPUs, cloud computing, and other HPC environments that make scaling to at least tens of millions of data points trivial~\citep{ZhuZSYFMBK16}.

We can now formally define the problems we solve.

\begin{problem}[Variable-Length Motif Pair Discovery]
\label{def:ValMotif}
Given a data series $T$ and a subsequence length-range $[\ell_{min},...,\ell_{max}]$, we want to find the data series motif pairs of all lengths in $[\ell_{min},...,\ell_{max}]$, occurring in $T$. 
%Formally, $\ell_{max}\ge \ell_{min}$.
\end{problem}

One naive solution to this problem is to repeatedly run the state-of-the art motif discovery algorithms for every length in the range. However, note that the size of this range can be as large as $\mathcal{O}(n)$, which makes the naive solution infeasible for even middle-size data series. We aim at reducing this $\mathcal{O}(n)$ factor to a small value.

Note that the motif pair discovery problem has been extensively studied in the last decade~\citep{YehZUBDDSMK16,ZhuZSYFMBK16,DBLP:journals/kais/MueenC15,DBLP:conf/icde/LiUYG15,DBLP:conf/sdm/MueenKZCW09,DBLP:conf/aciids/MohammadN14,6426960}.
The reason is that if we want to find a collection of recurrent subsequences in $T$, the most computationally expensive operation consists of identifying the motif pairs~\citep{ZhuZSYFMBK16}, namely, solving Problem~\ref{def:ValMotif}. 
Extending motif pairs to sets incurs a negligible additional cost (as we also show in our study).

Given 
%the $i^{th}$ 
a motif pair $\{T_{\alpha,\ell},T_{\beta,\ell}\}$, the data series motif set $S^{\ell}_{r}$, with radius $r\in \mathbb{R}$, is the set of subsequences of length $\ell$, which are in distance at most $r$ from either $T_{\alpha,\ell}$, or $T_{\beta,\ell}$. 
More formally:

\begin{definition}[Data series motif set]
Let $\{T_{\alpha,\ell},T_{\beta,\ell}\}$ be a motif pair of length $\ell$ of data series $T$. 
The motif set $S^\ell_r$ is defined as: $S^\ell_r= \{T_{i,\ell} | dist(T_{i,\ell} ,T_{\alpha,\ell})$ $< r \lor dist(T_{i,\ell} ,T_{\beta,\ell}) < r \}$.
\end{definition}

The cardinality of $S^\ell_r$, $|S^\ell_r|$, is called the frequency of the motif set.   

Intuitively, we can build a motif set starting from a motif pair.
Then, we iteratively add into the motif set all subsequences within radius $r$. 
We use the above definition to solve the following problem (optionally including a constraint on the minimum frequency for motif sets in the final answer).

\begin{problem}[Variable-Length Motif Sets Discovery]
\label{def:ValMotifSet}
Given a data series $T$ and a length range $[\ell_{min},\ldots,\ell_{max}]$, 
%and $K\in \mathbb{N}$, which is the maximum number of motif sets we want to compute, 
%and a function $f:\mathbb{R} \rightarrow \mathbb{R^+}$, called ranging function, 
%we want to find the set: $S^*= \{S^k_r, 1\le k \le K|
we want to find the set $S^*= \{S^\ell_r |
%r=f(x), \exists \ell'$ such that $\ell_{min}\le\ell'\le\ell_{max}, \forall T_{a,\ell} \in S_r$ such that $\ell'=\ell \}$. 
%(S^k_r$ is a motif set$)\wedge (K \le |T|-\ell_{min}+1 )\}$.
S^\ell_r$ is a motif set, $\ell_{min}\le\ell\le\ell_{max}\}$. 
%(\forall T_{a,\ell} \in S^k_r, \exists \ell', \ell_{min} \le\ell'\le\ell_{max},$ such that $\ell'=\ell) 
%($ something about the topk $) \}$. 
%In addition, we require that if $S^k_r,S^{k'}_{r'}\in S^* \Rightarrow S^k_r \cap S^{k'}_{r'} = \emptyset$.
In addition, we require that if $S^\ell_r,S'^{\ell'}_{r'}\in S^* \Rightarrow S^\ell_r \cap S'^{\ell'}_{r'} = \emptyset$.
%Given the namely the subsequences pairs having the $k$ smallest distances, a radius factor $D$ and a (optional) minimum frequency threshold $F$, we want to find all the motif sets, each one containing one of the best $k$ motif pairs. Formally, $\forall m \in M^k$, with $dist(m)$ their Euclidean distance, $\exists S$, a motif set of range $r=(D \times dist(m))$, such that $ m \in S$. Moreover, given $S,S'$ two motif sets, if $m \in S$, $ m' \in S'$ and $m,m' \in M^k$ $\Rightarrow S \cap S' = \emptyset$. Since we are in a variable length setting, if $m$ and $m'$ have different subsequence length we consider equality based on the offsets. If $F$ is defined, $\exists S \Leftrightarrow 2 \le F \le |S|$.
\end{problem}

By abuse of notation, we consider an intersection non-empty in the case where subsequences have different lengths, but the same starting position offset (e.g., $S^{200}_r =\{ T_{4,200}\}, S^{500}_{r'} =\{ T_{4,500}\} \Rightarrow S^{200}_r \cap S^{500}_{r'} \ne \emptyset$).

Thus, the variable-length motif sets discovery problem results in a set, $S^*$, of 
%the top-$K$ 
motif sets.
%, ranked in increasing distance. 
The constraint at the end of the problem definition restricts each subsequence to be included in at most one motif set.
Note that in practice we may not be interested in all the motif sets, but only in those with the $k$ smallest distances, leading to a $top$-$k$ version of the problem.
In our work, we provide a solution for the $top$-$k$ problem (though, setting $k$ to a very large value will produce all results).

%The Top k-Motif Sets, are the sets with the $k$ smallest ranges. The proximity of subsequences is used here as a quality measure. Moreover the range $r$ is not a straight user define parameters, but it is built considering the natural ranking of the data (distance of the best $k$ pairs) and then adjusted by the user input radius factor $D$. This allows a more fine grained motif discovery, given the fact that an appropriate range $r$ is typically hard to set a priori.  

\subsection{Discord Discovery}

In order to introduce the problem of discord discovery, we first define the notion of \emph{best match}, or \emph{nearest neighbor}.
	
\begin{definition}[$m^{th}$ best match]
%		The subsequence $T_{j,\ell}$ is called the $m^{th}$ (non-trivial) match of $T_{i,\ell}$, if there exist an array of index $A \in \mathbb{N}^{m-1}$, where for each pair $a,a'\in A \implies a \ne a'$, and $1\le a,a'\le |T-\ell+1|$. Moreover if $a\in A \iff dist(T_{i,\ell},T_{a,\ell})< dist(T_{j,\ell},T_{a,\ell})$ and $T_{a,\ell}$ is not a trivial match of $T_{i,\ell}$. For each $b \notin A $, where $1\le b \le |T-\ell+1|$, $dist(T_{i,\ell},T_{b,\ell}) > dist(T_{i,\ell},T_{j,\ell})$ and $T_{b,\ell}$ is not a trivial match of $T_{i,\ell}$. {\bf ??? I cannot follow, we need to discuss this def ???}
Given a subsequence $T_{i,\ell}$, we say that its $m^{th}$ best match, or Nearest Neighbor ($m^{th}$ NN) is $T_{j,\ell}$, if  $T_{j,\ell}$ has the $m^{th}$ shortest distance to $T_{i,\ell}$, among all the subsequences of length $\ell$ in $T$, excluding trivial matches. 
%We implicitly assume that the $m^{th}$ best match of a generic subsequence $T_{i,\ell}$, has always the same length $\ell$. 
\end{definition}

In the distance profile of $T_{i,\ell}$, the $m^{th}$ smallest distance, is the distance of the $m^{th}$ best match of $T_{i,\ell}$.
We are now in the position to formally define the discord primitives, we use in our work.
	 
\begin{definition}[$m^{th}$ discord~\citep{Keogh2005}]
	The subsequence $T_{i,\ell}$ is called the $m^{th}$ discord of length $\ell$, if its $m^{th}$ best match is the largest among the best match distances of all subsequences of length $\ell$ in $T$.   
\end{definition}

Intuitively, discovering the $m^{th} discord$ enables us to find an isolated group of $m$ subsequences, which are far from the rest of the data.  
Furthermore, we can rank the $m^{th} discords$, according to their $m^{th}$ best matches. 
This allows us to define the \topkmd discords.

\begin{definition}[\topkmd discord]
	%We call the $k$ subsequences, with the $k$ largest distances to their $m^{th}$ best matches, the \topkmd discords.   
	A subsequence $T_{i,\ell}$ is a $Top$-$k$ $m^{th}$-discord if it has the $k^{th}$ largest distance to its $m^{th}$ NN, among all subsequences of length $\ell$ of $T$. 
\end{definition}

\begin{figure}[h]
	\centering
	\includegraphics[trim={0 14.5cm 21cm 3cm},scale=1]{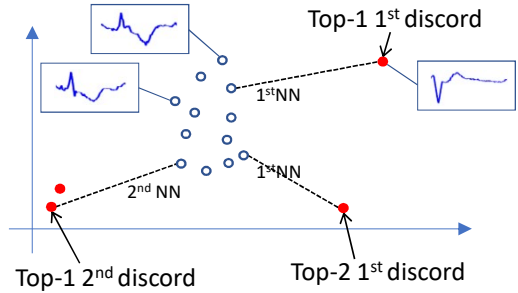}
	\caption{A dataset with 12 subsequences (of the same length $\ell$) depicted as points in 2-dimensional space. We report the \topkmd discords. They belong to groups of subsequences, whose cardinality depends on $m$. The index $k$ ranks the subsequences according their $m^{th}$ best match distances, in descending order.}
	\label{topkmdiscords}	
\end{figure}

In Figure~\ref{topkmdiscords}, we plot a group of \textit{12} subsequences (represented in a 2-dimensional space), and we depict three \topkmd discords (groups of red/dark circles). 
Remember that $m$ represents the number of anomalous subsequences in a discord group. 
On the other hand, $k$ ranks the discords and implicitly the groups, according to their $m^{th}$ best match distances, in descending order (e.g., $Top-1$ $1^{st}$ discord and $Top-1$ $2^{nd}$).     	
	
Given these definitions, we can formally introduce the following problem: 
\begin{problem}[Variable-Length \topkmd Discord Discovery]
\label{def:Variable-Length Discord discovery}
Given a data series $T$, a subsequence length-range $[\ell_{min},...,\ell_{max}]$ and the parameters $a,b \in \mathbb{N^{+}}$ we want to enumerate the \topkmd discords for each $k \in \{1,..,a\}$ and each $m \in \{1,..,b\}$, and for all lengths in $[\ell_{min},...,\ell_{max}]$, occurring in $T$.
\end{problem}

Observe that solving the Variable-Length \topkmd Discord Discovery problem is relevant to solving the Variable-Length Motif Set Discovery problem: in the former case we are interested in the subsequences with the most distant neighbors, while in the latter case we seek the subsequences with the most close neighbors.
Therefore, the Matrix Profile, which contains all this information, can serve as the basis to solve both problems.

\section{Comparing Motifs of Different Lengths}
\label{sec:lengthnorm}

%\commentRed{1 Reviewer said:
%There are some claims in the paper that appear to be insufficiently justified or possibly wrong. For example,
%argue that when comparing motifs of different lengths, sqrt(1/length) would provide "a near perfect invariant
%distance"; however, this is only anecdotally verified and the Figure 3 where this is argued is very poorly presented.
%The 2nd reviewer said that this example was well presented.
%}

Before introducing our solutions to the problems outlined above, we first discuss the issue of comparing motifs of different lengths.
This becomes relevant when we want to rank motifs of different lengths (within the given range), which is useful in order to identify the most prominent motifs, irrespective of their length.
In this section, we propose a length-normalized distance measure that the VALMOD algorithm uses in order to produce such rankings.

The increased expressiveness of VALMOD offers a challenge. 
Since we can \textit{discover} motifs of different lengths, we also need to be able to \textit{rank} motifs of different lengths. 
A similar problem occurs in string processing, and a common solution is to replace the edit-distance by the length-normalized edit-distance, which is the classic distance measure divided by the length of the strings in question \citep{Marzal:1993:CNE:628305.628518}. This correction would find the pair \{conca\textcolor{red}{t}e\textcolor[rgb]{0.21,0.57,0.21}{n}ation, conca\textcolor{red}{m}e\textcolor[rgb]{0.21,0.57,0.21}{r}ation\} \textit{more} similar than \{cat, cot\}, matching our intuition, since only 15\% of the characters are different in the former pair, as opposed to 33\% in the latter. 

Researchers have suggested this length-normalized correction for time series, but as we will show, the correction factor is incorrect.
To illustrate this, consider the following thought experiment. Imagine that some process in the system we are monitoring occasionally ``injects'' a pattern into the time series. As a concrete example, washing machines typically have a prototypic signature (as exhibited in the TRACE dataset~\citep{tracedataset}), but the signatures express themselves more slowly on a cold day, when it takes longer to heat the cooler water supplied from the city~\citep{6602387}. We would like all equal length instances of the signature to have approximately the same distance. As a consequence, we factorize the Euclidean distance by the following quantity: $sqrt(1/\ell)$, where $\ell$ is the length of the sequences. This aims to favor longer and similar sequences in the ranking process of matches that have different lengths.

\begin{figure}[!tb]
	\centering
	\includegraphics[trim={0cm 13.5cm 11cm 3cm},scale=0.70]{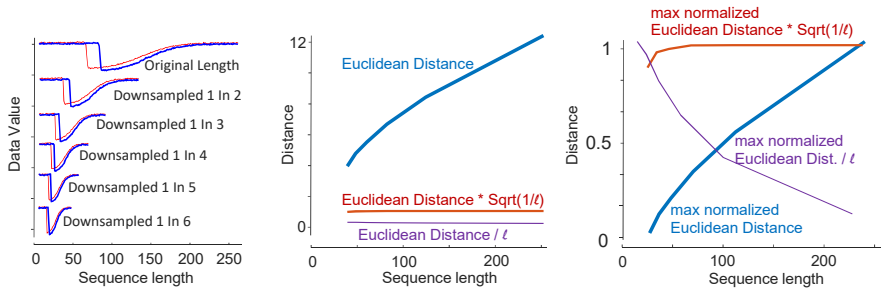}
	\caption{(\textit{left}) Two series from the TRACE dataset, as proxies for time series signatures at various speeds. (\textit{center}) The classic Euclidean distance is clearly not length invariant. (\textit{right}) After divide-by-max normalizing, it is clear that the length-normalized Euclidean distance has a strong bias toward the longest pattern.}
	\label{fig:varLengthCorrect}
	\vspace*{-0.5cm}
\end{figure}

In Figure~\ref{fig:varLengthCorrect}(\textit{left}) we show two examples from the TRACE dataset~\citep{tracedataset}, which will act as proxies for a variable length signature. We produced the variable lengths by down sampling. 
In Figure~\ref{fig:varLengthCorrect}(\textit{center}), we show the distances between the patterns as their length changes. 
With no correction, the Euclidean distance is obviously biased to the shortest length. 
The length-normalized Euclidean distance looks ``flatter'' and suggests itself as the proper correction. 
However, its variation over the sequence length change is not visible due to the small scale. 
In Figure~\ref{fig:varLengthCorrect}(\textit{right}), we show all of the measures after dividing them by their largest value. 
Now we can see that the length-normalized Euclidean distance has a strong bias toward the longest pattern. 
In contrast to the other two approaches, the $sqrt(1/length)$ correction factor provides a near perfect invariant distance over the entire range of values.

%\commentRed{
%Introducing the length-normalized Euclidean distance permits to define the second problem solved by our proposal. 
%	
%\begin{problem}[Top k-Motif Sets Discovery]
%	\label{def:ValMotifSet}
%  	Given the best $k$ motif pairs $M^k$, namely the subsequences pairs with the $k$ smallest length normalized distances, a radius factor $D$ and a minimum frequency threshold $F$, we want to find the set $S^k$, which groups all the motif sets, each one containing at least one of the best $k$ motif pairs. Formally, for each pair of subsequences $ m \in M^k$, with $dist(m)$ their straight Euclidean distance, $\exists S \in S^k$ of range $r$, such that $ m \in S$, $r = (D \times dist(m))$ and {\tiny $2 \le F \le |S|$}. Moreover, if $S,S' \in S^k \Rightarrow S \cap S' = \emptyset$. Since we may compare subsequences of different length, we consider equality based on the offset.
%\end{problem}
%}

\section{Proposed Approach for Motif Discovery}
\label{sec:approach}

% We can rely on a fast algorithm, STOMP, which runs in O($n^2$), computing the \textit{Matrix Profile} from which we may extract Motifs, along with other Data Series primitives, of a fixed subsequence length.%
%Concerning STOMP, 

\begin{figure}[tb]
	\centering
	\includegraphics[trim={0 4cm 0cm 3cm},scale=0.30]{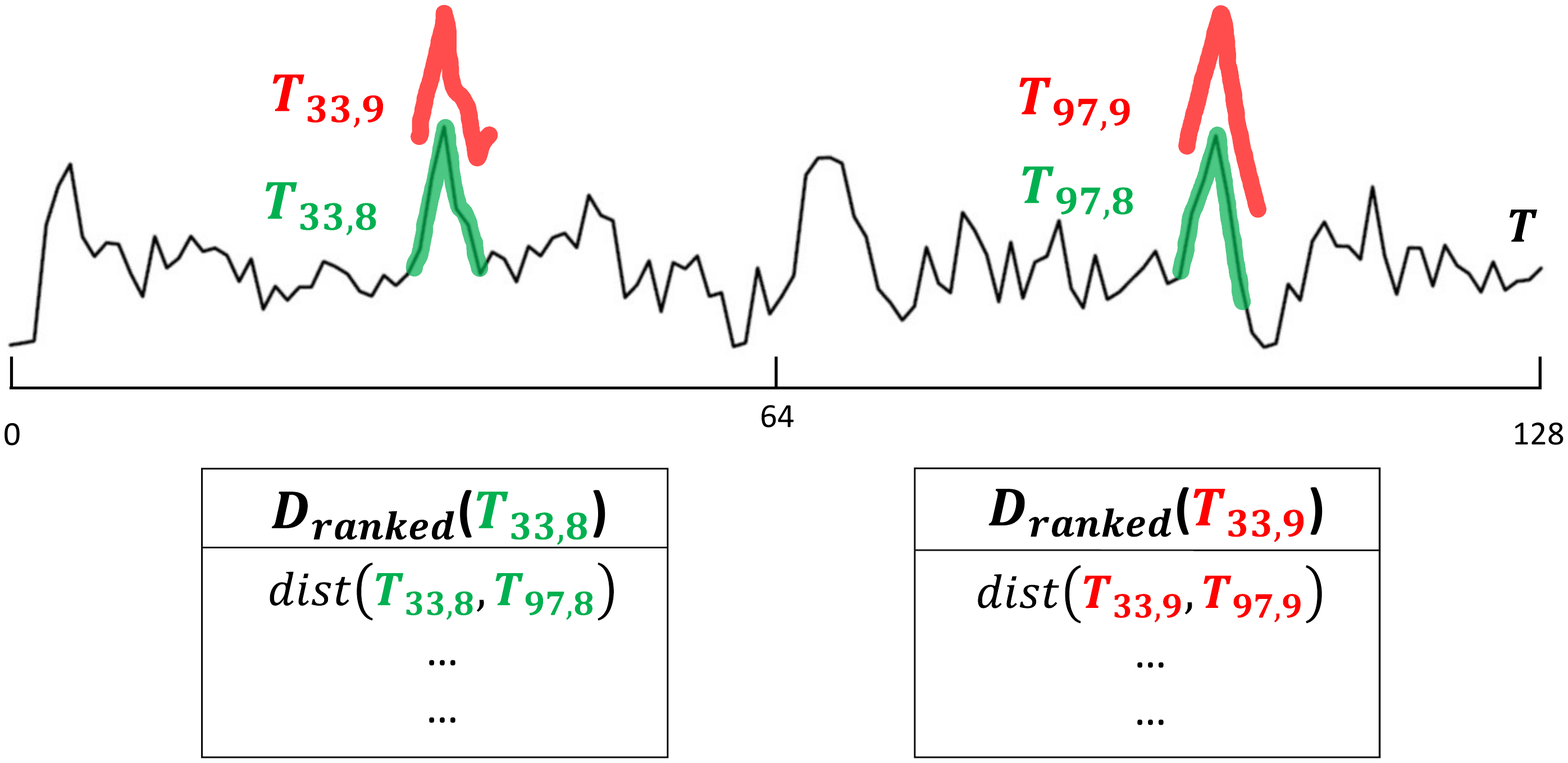}
	\caption{(\textit{top}) The top motifs of length 9 in an example data series have the same offsets as the top motifs of length 8. (\textit{bottom}) The sorted distance profiles of $T_{33,8}$ and $T_{33,9}$.}
	\label{dpEx}
\end{figure}

%\subsubsection{Key idea} 
%We call our algorithm \textit{Variable Length Motif Discovery} (\textit{VALMOD}). 
Our algorithm, VALMOD (Variable Length Motif Discovery), starts by computing the \textit{matrix profile} on the smallest subsequence length, namely $\ell_{min}$, within a specified range $[\ell_{min},\ell_{max}]$.
The key idea of our approach is to minimize the work that needs to be done for subsequent subsequence lengths ($\ell_{min}+1$, $\ell_{min}+2$, $\ldots$, $\ell_{max}$). In Figure~\ref{dpEx}, it can be observed that the motif of length 8 ($T_{33,8}-T_{97,8}$) has the same offsets as the motif of length 9 ($T_{33,9}-T_{97,9}$). 
Can we exploit this property to accelerate our computation?

It seems that if the nearest neighbor of $T_{i,\ell_{min}}$ is $T_{j,\ell_{min}}$, then probably the nearest neighbor of $T_{i,\ell_{min}+1}$ is $T_{j,\ell_{min}+1}$. 
For example, as shown in Figure~\ref{dpEx}(bottom), if we sort the distance profiles of $T_{33,8}$ and $T_{33,9}$ in ascending order, we can find that the nearest neighbor of $T_{33,8}$ is $T_{97,8}$, and the nearest neighbor of $T_{33,9}$ is $T_{97,9}$. 
%If we sort the distnace profiles of $T_{33,8} entries of the distance profile regarding $T_{33,8}$ in ascending order, we can obtain a new vector $D_{ranked}(T_{33,8})$. We can find the $p$ nearest neighbors of $D_{ranked}(T_{33,8})$ from the first $p$ entries of $D_{ranked}(T_{33,8})$. We can see that the nearest neighbor of $T_{33,8}$ is $T_{97,8}$, and the nearest neighbor of $T_{33,9}$ is $T_{97,9}$. 

%One can imagine that if the location the nearest neighbor of $T_{i,l}$ ($i=1, 2, ..., n-m+1$) remains the same as we increase $ell$, then we can obtain the matrix profile of length $l+k$ in $O(n)$ time ($k=1$, $2$, $\ldots$). 

\begin{figure}[tb]
	\centering
	\includegraphics[trim={1.5cm 2cm 0cm 0cm},scale=0.33]{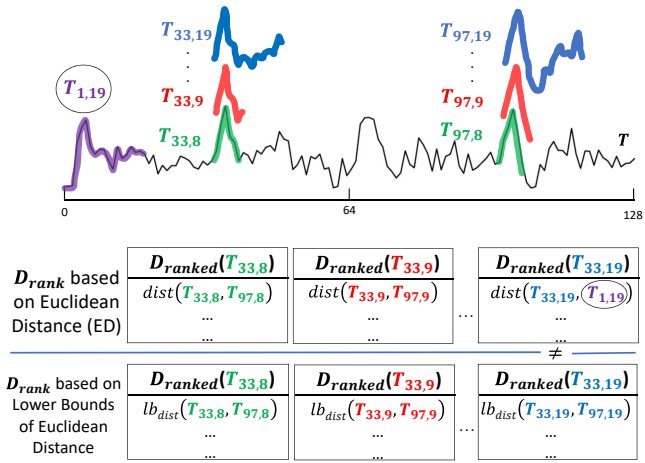}
	\caption{(\textit{top distance profiles}) Ranking by true distances leads to changes in the order of the pairs. (\textit{bottom distance profiles}) Ranking by lower bound distances maintains the same order of pairs over increasing lengths.}
	\label{LBRankingWish}
\end{figure}

One can imagine that if the location of the nearest neighbor of $T_{i,\ell}$ ($i=1, 2, ..., n-m+1$) remains the same as we increase $\ell$, then we could obtain the matrix profile of length $\ell+k$ in $\mathcal{O}(n)$ time ($k=1$, $2$, $\ldots$). 
However, this is not always true.
%Since the motif pair of length $l+k$ is simply ... If the the ranking of $D_{ranked}(T_{i,l})$ preserves for all $i$ and $l$, then once we know the motif of length $l$, we can know the motif of length $l+k$ ($k \in [1,2,…]$) in $O(n)$ time.
%We consider this fact as a non-negligible property of a distance profile.
%More precisely, we are assuming that a subsequence pair, which has the best distance in $D_{ranked}(T_{i,\ell_{min}+1})$, has a %closer distance rank in $D_{ranked}(T_{i,\ell_{min}})$, or at best the same. %\commentOr{Note for Themis by Michele: I arranged the following 2 sentences, it should be more clear.}
The location of the nearest neighbor of $T_{i,\ell}$ may not change as we slightly increase $\ell$, if there is a substantial margin between the first and second entries of $D_{ranked}(T_{i,\ell})$. 
But, as $\ell$ gets larger, %, and when $k$ is small.
%On the other hand, especially when the data are noisy or skewed, 
the nearest neighbor of $T_{i,\ell}$ is likely to change. %more often as $\ell$ grows. 
For example, as shown in Figure~\ref{LBRankingWish}, when the subsequence length grows to 19, the nearest neighbor of $T_{33,19}$ is no longer $T_{97,19}$, but $T_{1,19}$. %As a result, we find $T_{97,19}$ in the first entry of $D_{ranked}(T_{33,19})$.
%Note that considering the same subsequence pair in different distance profiles of different lengths corresponds to examining two pairs of subsequences with same offset (prefix) but different trailing points, as for the two series pairs in Figure~\ref{dpEx}.
We observe that the ranking of the distance profile values may change, even when the data is relatively smooth. 
When the data is noisy and skewed, this ranking can change even more often. 
Is there any other rank-preserving measure that we can exploit to accelerate the computation? 

%In order to devise an efficient algorithm, we need to maintain a certain subsequence pairs ranking. This needs to take place in distance profiles with different length and equal subsequence offsets.
%This is clearly not feasible if we rank our pairs according to the true Euclidean distances, since adding points to the sequences surely varies the distances and in turn, the ranking might change.   

%Now that we know the ranking of the distance profile values does not always preserve as the subsequence length $\ell$ increases, is there any other measure that we can exploit to accelerate the computation?

The answer is yes. 
Instead of sorting the entries of the distance profile, we create and sort a new vector, called the \emph{lower bound distance profile}. 
Figure~\ref{LBRankingWish}(bottom) previews the rank-preserving property of the lower bound distance profile. 
As we will describe later, once we know the distance between $T_{i,\ell}$ and $T_{j,\ell}$, we can evaluate a lower bound distance between $T_{i,\ell+k}$ and $T_{j,\ell+k}$, $\forall k \in [$1$,$2$,$3$,$\ldots$]$. 
The rank-preserving property of the lower bound distance profile can help us prune a large number of unnecessary computations as we increase the subsequence length.
% Nevertheless, if we rank according to a lower bounding measure, we can ensure the same subsequences distance ranking, as well as an increase in length. This property is shown in the second row of distance profiles of Figure~\ref{LBRankingWish}. 
%On the other hand, when the first row is ranked according to the true Euclidean distance, the best matches of the distance profiles are composed of subsequences with different offsets.

%\subsubsection{Intermediate Results Storage}

%In the pseudo-code, we denote a distance profile, computed for a generic subsequence $T_{i,\ell}$ with the signature $D(T_{i,\ell})$. It is crucial to note that we may rank the distances of $D(T_{i,\ell})$ in ascending order, obtaining a new structure: $D_{ranked}(T_{i,\ell})$. We can then use this structure in order to establish a relationship between $D_{ranked}(T_{i,\ell})$ and the same distance profile computed by increasing the length by one point, i.e., $D_{ranked}(T_{i,\ell+1})$. I

%\commentRed{1 Reviewer out of 3 said that the flow is not smooth, we should provide more steps between (2) and (3) }
 \subsection{The Lower Bound Distance Profile}
 \label{sec:lowwerbounddistanceprofile}
Before introducing the lower bound distance profile, let us first investigate its basic element: the lower bound Euclidean distance.

\begin{figure}[tb]
	\centering
	\includegraphics[trim={-1cm 3cm 0cm 3cm},scale=0.30]{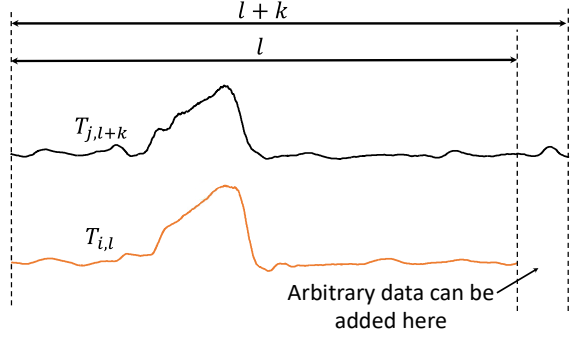}
	\caption{Increasing the subsequence length from $\ell$ to $\ell+k$.}
	\label{LBRanking}	
\end{figure}

%\noindent{\bfThe Lower Bound Euclidean Distance}
Assume that we already know the z-normalized Euclidean distance $d_{i,j}^\ell$ between two subsequences of length $\ell$: $T_{i,\ell}$ and $T_{j,\ell}$, and we are now estimating the distance between two longer subsequences of length $\ell+k$: $T_{i,\ell+k}$ and $T_{j,\ell+k}$. Our problem can be stated as follows: given $T_{i,\ell}$, $T_{j,\ell}$ and $T_{j,\ell+k}$ (but not the last $k$ values of $T_{i,\ell+k}$), is it possible to provide a lower bound function $LB(d_{i,j}^{\ell+k})$, such that $LB(d_{i,j}^{\ell+k}) \le d_{i,j}^{\ell+k}$?
This problem is visualized in Figure \ref{LBRanking} .

\begin{comment}
The nature of Distance Profile, which is a vector containing the distances between a collection of series to a single one, permits to specify a lower bounding distance which enjoys several properties we use to take a decision about the smallest element in $D^k_{Rank}(a+1,s_i)$ knowing $D(a,s_i)$.\\
We define the function $LB_{profile}(x^l_i,y^l_j)$, where $x$ and $y$ are two series with offset $i,j$ and length $l$.
Hence, the following properties always hold:
\begin{itemize}
\item $LB_{profile}(s_i^{l},s_{j}^{l})$ $\leq$ $Euclidean_{distance}(s_i^{l},s_{j}^{l})$
\item if $LB_{profile}(s_i^{l},s_{j}^{l}) < LB_{profile}(s_i^{l},s_{z}^{l}) \Rightarrow LBdist(s_{i}^{l'},s_{j}^{l'}) < LBdist(s_{i}^{l'},s_{z}^{l'})$ where $l' > l$
\end{itemize}

\commentOr{Here I let Yan explain better. I just put this "incipit" that eventually may be used. Michele}
\end{comment}

One may assume that we can simply set $LB(d_{i,j}^{\ell+k}) = d_{i,j}^\ell$ by assuming that the last $k$ values of $T_{i,\ell+k}$ are the same as the last $k$ values of $T_{j,\ell+k}$. However, this is not an answer to our problem, as we need to evaluate z-normalized Euclidean distances, which are not simple Euclidean distances. The mean and standard deviation of a subsequence can change as we increase its length, so we need to re-normalize both $T_{i,\ell+k}$ and $T_{j,\ell+k}$.
Assume that the mean and standard deviation of $T_{x,y}$ are $\mu_{x,y}$ and $\sigma_{x,y}$, respectively (i.e. $T_{j,\ell+k}$ corresponds to $\mu_{j,\ell+k}$ and $\sigma_{j,\ell+k}$). Since we do not know the last $k$ values of $T_{i,\ell+k}$, both $\mu_{i,\ell+k}$ and $\sigma_{i,\ell+k}$ are unknown and can thus be regarded as variables. We recall that $t_{i}$ denotes the $i^{th}$ point of a generic sequence $T$ (or a subsequence $T_{a,b}$), we thus we have the following:

\[ d_{i,j}^{\ell+k}  \ge  \substack{min \\ \mu_{i,\ell+k},\sigma_{i,\ell+k}}  \sqrt{\sum_{p=1}^{\ell}(\frac{t_{i+p-1}-\mu_{i,\ell+k}}{\sigma_{i,\ell+k}} - \frac{t_{j+p-1}-\mu_{j,\ell+k}}{\sigma_{j,\ell+k}})^2 } \]

\[= \substack{min \\ \mu_{i,\ell+k},\sigma_{i,\ell+k}} \frac{\sigma_{j,\ell}}{\sigma_{j,\ell+k}}  \sqrt{\sum_{p=1}^{\ell}(\frac{t_{i+p-1}-\mu_{i,\ell+k}}{\frac{\sigma_{i,\ell+k}\sigma_{j,\ell}}{\sigma_{j,\ell+k}}} - \frac{t_{j+p-1}-\mu_{j,\ell+k}}{\sigma_{j,\ell}})^2 } \]

Here, we substitute the variables $\mu_{i,\ell+k}$ and $\sigma_{i,\ell+k}$, respectively with $\mu'$ and $\sigma'$. Hence, we obtain:

\begin{equation}
\label{eq:lowerBoundFormula}
= \substack{min \\ \mu',\sigma'} \frac{\sigma_{j,\ell}}{\sigma_{j,\ell+k}}  \sqrt{\sum_{p=1}^{\ell}(\frac{t_{i+p-1}-\mu'}{\sigma'} - \frac{t_{j+p-1}-\mu_{j,\ell}}{\sigma_{j,\ell}})^2 }
\end{equation}

Clearly, the minimum value shown in Eq.~(\ref{eq:lowerBoundFormula}) can be set as $LB(d_{i,j}^{\ell+k})$.
We can obtain $LB(d_{i,j}^{\ell+k})$ by solving $\frac{\partial LB(d_{i,j}^{\ell+k})}{\partial \mu'}=0$ and $\frac{\partial LB(d_{i,j}^{\ell+k})}{\partial \sigma'}=0$:

\begin{equation}
	\label{eq:lowerBoundFinalFormula}
	LB(d_{i,j}^{\ell+k})=\begin{cases}
	\sqrt{{\ell}} \frac{\sigma_{j,\ell}}{\sigma_{j,\ell+k}} & \text{if $q_{i,j} \leq 0$}\\
	\sqrt{{\ell}(1- q^2_{i,j})} \frac{\sigma_{j,\ell}}{\sigma_{j,\ell+k}} & \text{otherwise}
	\end{cases}
\end{equation}

where $q_{i,j} = \frac{\sum_{p=1}^{\ell} \frac{(t_{j+p-1} t_{i+p-1})}{\ell} - \mu_{i,\ell}\mu_{j,\ell}}{\sigma_{i,\ell}\sigma_{j,\ell}}$.

$LB(d^{\ell+k}_{i,j})$ yields the minimum possible z-normalized Euclidean distance between $T_{i,\ell+k}$ and $T_{j,\ell+k}$, given $T_{i,\ell}$, $T_{j,\ell}$ and $T_{j,\ell+k}$ (but not the last $k$ values of $T_{i,\ell+k})$.
Now that we have obtained the lower bound Euclidean distance between two subsequences, we are able to introduce the lower bound distance profile.
%\vspace{0.1cm}\\
%\noindent{\bf The Ranked Lower Bound Distance Profile}

Using Eq.~(\ref{eq:lowerBoundFinalFormula}), we can evaluate the lower bound Euclidean distance between $T_{j,\ell+k}$ and every subsequence of length $\ell+k$ in $T$. By putting the results in a vector, we obtain the lower bound distance profile $LB(D_j^{\ell+k})$ corresponding to subsequence $T_{j,\ell+k}$: $LB(D_j^{\ell+k})$ = $LB(d_{1,j}^{\ell+k})$, $LB(d_{2,j}^{\ell+k})$, ...,$LB(d_{n-\ell-k+1,j}^{\ell+k})$.
If we sort the components of $LB(D_j^{\ell+k})$ in an ascending order, we can obtain the ranked lower bound distance profile: $LB_{ranked} (D_j^{\ell+k})= LB(d_{r_1,j}^{\ell+k}), LB(d_{r_2,j}^{\ell+k}),\\...,LB(d_{r_{n-\ell-k+1},j}^{\ell+k})$, where $LB(d_{r_1,j}^{\ell+k}) \le LB(d_{r_2,j}^{\ell+k}) \le ... \\\le LB(d_{r_{n-\ell-k+1},j}^{\ell+k})$.
%\vspace{0.1cm}\\

We would like to use this ranked lower bound distance profile to accelerate our computation. Assume that we have a best-so-far pair of motifs with a distance $dist_{BSF}$.
If we examine the $p^{th}$ element in the ranked lower bound distance profile and find that $LB(d_{r_p,j}^{\ell+k}) > dist_{BSF}$, then we do not need to calculate the exact distance for  $d_{r_p,j}^{\ell+k}, d_{r_{p+1},j}^{\ell+k},...,d_{r_{n-\ell-k+1},j}^{\ell+k}$ anymore, as they cannot be smaller than $dist_{BSF}$.
Based on this observation, our strategy is as follows. We set a small, fixed value for $p$. Then, for every $j$, we evaluate whether $LB(d_{r_p,j}^{\ell+k}) > dist_{BSF}$ is true: if it is, we only calculate $d_{r_1,j}^{\ell+k}, d_{r_2,j}^{\ell+k},...,d_{r_{p-1},j}^{\ell+k}$. If it is not,  we compute all the elements of $D_j^{\ell+k}$. We update $dist_{BSF}$ whenever a smaller distance value is observed. In the best case, we just need to calculate $\mathcal{O}(np)$ exact distance values to obtain the motif of length $l+k$.
Note that the order of the ranked lower bound distance profile is preserved for every $k$. That is to say, if $LB(d_{a,j}^{\ell+k}) \le LB(d_{b,j}^{\ell+k})$, then $LB(d_{a,j}^{\ell+k+1}) \le LB(d_{b,j}^{\ell+k+1})$. This is because the only component in Eq.~(\ref{eq:lowerBoundFinalFormula}) related to $k$ is $\sigma_{j,\ell+k}$. When we increase $k$ by 1, we are just performing a linear transformation for the lower bound distance: $LB(d_{i,j}^{\ell+k+1})=LB(d_{i,j}^{\ell+k}) \sigma_{j,\ell+k}/\sigma_{j,\ell+k+1}$. Therefore, we have $LB(d_{r_p,j}^{\ell+k+1})=LB(d_{r_p,j}^{\ell+k}) \sigma_{j,\ell+k}/\sigma_{j,\ell+k+1}$ , and the ranking is preserved for every $k$.

\subsection{The VALMOD Algorithm}

\begin{algorithm}[tb]
{\scriptsize 
	\KwIn{\textbf{DataSeries} $T$, \textbf{int} $\ell_{min}$ \textbf{int} $\ell_{max}$, \textbf{int} $p$}
	\KwOut{$VALMP$}
	\textbf{int} $nDP$ $\leftarrow$ $|T|-\ell_{min}+1$ \;
	$VALMP$  $\leftarrow$ \textbf{new} $VALMP$($nDP$)\;
	$VALMP.MP$ = \{$\bot$,...,$\bot$\}\;
	\textbf{MaxHeap}[] $listDP$ , \textbf{double []} $MP$, \textbf{int []} $IP$ \; 
	$listDP$, $MP$ , $IP$ $\leftarrow$ $ComputeMatrixProfile$($T$, $\ell_{min}$, $p$); \tcp{listDP contains p entries of each distance profile} \label{computeMatrixFirst}
	$VALMP$  $\leftarrow$ $updateVALMP$($VALMP$,$MP$,$IP$,$nDP$) \;
	\For{$i$ $\leftarrow$  $\ell_{min}+1$ \emph{\KwTo} $\ell_{max}$} 
	{
		\label{startForValmod}
		nDP $\leftarrow$ $|T|-i+1$ \;
		\tcp{compute SubMP and update listDP for the length i}
		\textbf{bool} bBestM, \textbf{double []} $SubMP$, $IP$ $\leftarrow$ $ComputeSubMP$($T$,$nDP$,$listDP$,$i$,$p$)\; 
		\label{ComputeSubMP} 
		\uIf{bBestM}
		{
			\tcp{SubMP surely contains the motif, update VALMP with it}
			$updateVALMP$($VALMP$,$SubMP$,$IP$,$nDP$)\;	
		}
		\Else
		{
			$listDP$,$MP$,$IP$ $\leftarrow$ $ComputeMatrixProfile$($T$,$i$,$p$)\;\label{computeMatrixSecond}
			\tcp{SubMP might not contain the motif, update VALMP computing MP}
			$updateVALMP$($VALMP$,$MP$,$IP$,$nDP$)\;
		}	
	}	\label{endForValmod}
}
\caption{$VALMOD$}
\label{algo1}
\end{algorithm}

\begin{algorithm}[tb]
	{\scriptsize 
		\KwIn{$VALMP$, \textbf{double []} $MPnew$, \textbf{int []} $IP$, $nDP$, $\ell$}
		\KwOut{$VALMP$}
		\For{$i$ $\leftarrow$ 1 \emph{\KwTo} $nDP$} 
		{
			\tcp{length normalize the Euclidean distance}
			\textbf{double} $lNormDist$ $\leftarrow$ $MPnew[i] * \sqrt{1/\ell}$\;
				\tcp{if the distance at offset i of VALMP, surely computed with previous lengths, is larger than the actual, update it}
			\If{($VALMP.distances[i] > lNormDist$ \textbf{or} $VALMP.MP[i]$ == $\bot$)}
			{
				$VALMP.distances[i]$ $\leftarrow$ $MPnew[i]$\;
				$VALMP.normDistances[i]$ $\leftarrow$ $lNormDist$\;
				$VALMP.lengths[i]$ $\leftarrow$ $\ell$\;	
				$VALMP.indices[i]$ $\leftarrow$ $IP[i]$\;	
			}
		}
	}
	\caption{$updateVALMP$}
	\label{algo1_2}
\end{algorithm}

We are now able to formally describe the VALMOD algorithm. 
The pseudocode for VALMOD is shown in Algorithm~\ref{algo1}.
With the call of $\textit{ComputeMatrixProfile()}$ in line \ref{computeMatrixFirst}, we build the matrix profile corresponding to $\ell_{min}$, and in the meantime store the smallest $p$ values of each distance profile in the memory. Note that the matrix profile is stored in the vector $MP$, which is coupled with the matrix profile index, $IP$, which is a structure containing the offsets of the nearest neighbor subsequences.  
We can easily find the motif corresponding to $\ell_{min}$ as the minimum value of $MP$. Then, in lines \ref{startForValmod}-\ref{endForValmod}, we iteratively look for the motif of every length within $\ell_{min}+$1 and $\ell_{max}$. The $ComputeSubMP$ function in line \ref{ComputeSubMP} attempts to find the motif of length $i$ only by evaluating a subset of the matrix profile corresponding to subsequence length $i$. Note that this strategy, which is based on the lower bounding technique introduced in Section \ref{sec:lowwerbounddistanceprofile}, might not be able to capture the global minimum value within the matrix profile. In case that happens (which is rare), the Boolean flag $bBestM$ is set to false, and we compute the whole matrix profile with the $computeMatrixProfile$ procedure in line \ref{computeMatrixSecond}.

The final output of $VALMOD$ is a vector, which is called  $VALMP$ (\textit{variable length matrix profile}) in the pseudo-code. If we were interested in only one fixed subsequence length, \textit{VALMP} would be the matrix profile normalized by the square root of the subsequence length. If we are processing various subsequence lengths, then as we increase the subsequence length, we update \textit{VALMP} when a smaller length-normalized Euclidean distance is observed. 

Algorithm \ref{algo1_2} shows the routine to update the $VALMP$ structure. The final $VALMP$ consists of four parts. The $i^{th}$ entry of the $normDistances$ vector stores the smallest length-normalized Euclidean distance values between the $i^{th}$ subsequence and its nearest neighbor, while the $i^{th}$ place of vector $distances$ stores their straight Euclidean distance. 
The location of each subsequence's nearest neighbor is stored in the vector $indices$. The structure $lengths$ contains the length of the $i^{th}$ subsequences pair. 

In the next two subsections, we detail the two sub-routines, $computeMatrixProfile$ and the $ComputeSubMP$.%, of $VALMOD$.

\subsection{Computing The Matrix Profile}

\begin{algorithm}[tb]
{\scriptsize 
	\KwIn{\textbf{DataSeries} $T$, \textbf{int} $\ell$, \textbf{int} $p$ }
	\KwOut{$MP$, $listDP$}
	\textbf{int} $nDP$ $\leftarrow$ $|T|$-$\ell$+1\;
	\textbf{double []} $MP$ $\leftarrow$ double [$nDP$]\;
	\textbf{int []} $IP$ $\leftarrow$ int [$nDP$]\;
	\textbf{MaxHeap[]} $listDP $= \textbf{new MaxHeap}(p)[$nDP$]\;
	\tcp{compute the dot product vector QT for the first distance profile}
	\textbf{double []} $QT$ $\leftarrow$ $SlidingDotProduct$($T_{1,\ell}$, $T$)\;\label{FFT} 
	\tcp{compute sum and squared sum of the first subsequence of length $\ell$}
	$s$ $\leftarrow$ $sum$($T_{1,\ell}$);
	$ss$ $\leftarrow$ $squaredSum$($T_{1,\ell}$)\; \label{sumSsum}
	\tcp{compute the first distance profile with distance formula (Eq.(\ref{eq:equationDistance})) and store the minimum distance in MP and the offset of the nearest neighbor in IP}
	D($T_{i,\ell}$) $\leftarrow$ $CalcDistProfile$($QT$,$T_{i,\ell}$, $T$, $s$, $ss$)\;\label{CDp} 
	$MP[1]$, $IP[1]$ $\leftarrow$ \textbf{min}(D($T_{i,\ell}$))\; 	\label{matIndProf}
	
	\tcp{iterate over the subsequences of T}
	\For{i $\leftarrow$ 2 \emph{\KwTo} $nDP$} 
	{ 
		\tcp{update the dot product vector QT for the $i^{th}$ subsequence}
		\For{j $\leftarrow$ $nDP$ \emph{\textbf{down to}} 2} 
		{
			$QT[j]$$\leftarrow$$QT[j-1] - T[j-1] \times T[i-1] + T[j+ \ell - 1] \times T[i + \ell -1]$ \label{overlapping}\; 
		}
		\tcp{update sum and squared sum of the $i^{th}$ subsequence  }
		$s$ $\leftarrow$  $s - T[i-1] + T[\ell+i-2] $\;
		$ss$ $\leftarrow$  $ss - T[i-1]^2 + T[\ell+i-2]^2 $\;		
		D($T_{i,\ell}$) $\leftarrow$ $CalcDistProfile$($QT$,$T_{i,\ell}$, $T$, $s$, $ss$)\;\label{computeDP}
		$MP[i]$, $IP[i]$ $\leftarrow$ \textbf{min}($D(T_{i,\ell}$))\;\label{matIndProfAgain}
		\tcp{Store in listDP[i] the p entries e with smallest lower bounding distance}
		\textbf{int} c $\leftarrow$ 0\;
		\For{\textbf{each entry} $e$ $in$ D($T_{i,\ell}$)} 	
		{\label{loopK}
			\tcp{Compute the lower bound for the length $\ell+1$}
			$e.LB$  $\leftarrow$ $compLB$($\ell$, $\ell+1$, $QT[c]$, $e.s1$, $e.s2$, $e.ss1$, $e.ss2$)\; \label{compLB}
			\tcp{save the entry only if is smaller than the max lb so far or if listDP[i] contains fewer than p elements}
			\If{$e.LB < max(listDP[i])$ \textbf{or} $|listDP[i]| < p$ }
			{
				$insert$($listDP[i]$, $e$)\;\label{insertSort}
			}
			c$\leftarrow$ $c+1$\;
			
		\label{loopK2}}	
	}	
	}
	\caption{\textit{$ComputeMatrixProfile$}\label{computeMatrix}}
\end{algorithm}			

The routine $ComputeMatrixProfile$ (Algorithm~\ref{computeMatrix}) computes a matrix profile for a given subsequence length, $\ell$. It essentially follows the STOMP algorithm~\citep{ZhuZSYFMBK16}, except that we also calculate the lower bound distance profiles in line \ref{loopK}. In line~\ref{FFT}, the dot product between the sequence $T_{1,\ell}$ and the others in $T$ is computed in \textit{frequency domain} in $\mathcal{O}(nlogn)$ time, where $n=|T|$. 
%Since we are considering overlapping subsequences, please note that, for the remaining sequences, 
The dot product is computed in constant time in line \ref{overlapping} by using the result of the previous overlapping subsequences.

In line~\ref{CDp} we measure each z-normalized Euclidean distance, between $T_{i,\ell}$ and the other subsequence of length $\ell$ in $T$, avoiding trivial matches. The distance measure formula used is the following~\citep{DBLP:conf/icdm/MueenHE14,YehZUBDDSMK16,ZhuZSYFMBK16}:

\begin{equation}
\label{eq:equationDistance}
dist(T_{i,\ell},T_{j,\ell}) =\sqrt{ 2\ell ( 1 - \frac{QT_{i,j} - \ell \mu_i \mu_j} {  \ell \sigma_i \sigma_j } )} %\times \sqrt{1/\ell}
\end{equation}

In Eq.~(\ref{eq:equationDistance}) \ $QT_{i,j}$ represents the dot product of the two sub-series with offset $i$ and $j$ respectively. It is important to note that, we may compute $\mu$ and $\sigma$ in constant time by using the \textit{running} plain and squared sum, namely $s$ and $ss$ (initialized in line \ref{sumSsum}). It follows that $\mu = s/\ell$  and $\sigma = \sqrt{(ss/\ell) - \mu^2}$.

In lines~\ref{matIndProf} and \ref{matIndProfAgain}, we update both the matrix profile and the matrix profile index, which holds the offset of the closest match for each $T_{i,l}$.  

Algorithm~\ref{computeMatrix} ends with the loop in line \ref{loopK}, which evaluates the lower bound distance profile and stores the $p$ smallest lower bound distance values in $listDP$. In line~\ref{compLB}, the procedure $compLB$ evaluates the lower bound distance profile introduced in Section~\ref{sec:lowwerbounddistanceprofile} using Eq.~(\ref{eq:lowerBoundFinalFormula}). The structure $listDP$ is a Max Heap with a maximum capacity of $p$.
%(inserting an element in a full heap triggers the maximum element replacement).
%Popping or getting the entry with min/max lower bounding distance is done in constant time, whereas updating them, in line \ref{insertSort} takes $O(logp)$ time. 
Each entry $e$ of the distance profile in line~\ref{loopK} is a tuple containing the Euclidean distance between a subsequence $T_{j,\ell}$ and its nearest neighbor, the location of that nearest neighbor, the lower bound Euclidean distance of the pair, the dot product of them, and the plain and squared sum of $T_{j,\ell}$.
In Figure~\ref{Example}(b), we show an example of the distance profile in line~\ref{loopK}. The distance profile is sorted according to the lower bound Euclidean distance values (shown as LB in the figure). The entries corresponding to the $p$ smallest LB values are stored in memory to be reused for longer motif lengths.

We note that this routine is called at least once, for the first subsequence length of the range, namely $\ell=\ell_{min}$.
In the worst case, it is executed for each length in the range.% (though, this never occurred in our experiments).  

\noindent{\bf Complexity Analysis.} 
In line~\ref{computeDP} of Algorithm~\ref{computeMatrix}, the time cost to compute a single distance profile is $\mathcal{O}(n)$, where $n$ is the number of subsequences of length $\ell$.
Therefore computing the $n$ distance profiles takes $\mathcal{O}(n^2)$ time. 
In line~\ref{loopK}, computing the lower bounds of the smallest $p$ entries of each distance profile takes $\mathcal{O}(n\log(p))$ additional time. 
The overall time complexity of the~$ComputeMatrixProfile$ routine is thus $\mathcal{O}(n^2\log(p))$.
%The overall time complexity of the~$ComputeMatrixProfile$ routine is $O(n^2logp)$, where $n$ is the number of the computed distance profiles.

%\commentRed{1 Reviewer out of 3 said that the flow is not smooth }
%\commentRed{correct error in Figure 8 (lost sources) }
\subsection{Matrix Profile for Subsequent Lengths}

\begin{algorithm}[tb]
{\scriptsize 
	\KwIn{\textbf{DataSeries} $T$, \textbf{int} $nDp$, \textbf{MaxHeap[]} $listDP$, \textbf{int} $newL$, \textbf{int} $p$}
	\KwOut{bBestM, $SubMP$, $IP$}
	\textbf{double[]} $SubMP$ $\leftarrow$ \textbf{double[$nDp$]}\;
	\textbf{int[]} $IP$ $\leftarrow$ \textbf{int[$nDp$]}\;
	\textbf{double} $minDistAbs$ $\leftarrow$  $\inf$, \textbf{double} $minLbAbs$ $\leftarrow$  $\inf$\;
	\textbf{List $\langle$ int,double $\rangle$ } $nonValidDP$\; 
	\tcp{iterate over the partial distance profiles in listDP}
	\For{$i$ $\leftarrow$ $1$ \emph{\KwTo} $nDp$ } 
	{\label{loopDP}
		\textbf{double} $minDist$ $\leftarrow$ $\inf$\;
		\textbf{int} $ind$ $\leftarrow$ $0$\;
		\textbf{double} $maxLB$ $\leftarrow$ \textbf{popMax}($listDP[i]$)\;
		\tcp{update the partial distance profile for the length newL (true Euclidean and lower bounding distance )}
		\For{\textbf{each entry} $e$ $in$ $listDP[i]$}
		{\label{loopsingleDP}
			$e.dist$, $e.LB$ $\leftarrow$ $updateDistAndLB(e,newL)$\; \label{constantDistLB}
			$minDist$  $\leftarrow$ \textbf{min}($minDist$,$e.dist$)\;
			\If{$minDist == e.dist$}{$ind=e.offset$\;}
		}\label{partDPAvailable}
		\tcp{check if the min (minDist) of this partial distance profile is the min of the complete distance profile}
		\uIf{$minDist < maxLB$} 
		{\label{checkCOrrectMin}
			\tcp{minDist is the real min; valid distance profile}
			$minDistABS$ $\leftarrow$ \textbf{min}($minDistAbs$,$minDist$)\; \label{updateMinDist}
			$SubMP[i]$ = $minDist$\; \label{updateMatrixProfile}
			$IP[i]$ =$ind$\;
		}
		\Else 
		{\label{nonvalid}
			\tcp{minDist is not the real min; non-valid distance profile}
			$minLbAbs$  $\leftarrow$ \textbf{min}($minLbAbs$, $maxLB$))\;\label{updateLBMin}
			$SubMP[i]$ = $\bot$\; 
			$nonValidDP.add(\langle i,maxLB \rangle )$ \label{nonValidDP}
		} 
		
	}
	\textbf{bool} $bBestM$ $\leftarrow$ ($minDistABS < minLbAbs$) \; \label{bBestMFirst}
	\tcp{if SubMP does not contain the motif distance (bBestM = false), compute the whole non-valid distance profiles, if it is faster then computeMatrixProfile (nDp / 2 = true)}
	\If{$!bBestM$ \textbf{and} $nonValidDP.size() < (\frac{n \log(p)}{\log(n)}) $}
	{\label{checkNonValidDP}
		\For{\textbf{each pair} $<ind,lbMax>$ \textbf{in} $nonValidDP$}
		{
			\If{$lbMax < minDistABS$}
			{
				$QT$ $\leftarrow$ $SlidingDotProduct$($T_{ind,\ell}$, $T$)\; \label{MASSscratch}
				\textbf{double} $s$ $\leftarrow$ $sum$($T_{ind,\ell}$);
				\textbf{double} $ss$ $\leftarrow$ $squaredSum$($T_{ind,\ell}$)\; 
				$D(T_{ind,\ell}) \leftarrow CalcDistProf$($QT$,$T_{ind,\ell}$, $T$, $s$, $ss$)\;
				$SubMP[ind]$, 	$IP[ind]$ = \textbf{min}($D(T_{ind,\ell})$)\; 
			    $insert(listDP[ind],D(T_{ind,\ell}))$\;
			}
		}
		$bBestM$ $\leftarrow$ 1\;	
	} 
	\caption{\textit{ComputeSubMP} \label{bestEffort}}
}
\end{algorithm}			

We are now ready to describe our \textit{ComputeSubMP} algorithm, which allows us to find the motifs for subsequence lengths greater than $\ell$ in linear time.

The input of \textit{ComputeSubMP}, whose pseudo-code is shown in Algorithm~\ref{bestEffort}, is the vector $listDp$ that we built in the previous step.
In line~\ref{loopDP}, we start to iterate over the $p \times n $ elements of $listDp$ in order to find the motif pair of length $newL$, using a procedure that is faster than Algorithm~\ref{algo1}, leading to a complexity that is now linear in the best case.
% (as the experiments show, this is often the case). 
Since $listDP$ potentially contains enough elements to compute the whole matrix profile, it can provide more information than just the motif pair. 

In the loop of line~\ref{loopsingleDP}, we update all the entries of $listDP[i]$ by computing the Euclidean and lower bound distance for the length $newL$. This operation is valid, since the ranking of each $listDP[i]$ is maintained as the lower bound gets updated. Moreover, this latter computation is done in constant time (line \ref{constantDistLB}), since the entries contain the statistics (i.e. sum, squared sum, dot product) for the length $newL-1$. Also note that the routine $updateDistAndLB$ avoids the trivial matches, which may result from the length increment. 

Subsequently, the algorithm checks in line~\ref{checkCOrrectMin} if $minDist$ is smaller than or equal to $maxLB$, the largest lower bound distance value in $listDP[i]$. If this is true, $minDist$ is the smallest value in the whole distance profile. In lines~\ref{updateMinDist} and~\ref{updateMatrixProfile}, we update the best-so-far distance value and the matrix profile.
%, since this minimum is an exact point of the matrix profile. 
On the other hand, we update the smallest max lower bounding distance in line~\ref{updateLBMin}, recording also that we do not have the true min for the distance profile with offset $i$ (line~\ref{nonValidDP}). Here, we may also note that even though the local true min is larger than the max lower bound (i.e., the condition of line~\ref{checkCOrrectMin} is not true), $minDist$ may still represent an approximation of the true matrix profile point.

When the iteration of the partial distance profiles ends (end of for loop in line~\ref{loopDP}), the algorithm has enough elements to know if the matrix profile computed contains the real motif pair.
In line~\ref{bBestMFirst}, we verify if the smallest Euclidean distance we computed ($minDistABS$) is less than $minLbAbs$, which is the minimum lower bound of the \emph{non-valid} distance profiles. 
We call non-valid all the partial distance profiles, for which the maximum lower bound distance (i.e., the $p$-th largest lower bound of the distance profile) is smaller than the minimum true distance (line~\ref{nonvalid}); otherwise, we call them valid (line~\ref{checkCOrrectMin}).

As a result of the ranking preservation of the lower bounding function, if the above criterion holds, we know that each true Euclidean distance in the non-valid distance profiles must be greater than $minDistABS$.
In line~\ref{checkNonValidDP}, the algorithm has its last opportunity to exploit the lower bound in the distance profiles, in order to avoid computing the whole matrix profile.
If $bBestM$ is false (the motif has not been found), we start to iterate through the non-valid distances profiles. 
% is less than half of the total distance profiles.
We perform this iteration, when their number is not larger than $\frac{n \log(p)}{\log(n)}$. This condition guarantees that Algorithm~\ref{bestEffort} is faster than Algorithm~\ref{computeMatrix}.

We present here two examples that explain the main procedures of $VALMOD$.  

\begin{figure}[tb]
	\centering
	\includegraphics[trim={3cm 6cm 0cm 6cm}, scale=0.42]{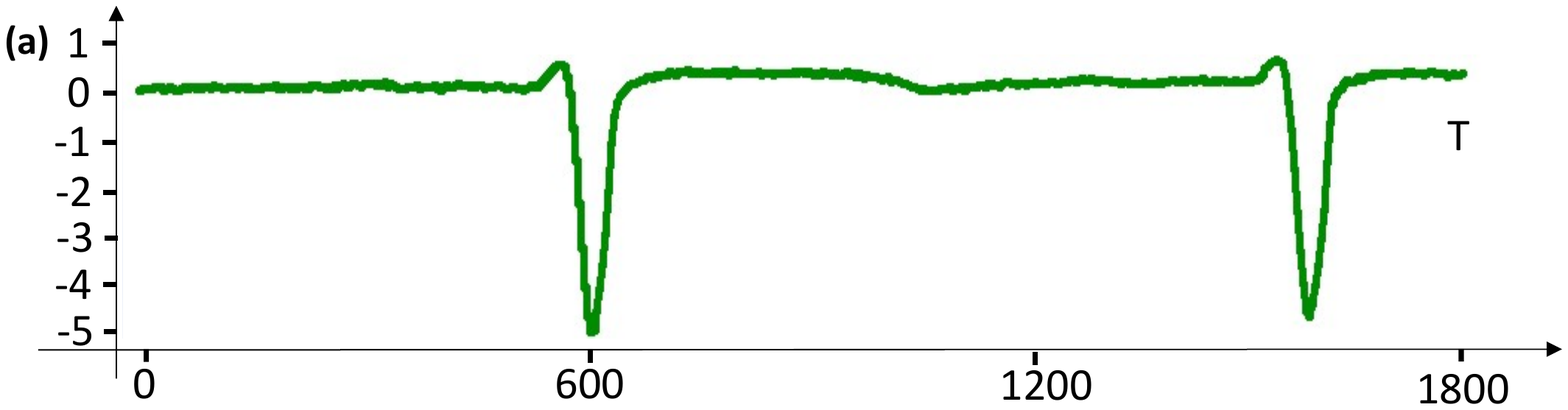}
	\includegraphics[trim={3cm 6cm 0cm 9cm}, scale=0.42]{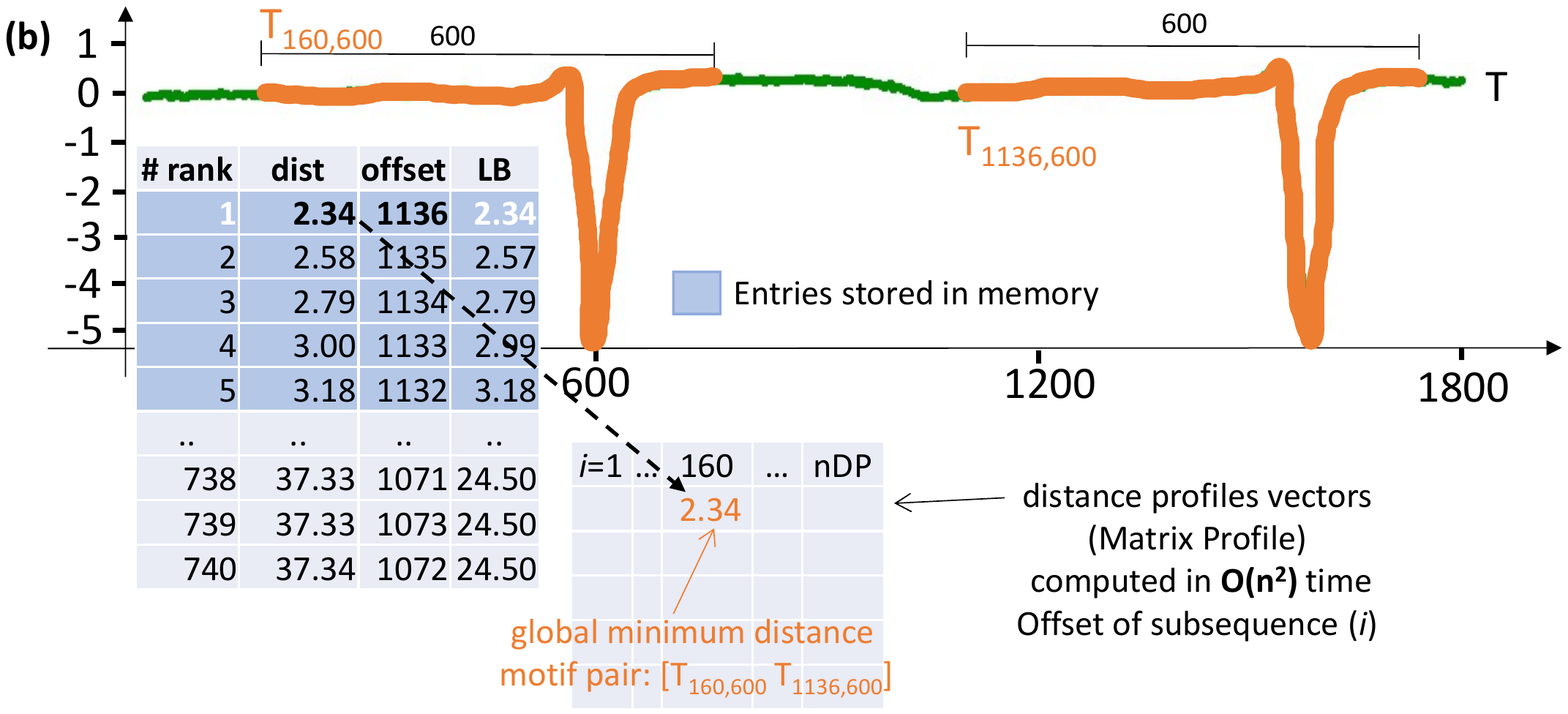}
	\caption{(\textit{a}) Input time series, (\textit{b}) Compute matrix profile snapshot: (on the left) distance profile of the subsequence $T_{160,600}$ which is part of the motif. }
	\label{Example} 
\end{figure} 

\begin{example}
In Figure~\ref{Example}, we show a snapshot of a VALMOD run. 
In Figure~\ref{Example}(a), VALMOD receives as input a data series of length 1800. 
In Figure~\ref{Example}(b), the matrix profile for subsequence length $\ell = 600$ is computed (Algorithm~\ref{computeMatrix}). 
On the left, we depict the distance profile regarding $T_{160,600}$, and rank it according to the lower bound (LB) distance values. 
Although we are computing the entire distance profile, we store only the first $p=5$ entries in memory.
\end{example}

\begin{figure}[tb]
	\centering
	\includegraphics[trim={3cm 7cm 0cm 6cm},scale=0.42]{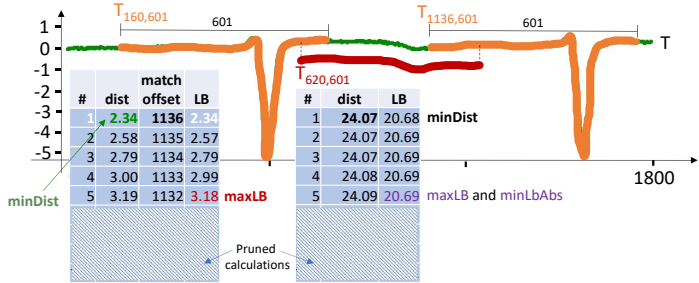}
	\caption{Compute Sub Matrix profile: the partial distance profile of $T_{160,601}$ contains the motif's subsequences distance.}
	\label{Example2}
	\vspace*{-0.5cm}
\end{figure}

\begin{example}
Figure~\ref{Example2} shows the execution of $ComputeSubMP$ (Algorithm~\ref{bestEffort}), taking place after the step illustrated in Figure~\ref{Example}(b). 
In this picture, we show the distance profile of a subsequence belonging to the motif pair, for subsequence length $\ell=601$. 
This time it is built by computing $p=5$ distances (left side of the picture).
We can now make the following observations:\\
(a) In the distance profile of the subsequence $T_{160,601}$ (left array): $minDist = 2.34 < maxLB = 3.18$ $\iff$ the value $2.34$ is both a local and a global minimum (among all the distance profiles).\\
(b) Considering the partial distance profile of subsequence $T_{620,601}$ (right array), we do not know if its $minDist$ is its real global minimum, since 20.69 ($maxLB$) $<$ 24.07 ($minDist$).\\
(c) We know, that 20.69 ($maxLB$ of the distance profile of subsequence $T_{620,601}$) is the $minLbAbs$, or in other words, the smallest $maxLB$ distance among all the partial distance profiles in which $maxLB<minDist$ holds.\\
(d) We know that there are no true Euclidean distances (among those computed) smaller than $2.34$. Since  $minDist = 2.34 < minLbAbs = 20.69$ $\iff$ 2.34 is the distance of the motif $\{T_{160,601}; T_{1136,601}\}$.
\end{example}

%\noindent{\bf Complexity Analysis.} 
%\commentBlue{In the best case, $ComputeSubMP$ can find the motif pair in $O(np)$ time, where $n$, is the total number of distance profiles.
%This means that, no distance profile computation takes place, since the condition in line~\ref{bBestMFirst} is satisfied.
%Otherwise, if we need to iterate over the non-valid distances profiles for finding the answer (which occurs rarely in practice), the time complexity reaches its worst case, $O(nClogn)$, with $C=n/p$.
%This is asymptotically faster than re-executing \textit{ComputeMatrixProfile}, which takes $O(n^2logp)$ time.
%
%\noindent{\bf Complexity Analysis.} 
%In the best case, $ComputeSubMP$ takes $O(np)$ time, whereas the worst case time complexity is $O(nClogn)$, where $n$ is the number of the computed distance profiles, and $C=n/p$.

\noindent{\bf Complexity Analysis.} In the best case, $ComputeSubMP$ can find the motif pair in $\mathcal{O}(np)$ time, where $n$ is the total number of distance profiles.
This means that no distance profile computation takes place, since the condition in line~\ref{bBestMFirst} of Algorithm~\ref{algo1_3} is satisfied.
Otherwise, if we need to iterate over the non-valid distance profiles for finding the answer,
%(which occurs rarely in practice), 
the time complexity reaches its worst case, $\mathcal{O}(nC\log(n))$, with $C$ denoting the number of non-valid distance profiles that are recomputed. When $C < \frac{n \log(p)}{\log(n)}$, the algorithm is asymptotically faster than re-executing \textit{ComputeMatrixProfile}, which takes $\mathcal{O}(n^2\log(p))$ time.

Note that, each non-valid distance profile (starting in line~\ref{MASSscratch}) is computed by using the primitives introduced in the \textit{ComputeMatrixProfile} algorithm, only if its maximum lower bound is less than the smallest true distance $minDistABS$. 
This indicates that the distance profile for length $newL$ may contain not yet computed distances smaller than $minDistABS$, which is our \textit{best-so-far}.
Therefore, the overall complexity of VALMOD is $\mathcal{O}(n^2\log(p) + (\ell_{max}-\ell_{min})np)$ in the best case, whereas the worst case time complexity is $\mathcal{O}((\ell_{max}-\ell_{min})n^2\log(p))$. 
Clearly, the $n^2\log(p)$ factor dominates, since ($\ell_{max}-\ell_{min}$) acts as a constant. 
%Nevertheless, the length range is not negligible, w.r.t the time performance, when we need to run a quadratic routine over it. 
%If the worst case occurs often, then the performance will degrade. 
%However, this is not the case, as we show in the experimental evaluation.

\section{Finding Motif Sets}
\label{sec:motifsets}

We finally extend our technique in order to find the variable-length motif sets.
In that regard, we start to consider the $top$-$k$ motif pairs, namely the pairs having the $k$ smallest length-normalized distances. The idea is to extend each motif pair to a motif set considering the subsequence's proximity as a quality measure, thus favoring the motif sets, which contain the closest subsequence pairs. 
Moreover, for each top-K motif pair ($T_{a,\ell}$,$T_{b,\ell}$), we use a radius $r = D * dist(T_{a,\ell},T_{b,\ell})$, when we extend it to a motif set. 
We call the real variable $D$ \textit{radius factor}. 
This choice permits us to tune the radius $r$ by the user defined radius factor, considering also the characteristics of the data. 
Setting a unique and non data dependent radius for all motif sets, would penalize the results of exploratory analysis.

First, we introduce Algorithm~\ref{algo1_3}, a slightly modified version of the $updateValmp$ routine (Algorithm~\ref{algo1_2}). 
The new algorithm is called $updateVALMPForMotifSets$, and its main goal is to keep track of the best $k$ subsequence pairs (motif pairs) according to the $VALMP$ ranking, and the corresponding partial distance profiles. 
The idea is to later exploit the lower bounding distances for pruning computations, while computing the motif sets. 

In lines~\ref{startPair} to~\ref{endPair}, we build a structure named $pair$, which carries the information of the subsequences pairs that appear in the $VALMP$ structure. 
During this iteration, we leave the fields $partDP1$ and $partDP2$ empty, since they will be later initialized with the partial distance profiles, if their $pair$ is in the top $k$ of $VALMP$. 
In order to enumerate the best $k$ pairs, we use the global maximum heap $heapBestKPairs$ in line~\ref{inserPairHeap}. 
Then, we assign (or update) the corresponding partial distance profiles (line~\ref{updateDistProfPartial}) to each pair.

\begin{algorithm}[tb]
	{\scriptsize 
	
	\KwIn{$VALMP$, \textbf{double []} $MPnew$, \textbf{int []} $IP$, $nDP$, $\ell$, \textbf{MaxHeap[]} $listDP$,}
	\KwOut{$VALMP$}
	\For{$i$ $\leftarrow$ 1 \emph{\KwTo} $nDP$} 
	{
		\tcp{length normalize the Euclidean distance}
		\textbf{double} $lNormDist$ $\leftarrow$ $MPnew[i] * \sqrt{1/\ell}$\;
		\tcp{if the distance at offset i of VALMP, surely computed with previous lengths, is larger than the actual, update it}
		\If{($VALMP.distances[i] > lNormDist$ \textbf{or} $VALMP.MP[i]$ == $\bot$)}
		{
			\textbf{entry} $pair$\;\label{startPair}
			$pair$.$off1$ $\leftarrow$ i,  $pair$.$off2$ $\leftarrow$ $IP[i]$ \; 
			$pair$.$distance$ $\leftarrow$ $MPnew[i]$, $pair$.$\ell$ $\leftarrow$ $\ell$\;  
			$pair$.$partDP1$ $\leftarrow$  $\bot$, $pair$.$partDP2$ $\leftarrow$ $\bot$ \; \label{endPair}
			$insert(heapBestKPairs,pair)$\;\label{inserPairHeap}
			$VALMP.distances[i]$ $\leftarrow$ $MPnew[i]$\;
			$VALMP.normDistances[i]$ $\leftarrow$ $lNormDist$\;
			$VALMP.lengths[i]$ $\leftarrow$ $\ell$\;	
			$VALMP.indices[i]$ $\leftarrow$ $IP[i]$\;	
		}
	}
	\For{\textbf{each} $pair$ \textbf{in} $heapBestKPairs$}
	{\label{updateDistProfPartial}
		\If{($pair.partDP1$== $\bot$)}
		{
			$pair$.$partDP1$ $\leftarrow$ $listDP[pair$.$off1]$\;
			$pair$.$partDP2$ $\leftarrow$ $listDP[pair$.$off2]$\;
		}
	} 
	\caption{$updateVALMPForMotifSets$}
	\label{algo1_3}
	}
\end{algorithm}

We are now ready to present the variable length motif sets discovery algorithm (refer to Algorithm~\ref{computeVLMotifSets}). 
Starting at line~\ref{iteratePairs}, the algorithm iterates over the best pairs. 
For each one of those, we need to check if the search range is smaller than the maximum lower bound distances of both partial distance profiles. 
If this is true, we are guaranteed to have already computed all the subsequences in the range.
Therefore, in lines~\ref{sortToVector} and~\ref{sortToVector2} we filter the subsequences in the range, sorting the partial distance profile according to the offsets. 
This operation will permit us to find the trivial matches in linear time. 

\begin{algorithm}[tb]
	{
		\scriptsize 
		\KwIn{\textbf{DataSeries} $T$, \textbf{MaxHeap} $heapBestKPairs$, \textbf{double} $D$ }%, \textbf{int} $\ell$ } %\textbf{int} $F$,
		\KwOut{\textbf{Set} $S^{*}$}
		\For{\textbf{each} $pair$ \textbf{in} $heapBestKPairs$}
		{\label{iteratePairs}
			
			\textbf{double} $r$ $\leftarrow$ $pair$.$distance$ * $D$ \;
			\textbf{double} $maxLB1$ $\leftarrow$ \textbf{popMax}($pair$.$partDP1$)\;
			\textbf{double} $maxLB2$ $\leftarrow$ \textbf{popMax}($pair$.$partDP2$)\;
			$D(T_{pair.off1,pair.\ell})$ $\leftarrow$ $\emptyset$, $D(T_{pair.off2,pair.\ell})$ $\leftarrow$ $\emptyset$ \;
			\uIf{$maxLB1 > r$} 
			{\label{checkLBRangeDP1}
				\tcp{sort according  the offset, the partial distance profile contains all the elements in the range}
				$D(T_{pair.off1,pair.\ell})$ $\leftarrow$ $sortAndFilterRange$($r$,$pair$.$partDP1$.toVector());\label{sortToVector}
			}
			\Else
			{
				\tcp{re-compute the mat}
				\textbf{double} $s$ $\leftarrow$ $sum$($T_{ind,\ell}$)\;\label{computeDPfromScratch1}
				\textbf{double} $ss$ $\leftarrow$ $squaredSum$($T_{ind,\ell}$)\; 
				$D(T_{pair.off1,pair.\ell}) \leftarrow CalcDistProfInRange$($r$,$QT$,$T_{pair.off1,pair.\ell}$, $T$, $s$, $ss$)\;\label{computeDPfromScratch2}
			}
			\uIf{$maxLB2 > r$} 
			{\label{checkLBRangeDP2}
				$D(T_{pair.off2,\ell})$ $\leftarrow$ $sortAndFilterRange$($r$,$pair$.$partDP2$.toVector());\label{sortToVector2}
			}
			\Else
			{
				\textbf{double} $s$ $\leftarrow$ $sum$($T_{ind,\ell}$)\;\label{computeDPfromScratch1_2}
				\textbf{double} $ss$ $\leftarrow$ $squaredSum$($T_{ind,\ell}$)\; 
				$D(T_{pair.off2,pair.\ell}) \leftarrow CalcDistProfInRange$($r$,$QT$,$T_{pair.off1,pair.\ell}$, $T$, $s$, $ss$)\;\label{computeDPfromScratch2_2}
			}
		
			\textbf{Set} $S^{pair.\ell}_{r}$ $\leftarrow$ $mergeRemoveTM(D(T_{pair.off1,\ell}),D(T_{pair.off2,\ell}))$\;\label{computeMotifSet}
			%\If{$|motifSet| \ge F $}
			%{
				$S^{*}.add(S^{pair.\ell}_{r})$\; \label{addValidMotifSet}
			%}
		}
	}
	
	\caption{$computeVarLengthMotifSets$}
	\label{computeVLMotifSets}
\end{algorithm}

On the other hand, if the search range is larger than the maximum lower bound distances of both partial distance profiles, we have to re-compute the entire distance profile (lines~\ref{computeDPfromScratch2} and~\ref{computeDPfromScratch2_2}), to find all the subsequences in the range.
Once we have the distance profile pairs, we need to merge them and remove the trivial matches (line~\ref{computeMotifSet}). 
%If the resulting set size is greater or equal than the desired frequency, it is a valid motif set (line~\ref{addValidMotifSet}). 
Each time we add a subsequence in a motif set, we remove it from the search space: this guarantees the empty intersection among the sets in $S^*$.

\noindent{\bf Complexity Analysis.}
The complexity of the $updateVALMPForMotifSets$ algorithm is $\mathcal{O}(n\log(k))$, where $n$ is the length of the $VALMP$ structure, which is linearly scanned and updated. 
$\mathcal{O}(\log(k))$ time is needed to  
retain the $k$ best pairs of $VALMP$, using the heap structure in line~\ref{inserPairHeap}.
The final algorithm $computeVarLengthMotifSets$ takes $\mathcal{O}(k \times p \times \log(p))$ time, in the best case. 
This occurs when, after iterating the $k$ pairs in $heapBestKPairs$, each partial distance profile of length $p$, contains all the elements in the range $r$. 
In this case, we just need an extra $\mathcal{O}(p\log(p))$ time to sort its elements (line~\ref{sortToVector} and~\ref{sortToVector2}).  
On the other hand, the worst case time is bounded by $\mathcal{O}(k \times n \times \log(n))$, where $n$ is the length of the input data series $T$. 
In this case, the algorithm needs to recompute $k$ times the entire distance profile (line~\ref{computeDPfromScratch2} and~\ref{computeDPfromScratch2_2}), at a unit cost of $\mathcal{O}(n\log(n))$ time.

\section{Discord Discovery}

We now describe our approach to solving the Variable-Length \topkmd Discord Discovery problem. 
First, we explain some useful notions, and we then present our discord discovery algorithm.

\subsection{Comparing Discords of Different Lengths}

Before introducing the algorithm that identifies discords (from the $Top$-$1$ $1^{st}$ to the \topkmd one), we define the data structure that allows us to accommodate them. 
We can represent this structure as a $k \times m$ matrix, which contains the best match distance and the offset of each discord. 

More formally, given a data series $T$, and a subsequence length $\ell$ we define:
$dkm_{\ell} = \begin{bmatrix}
\langle d,o \rangle_{1,1} & .. & \langle d,o \rangle_{1,m} \\
.. & .. & ..\\
\langle d,o \rangle_{k,1} & .. &\langle d,o \rangle_{k,m}
\end{bmatrix}$, where a generic pair $\langle d,o \rangle_{i,j}$ contains the offset $o$ and the corresponding distance $d$ of the $Top$-$i$ $j^{th}$ discord of length $\ell$ ($1 \le i \le k$ and $ 1 \le j \le m$).
In $dkm_{\ell}$, rows rank the discords according to their positions ($m^{th}$ discords), and the columns according to their best match distance ($Top$-$k$).
For each pair $\langle d,o \rangle_{a,b}$,  $\langle d',o' \rangle_{a',b'} \in dkm_{\ell}$, we require that $T_{o,\ell}$ and $T_{o',\ell}$ are not trivial matches. 
	
Since we want to compute $dkm_{\ell}$ for each length in the range $[\ell_{min},\ell_{max}]$, we also need to rank discords of different lengths. In that regard, we want to obtain a unique matrix that we denote by $dkm_{\ell_{min},\ell_{max}}$.
Therefore, we can represent a discord  by the triple $\langle d^*,o^*,\ell^* \rangle_{i,j}$ $\in$ $dkm_{\ell_{min},\ell_{max}}$, where $d^*$ is the $i^{th}$ greatest length normalized $j^{th}$ best match distance. More formally: $$d^* = \max\{\frac{d}{\sqrt{\ell{min}}} : d \in dkm_{\ell_{min}}[i][j]),..., \frac{d}{\sqrt{\ell{max}}} : d \in dkm_{\ell_{max}}[i][j] \}$$ Each triple is also composed by the offset $o^*$ and the length $\ell^{*}$ of the discord, where $\ell_{min}\le \ell^{*}\le \ell_{max}$.

As in the case of motifs discovery, we length-normalize the discord distances, while constructing the $dkm_{\ell_{min},\ell_{max}}$ ranking. 
Thus, we multiply each distance by the $1/\sqrt{\ell}$ factor.
In this case, the length normalization aims to favor the selection of shorter discords.
Therefore, if we compare two \topkmd discord subsequences of different lengths, but equal best match distances, the shorter subsequence is the one with the highest point-to-point dissimilarity to its best match.
This is guaranteed by dividing each distance by the discord length.
Consequently, we promote the shorter subsequence as the more anomalous one.

%By multiplying by the $1/\sqrt{\ell}$ ratio each distance, we want to favor the selection of shorter discords.
%This strategy is based on the following fact: if we compare two \topkmd discord subsequences of different lengths, but equal best match distances, the shorter subsequence is the one with the highest point-to-point dissimilarity to its best match.

\subsection{Discord Discovery Algorithm}

We now describe our algorithm for the \topkmd discords discovery problem.
We note that we can still use the lower bound distance measure, as in the motif discovery case.
This allows us to efficiently build $dkm_{\ell}$, for each $\ell$ in the $[\ell_{min},\ell_{max}]$ range, incrementally reusing the distances computation performed. The final outcome of this procedure is the  $dkm_{\ell_{min},\ell_{max}}$ matrix, which contains the variable length discord ranking.
In this part, we introduce and explain the algorithms, which permit us to efficiently obtain $dkm_{\ell}$ for each length.
We report the whole procedure in Algorithm~\ref{computeDkmAlllength}.

	\begin{algorithm}[tb]
		\KwIn{\textbf{DataSeries} $T$, \textbf{int} $\ell_{min}$, \textbf{int} $\ell_{max}$, \textbf{int} $k$, \textbf{int} $m$ , \textbf{int} $p$}
		\KwOut{\textbf{Matrix} $dkm_{\ell_{min},\ell_{max}}$}	
		\textbf{MaxHeap[]} $listDP$=$ComputeMatrixProfile(T,\ell_{min},p)$\; \label{computeMatrixProfileLmin}
		\textbf{int} $nDp$ = $(|T|-\ell_{min})+1$\;
		\textbf{Matrix} $dkm_{\ell_{min},\ell_{max}}=\{ \{\langle -\infty,-\infty,-\infty \rangle,...,\langle -\infty,-\infty ,-\infty \rangle \},..., \{...\} \}$\;
		\textbf{Matrix} $dkm_{\ell_{min}}=\{ \{\langle -\infty,-\infty \rangle,...,\langle -\infty,-\infty \rangle \},..., \{...\} \}$\;
		\If{$p>=m$}
		{
		\tcp{iterate the partial distance profiles in listDP\\and compute $dkm_{\ell_{min}}$}
		\For{$i$ $\leftarrow$ $1$ \emph{\KwTo} $nDp$ } 
		{\label{iteratePartialDistanceProfile}
			\uIf {$T_{i,\ell_{min}}$ has no Trivial matches in $dkm_{\ell_{min}}$}
			{
				$UpdateFixedLengthDiscords$($dkm_{\ell_{min}}$, $listDP[i]$,$i$,$k$,$m$ )\;\label{updateRankingSingleLength}
			}	
		}
		$UpdateVariableLengthDiscords$($dkm_{\ell_{min}}$, $dkm_{\ell_{min},\ell_{max}}$,$k$,$m$ )\; \label{updateVariableLengthDiscords}
		\tcp{compute $dkm_{\ell_{nextL}}$ for each length, pruning distance computations}
		\For{$nextL$ $\leftarrow$ $\ell_{min}+1$ \emph{\KwTo} $\ell_{max}$ } 
		{\label{iteratenexLengths}
			\textbf{ \textbf{Matrix} $dkm_{nextL}=\{ \{\langle -\infty,-\infty \rangle,...,\langle -\infty,-\infty \rangle \},..., \{...\}\}$\;}
			$nDp$ = $(|T|-nextL)+1$\;
			$dkm_{\ell_{nextL}}$=$Topkm$\_$nextLength$( $T$,$nDp$,$listDP$,$nextL$,$k$,$m$)\; \label{nextLengthDiscord}
			$UpdateVariableLengthDiscords$($dkm_{\ell_{nextL}}$, $dkm_{\ell_{min},\ell_{max}}$,$k$,$m$)\;
		}
		}		 
		\caption{$Topkm$\_$DiscordDiscovery$ (\textit{Compute \topkmd Discords of variable lengths}) \label{computeDkmAlllength}}
	\end{algorithm}	
	
	\noindent{\bf Smallest Length Discords.} We start to find discords of length $\ell_{min}$, namely the smallest subsequence length in the range. We can thus run Algorithm~\ref{computeMatrix} in line~\ref{computeMatrixProfileLmin}, which computes the list of partial distance profiles of each subsequence of length $\ell_{min}$ ($listDP$), in the input data series $T$. 
	Each partial distance profile contains the $p$ smallest nearest neighbor distances of each subsequence. To that extent, we set $p \ge m$ in Algorithm~\ref{computeMatrix} ($ComputeMatrixProfile$).
	
	We then iterate the subsequences of $T$ in line~\ref{iteratePartialDistanceProfile}, using the index $i$.
	For each subsequence $T_{i,\ell{min}}$ that has no trivial matches in $dkm_{\ell_{min}}$, we invoke the routine $UpdateFixedLengthDiscords$ (line~\ref{updateRankingSingleLength}), which checks if $T_{i,\ell{min}}$ can be placed in $dkm_{\ell_{min}}$ as a discord.
	When $dkm_{\ell_{min}}$ is built, we update the variable length discords ranking ($dkm_{\ell_{min},\ell_{max}}$ matrix in line~\ref{updateVariableLengthDiscords}), using the procedure $UpdateVariableLengthDiscords$.
	
	In the loop of line~\ref{iteratenexLengths}, we iterate the discord lengths greater than $\ell_{min}$. Since we want to prune the search space, we consider the list of distance profiles in $listDP$, which also contains the lower bound distances of the $p$ ($p>m$) nearest neighbors of each subsequence. In that regard, we invoke the routine $Topkm$\_$nextLength$ (line~\ref{nextLengthDiscord}).
	Before we introduce the details, we describe the two routines we introduced, which allow to rank the discords.

	\noindent{\bf Ranking Fixed Length Discords.} In algorithm~\ref{algoUpdateKMDiscords}, we report the pseudo-code of the routine $UpdateFixedLengthDiscords$.
	This algorithm accepts as input the matrix $dkm_{\ell}$ to update, and a partial distance profile of the subsequence with offset $off$.
	It starts iterating the rows of $dkm_{\ell_{min}}$ in reverse order (line~\ref{iterate_rows}). 
	This is equivalent to considering the discords from the $m^{th}$ one to the $1^{st}$. 
	Hence, at each iteration we get the $j^{th}$ nearest neighbor of $T_{off,\ell_{min}}$ from its partial distance profile in line~\ref{getJth}.
	Subsequently, the loop in line~\ref{iterate_column} checks if the $j^{th}dist$ is among the $k$ largest ones in the $j^{th}$ column of $dkm_{\ell_{min}}$. If it is true, the smallest elements in the column are shifted (line~\ref{shift_ranking}) and $T_{off,\ell_{min}}$ is inserted as the $Top$-$i$ $j^{th}$ discord (line~\ref{udpateRanking}).

	\begin{algorithm}[tb]
		\KwIn{\textbf{Matrix} $dkm_{\ell}$, \textbf{MaxHeap} $minMDist$, \textbf{int} $off$, \textbf{int} $k$, \textbf{int} $m$}
		\For{$j$ $\leftarrow$ $m$ \emph{\textbf{down to}} $1$}
		{\label{iterate_rows}
			\textbf{double} $j^{th}dist$ $\leftarrow$ $minMDist.getMax(j)$\;  \label{getJth}
			\For{$i$ $\leftarrow$ $1$ \emph{\textbf{\KwTo}} $k$}
			{\label{iterate_column}
				$<d,o>_{i,j}=dkm_{newL}[i][j]$\;
				\uIf{ $j^{th}dist > d$} 
				{\label{updateDiscord}
					$shiftRankingTopK$($dkm_{newL}[i][j]$)\label{shift_ranking}\;
					\tcp{update the ranking with the new $Top$-$i$ $j^{th}$ discord $T_{off,\ell}$}
					$dkm_{\ell}[i][j]$ $\leftarrow$ $\langle j^{th}dist,off\rangle$\label{udpateRanking}\;
					\textbf{return};\
				}
			}
		}
		%	}	
		\caption{$UpdateFixedLengthDiscords$ (\textit{Update $dkm_{\ell}$)} \label{algoUpdateKMDiscords}}
		
	\end{algorithm}
	
	\noindent{\bf Ranking Variable Length Discords.} Once we dispose of the matrix $dkm_{\ell}$, we can invoke the procedure $UpdateVariableLengthDiscords$ for each length $\ell \in \{\ell_{min},...,\ell_{max}\}$ (Algorithm~\ref{algoUpdateVLKMDiscords}), in order to incrementally produce the final variable length discord ranking we store in $dkm_{\ell_{min},\ell_{max}}$.
	This algorithm accepts as input and iterates over the matrix $dkm_{\ell_{min},\ell_{max}}$. A position (discord) is updated if the length normalized best match distance of the discord in the same position of $dkm_{\ell}$ is larger (line~\ref{updatePosition}).
	
	\begin{algorithm}[tb]
		\KwIn{\textbf{Matrix} $dkm_{\ell_{min},\ell_{max}}$, \textbf{Matrix} $dkm_{\ell}$, \textbf{int} $k$, \textbf{int} $m$}

		\For{$i$ $\leftarrow$ $1$ \emph{\textbf{\KwTo}} $k$}
		{\label{iterate_columnVarLength}
			\For{$j$ $\leftarrow$ $1$ \emph{\textbf{\KwTo}} $m$}
			{\label{iterate_rowsVarLength}
				$<d,o>_{i,j}=dkm_{newL}[i][j]$\;
				$<d^*,o^*,l^*>_{i,j}=dkm_{\ell_{min},\ell_{max}}[i][j]$\;
				\tcp{if length normalized distance is greater or equal for length $\ell$, update the rank. }
				\uIf{($(d  / \sqrt{\ell})$ $>=$ $d^*$ )}
				{
					$dkm_{\ell_{min},\ell_{max}}[i][j] =$ $\langle (d  / \sqrt{\ell}), o, \ell \rangle$ \label{updatePosition}
				}	
			}
		}

		\caption{$UpdateVariableLengthDiscords$ (\textit{Update $dkm_{\ell_{min},\ell_{max}}$)} \label{algoUpdateVLKMDiscords}}
	\end{algorithm}

	%	Subsequently, we aim at pruning the search space for the discords of length $newL>\ell_{min}$. Hence, we consider the partial distance profiles stored in the trailing part of Algorithm~\ref{computeMatrix}. Observe that the number of elements stored in the partial distance profile must be at least $m$, since to enumerate \topkmd discords we need to know at least the $m$ largest nearest neighbors for each subsequence.
	%	
	%	Algorithm~\ref{algoUpdateKMDiscords} shows the pseudo-code of the routine, which takes as input a partial distance profile $minMDist$ that contains the distances to the $m^{th}$ nearest neighbor of a generic subsequence $T_{off,newL}$, and updates the $dkm_{newL}$ matrix.

	%As a first step, this routine checks if $T_{i,\ell}$, namely the potential discord, has a trivial match in $dkm_{\ell}$ (line~\ref{checkTMUpdate}). If this is verified, the update does not take place. Otherwise 
	%	
	%	The procedure starts to iterate $dkm_{\ell_{min}}$ from the $m^{th}$ column in line~\ref{iterate_rows}, since the algorithm starts in turn to pop the discords with the largest $m^{th}$ distances from the partial distance profile (line~\ref{updatePopMth}). When iterating a matrix column (loop of line~\ref{iterate_column}), if a distance is larger than an element of the column (line~\ref{updateDiscord}), the other discords in the column with smaller distances are shifted in the lower positions (line~\ref{shift_ranking}), and the update takes place (line~\ref{udpateRanking}). In this manner, the \topkmd ranking for each column is preserved. 
	%	

	\noindent{\bf Greater Length Discords.} In Algorithm~\ref{computeKMDiscordAlgo}, we show the pseudo-code of the routine $Topkm$\_$nextLength$. It starts performing the same loop of line~\ref{ComputeSubMP} in Algorithm~\ref{algo1}, iterating over the partial distance profiles (line~\ref{loopDPDiscord}), and updating the true Euclidean distances for the new length ($newL$) and the lower bounds (line~\ref{constantDistLB_discord}) for the subsequent length ($newL+1$). 
	Since we need to know the distances from each subsequence to their $m$ nearest neighbors, for each subsequence $T_{i,newL}$ that does not have trivial matches in $dkm_{newL}$, we check if the $m^{th}$ smallest distance is smaller than the maximum lower bound in the partial distance profile (line~\ref{checkMExact}). If this is true, we have the guarantee that the partial distance profile $minMDist$ contains the exact $m$ nearest neighbor Euclidean distances.
	Hence, in line~\ref{updateRankingDiscord}, we can update the matrix $dkm_{newL}$.
	On the other hand, if the distances are not verified to be correct, we keep $minMDist$ in memory, which becomes a non-valid partial distance profile, along with the offset of the corresponding subsequence (line~\ref{nonValidminMDist}).
	Once we have considered all the partial distance profiles, we need to iterate the non-valid partial distance profiles (line~\ref{checknonValidminMDist}).
	
	We therefore recompute those that contain at least one true Euclidean distance greater than the distances in the last row of $dkm_{newL}$. The correctness of this choice is guaranteed by the fact that the distances of a non-valid partial distance profile can be only larger than the non-computed ones.
	Hence, if the condition of line~\ref{checkOnlyGreater} is not verified, no updates in $dkm_{newL}$ can take place.  
	Otherwise, we recompute the non-valid distance profile starting at line~\ref{MASSscratch_discords} from scratch. Note that when we re-compute a distance profile, we globally update the corresponding position of the partial distance profiles $listDP$ (line~\ref{updatelistDP}) and $dkm_{newL}$ in the vector as well (line~\ref{updateDiscFormScratch}).

	\begin{algorithm}[tb]
		{\scriptsize
			{\KwIn{\textbf{DataSeries} $T$, \textbf{int} $nDp$, \textbf{MaxHeap[]} $listDP$, \textbf{int} $newL$, \textbf{int} $k$, \textbf{int} $m$, \textbf{int} $p$}
				\KwOut{\textbf{Matrix} $dkm_{newL}$}	
				\textbf{Matrix} $dkm_{newL}=\{ \{\langle -\infty,-\infty \rangle,...,\langle -\infty,-\infty \rangle \},..., \{...\} \}$\;
				\textbf{List $\langle$\textbf{MaxHeap,int}$\rangle$ } $nonValidMindistList$\; 
				\tcp{iterate over the partial distance profiles in listDP}
				\For{$i$ $\leftarrow$ $1$ \emph{\KwTo} $nDp$ } 
				{\label{loopDPDiscord}
					\textbf{MaxHeap} $minMDist$ $\leftarrow$ \textbf{new MaxHeap}($p$)\;
					\textbf{double} $minDist$ $\leftarrow$ $\inf$\;
					\textbf{int} $ind$ $\leftarrow$ $0$\;
					\textbf{double} $maxLB$ $\leftarrow$ \textbf{popMax}($listDP[i]$)\;
					\tcc{update the partial distance profile for the length newL (true Euclidean and lower bounding distance )}
					\For{\textbf{each entry} $e$ $in$ $listDP[i]$}
					{\label{loopsingleDP_discord}
						$e.dist$, $e.LB$ $\leftarrow$ $updateDistAndLB(e,newL)$\; \label{constantDistLB_discord}
						%$minDist$  $\leftarrow$ \textbf{min}($minDist$,$e.dist$))\;
						\tcp{the m shortest neighbor distances are stored in minMDist}
						$minMDist$.push($e.dist$)\;
						%\If{$minDist == e.dist$}{$ind=e.offset$\;}
					}
					\tcp{check if the $m^{th}$ shortest distance of this partial distance profile is the true $m^{th}$ shorthest.}
					$mDist$ = $minMDist.getMax(m)$\;
					\uIf {$T_{i,newL}$ has no Trivial matches in $dkm_{newL}$}
					{
						\uIf{$mDist < maxLB$} 
						{\label{checkMExact}
							\tcc{the discord ranking can be updated, without computing the whole distance profile }
							$UpdateFixedLengthDiscords$($dkm_{newL},minMDist$,$i$,$k$,$m$)\label{updateRankingDiscord}\;
						}
						\Else 
						{
							\label{nonvalid_discords}
							\tcc{minMDist might not be exact, store the partial distance profile in memory. }
							$nonValidMindistList$.add($<minMDist$,$i>$)\; \label{nonValidminMDist}
						}
					}
				}
				\For{\textbf{each} $<minMDist$, $i>$ \textbf{in} $nonValidMindistList$}	
				{\label{checknonValidminMDist}
					\uIf {$T_{i,\ell}$ has no Trivial matches in $dkm_{\ell}$}
					{
						\For{$j$ $\leftarrow$ $m$ \emph{\textbf{down to}} $1$}
						{
							$mDist$ = $minMDist.getMax(j)$\;
							\uIf{$mDist > dkm_{newL}[k][j].d$ }
							{\label{checkOnlyGreater}
								$QT$ $\leftarrow$ $SlidingDotProduct$($T_{i,newL}$, $T$)\; \label{MASSscratch_discords}
								\textbf{double} $s$ $\leftarrow$ $sum$($T_{ind,\ell}$);
								\textbf{double} $ss$ $\leftarrow$ $squaredSum$($T_{i,newL}$)\; 
								$D(T_{ind,\ell}) \leftarrow CalcDistProfAndLB$($QT$,$T_{i,newL}$, $T$, $s$, $ss$)\;
								$UpdatePartialDistanceProfile(listDP[i],D(T_{ind,\ell})$\label{updatelistDP}) \;
								$UpdateFixedLengthDiscords$($dkm_{newL},listDP[i]$,$i$,$k$,$m$)\;\label{updateDiscFormScratch}
								\textbf{break}\;
							} 
						}
					}
				}
			}
		} % small font size
		\caption{ $Topkm$\_$nextLength$ (Compute \topkmd Discords of greater lengths)\label{computeKMDiscordAlgo}}
	\end{algorithm}

\noindent{\bf Complexity Analysis.} The time complexity of Algorithm~\ref{computeDkmAlllength} ($Topkm$\_$DiscordDiscovery$) mainly depends on the use of $ComputeMatrixProfile$ algorithm, which always takes $\mathcal{O}(n^2\log(p))$ to compute the partial distance profiles for the $n$ subsequences of length $\ell_{min}$ in $T$.

In order to compute the exact \topkmd discord ranking in $dkm_{\ell}$, the routine $UpdateFixedLengthDiscords$ takes $\mathcal{O}(km)$ time in the worst case. Recall that this latter algorithm is called only for subsequences that do not have trivial matches in $dkm_{\ell}$.
Checking if two subsequences are trivial matches takes constant time, if for each $dkm_{\ell}$ update, we store the $\ell$ trivial match positions.
Given a series $T$, and the discord (subsequence) length $\ell$, we can represent by $S = \frac{|T|}{l/2}$ , the number of subsequences that are not trivial matches with one another. 
Therefore, updating the discord rank of each length has a worst case time complexity of $\mathcal{O}((\ell_{max} - \ell_{min}) \times S \times \ell \times k \times m \times \log(m))$, where the $\log(m)$ factor represents the time to get the $m^{th}$ largest distance in the partial distance profile (line~\ref{getJth} of Algorithm~\ref{algoUpdateKMDiscords}). 
%In all the common settings of discord discovery, the overall complexity is linear in terms of $|T|$. {\bf what exactly is the last sentence? which are the common settings? do we need this sentence? ???}
Similarly, the construction of the variable length discord ranking in $dkm_{\ell_{min},\ell_{max}}$ takes: $\mathcal{O}((\ell_{max} - \ell_{min}) \times k \times m)$.

Observe also that the time performance of the $Topkm$\_$nextLength$ algorithm depends on the Euclidean distance computations pruning. If all the partial distance profiles contain the correct nearest neighbor's distances, computing the discords of each length greater than $\ell_{min}$ takes $\mathcal{O}(n \times p \times \log(m))$ time, with $n$ equal to the number of subsequences in $T$. 
The worst case takes place when for each subsequence that can update $dkm_{\ell}$ (i.e., $S$), the complete distance profile is re-computed (Algorithm~\ref{computeKMDiscordAlgo}, line~\ref{MASSscratch_discords}); in this case the algorithm takes $\mathcal{O}(n^2 \times \log(n) \times p \times \log(m))$.
%In our experimental evaluation, we show that in all the cases we tested, the percentage of the recomputed distance profiles is indeed very low.        

%\\
%	%\tcp{if SubMP does not contain the motif distance (bBestM = false), compute the whole non valid distance profiles, if it is faster then computeMatrixProfile (nDp / 2 = true)}
%%\If{$!bBestM$ \textbf{and} $nonValidDP.size() < (nDp / p) $}

\section{Experimental Evaluation}
\label{sec:experiments}

\subsection{Setup}

%\subsection{Configuration} 
We implemented our algorithms in C (compiled with gcc 4.8.4), and we ran them in a machine with the following hardware: Intel Xeon E3-1241v3 (4 cores - 8MB cache - 3.50GHz - 32GB of memory)\footnote{In order to validate the time performance results, we repeated our experiments on a second machine with different characteristics (Intel Xeon E5-2650 v4, 24 cores - 30MB cache - 2.20GHz, 250GB of memory), where we observed the same trends.}. 
All of the experiments in this paper are reproducible. 
In that regard, the interested reader can find the analyzed datasets and source code on the paper web page~\citep{paperWebpage}.

%\subsection{Datasets And Benchmarking Details}
%\vspace{0.1cm}
\noindent{\bf Datasets And Benchmarking Details.} 
To benchmark our algorithm, we use five datasets: 
 \begin{itemize}
 	\item GAP, which contains the recording of the global active electric power in France for the period 2006-2008. This dataset is provided by EDF (main electricity supplier in France)~\citep{Dua:2019};
 	\item CAP, the Cyclic Alternating Pattern dataset, which contains the EEG activity occurring during NREM sleep phase~\citep{CAP:dataset};
 	\item ECG and EMG signals from stress recognition in automobile drivers~\citep{stressDriver};
 	\item ASTRO, which contains a data series representing celestial objects~\citep{ltv}.
 \end{itemize}

%{\bf ??? change this, if we include results with more than 1M points ???}
%The detailed characteristics of these datasets are shown in Table~\ref{datainfo}, in the appendix.

Table~\ref{datainfo} summarizes the characteristics of the datasets we used in our experimental evaluation.
For each dataset, we report the minimum and maximum values, the overall mean and standard deviation, and the total number of points.

\begin{table}[tb]
	\centering
	\includegraphics[trim={0cm 14cm 19cm 3cm},scale=0.8]{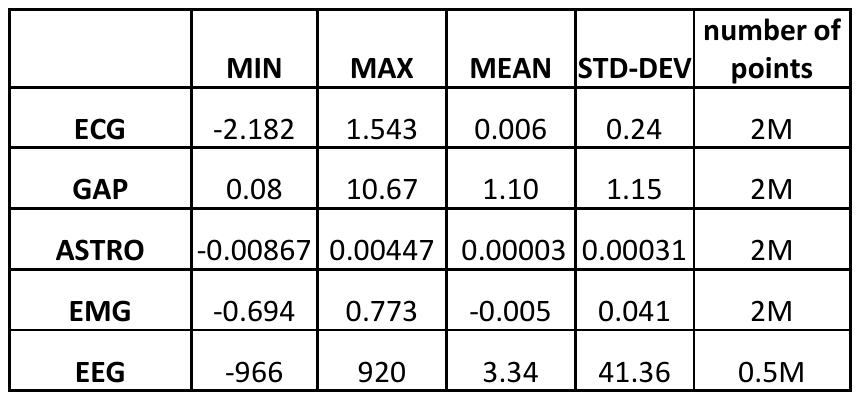}
	\caption{Characteristics of the datasets used in the experimental evaluation.}
	\label{datainfo}
\end{table}

The (CAP),(ECG) and (EMG) datasets are available in~\citep{PhysioNet}. 
We use several prefix snippets of these datasets, ranging from 0.1M to 1M of points.

In order to measure the scalability of our motif discovery approach, we test its performance along four dimensions, which are depicted in Table~\ref{tableParameters}. 
Each experiment is conducted by varying the parameter of a single column, while for the others, the default value (in bold) is selected. 
In our benchmark, we have two types of algorithms to compare to VALMOD. 
The first are two state-of-the-art motif discovery algorithms, which receive a single subsequence length as input: QUICKMOTIF~\citep{DBLP:conf/icde/LiUYG15} and STOMP~\citep{YehZUBDDSMK16}. 
In our experiments, they have been run iteratively to find all the motifs for a given subsequence length range. 
The other approach in the comparative analysis is MOEN~\citep{DBLP:journals/kais/MueenC15}, which accepts a range of lengths as input, producing the best motif pair for each length. 

  \begin{table}[tb]
  
  	\centering
  	\begin{tabular}{|c|c|c|c|}
  		\hline
  		
  		\begin{minipage}[c]{0.2\columnwidth}%
  			\centering
  			Motif length ($\ell_{min} $)
  		\end{minipage}
  		& 	
  		
  		\begin{minipage}[c]{0.2\columnwidth}%
  			\centering
  			Motif range ($\ell_{max} - \ell_{min}$) 
  		\end{minipage}
  		&  	
  		\begin{minipage}[c]{0.2\columnwidth}%
  			\centering
  			Data series size (points)
  		\end{minipage} 	
  		&  	
  		\begin{minipage}[c]{0.2\columnwidth}%
  			\centering
  			p (elements of distance profiles stored)
  		\end{minipage} \\
  		\hline
  		 256 & \textbf{100} & 0.1 M & 5 \\ 
  		\hline
  		512 & 150 & 0.2 M & 10 \\
  		\hline
  		\textbf{1024} & 200 & \textbf{0.5 M} & 15 \\	
  		\hline
  		2048 & 400 & 0.8 M & 20  \\	
  		\hline
  		4096 & 600 & 1 M & \textbf{50} , 100 , 150\\	
  		\hline	
  	
  	\end{tabular}

	\vspace*{0.5cm}
	\caption{\label{tableParameters}Parameters of VALMOD benchmarking (default values shown in bold).} 
	%Each one of the four experiments in the scalability section was ran varying one column and picking, for the other dimensions, the default parameter (bold).}
	%	\vspace*{-0.5cm}
  \end{table}

\begin{figure}[tb]
	\centering
	\includegraphics[trim={2cm 0cm 1cm 0cm},scale=0.40]{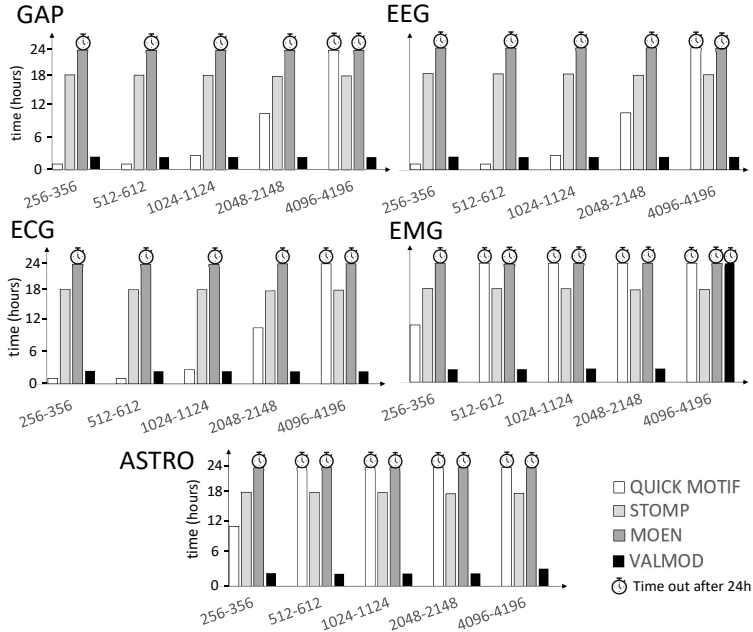}
	\caption{Scalability for various motif length ranges.}
	\label{Scalability1}
\end{figure}

For VALMOD, we report the total time, including the time to build the matrix profile (Algorithm~\ref{computeMatrix}).
The runtime we recorded for all the considered approaches is the average of five runs. 
Prior to each run we cleared the system cache.

\subsection{Motif Discovery Results}

\noindent{\bf Scalability Over Motif Length.} 
In Figure~\ref{Scalability1}, we depict the performance results of the \textit{four} motif discovery approaches, when varying the motif length. 
We note that the performance of VALMOD remains stable over the five datasets. 
On the other hand, we observe that a pruning strategy based on a summarized version of the data is sensitive to subsequence length variation. 
This is the case for QUICK MOTIF, which operates on PAA (Piecewise Aggregate Approximation) discretized data. 
Figure~\ref{Scalability1} shows that the performance of QUICK MOTIF varies significantly as a function of the motif length range, growing rapidly as the range increases, and failing to finish within a reasonable amount of time in several cases.

Moreover, we argue that our proposed lower bounding measure enables our method to improve upon MOEN, which clearly does not scale well in this experiment (see Figure~\ref{Scalability1}).
The main reason for this behavior is that the effectiveness of the lower bound of MOEN decreases very quickly as we increase the subsequence length $\ell$. 
When we increase the subsequence length by 1, MOEN multiplies the lower bound by a value smaller than 1 (\citep{DBLP:journals/kais/MueenC15}, Section IV.B), thus making it less tight. 
In contrast, the lower bound of VALMOD does not always decrease (refer to Eq.~(\ref{eq:lowerBoundFinalFormula})): $\frac{\sigma_{j,l}}{\sigma_{j,l+k}}$ may be larger than 1. 
Consequently, the lower bound of VALMOD can remain effective (i.e., tight) even after several steps of increasing the subsequence length.

%The reason resides behind the different scope of the parameters used in the lower bound the distances, respectively in our approach and in MOEN. 

%\commentRed{Here, we also need to fairly point out that VALMOD has a greater memory complexity than MOEN. More precisely, the difference resides in the $p$ factor. This trade-off, represents another reason for the sharp improvement. Nevertheless, as we may see in the next sections, our empirical evaluation suggests that $p$ is typically a small number, which guarantees a linear memory complexity. }

\begin{figure}[tb]
 	\centering
 	\includegraphics[trim={0cm 0cm 0cm 0cm},scale=0.38]{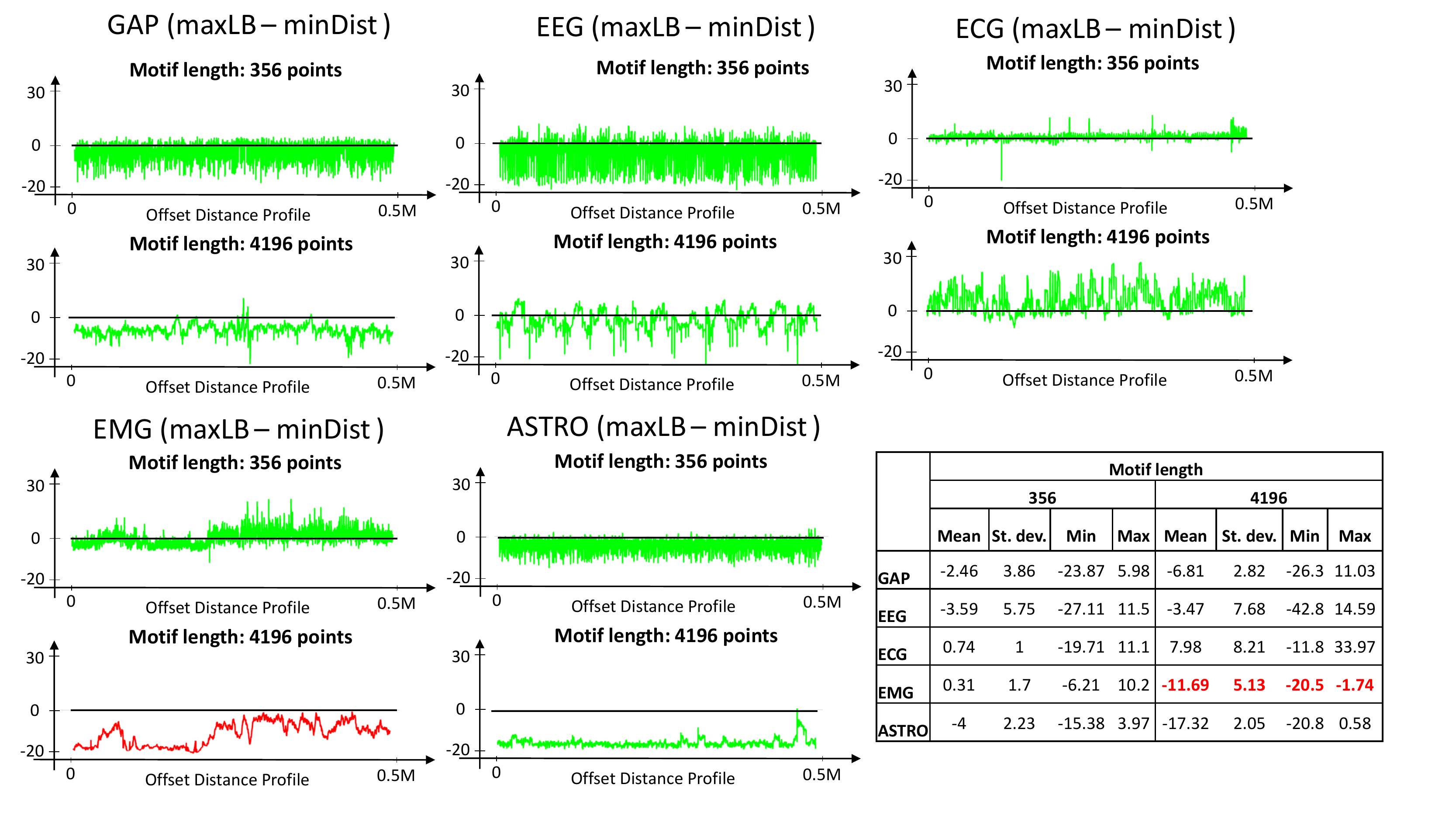}
 	\caption{The difference between the max lower bounding distance (maxLB) and the min Euclidean distance of partial distance profiles for all the datasets. Subsequence lengths: 356/4196. (We report the results for the EMG dataset in red, which corresponds to VALMOD's worst case for lengths 4096-4196, as shown in Figure~\ref{Scalability1}.)}
 	\label{EB-LB}
 \end{figure}

Concerning the VALMOD performance, we note a sole exception that appears for the noisy EMG data (Figure~\ref{Scalability1}), for a relatively high motif length range (4096-4196).
The explanation for this behavior is that the lower bounding distance used by VALMOD is coarse, or in other words, it is not a good approximation of the true distance.
Figure~\ref{EB-LB} shows the difference between the greater lower bounding distance ($maxLB$) and the smaller true Euclidean distance for each distance profile. 
%We use the subsequence lengths \textit{356} and \textit{4196}, which are the smallest and largest, respectively, subsequence lengths in this experiment.
We use the subsequence lengths \textit{356} and \textit{4196}, which are respectively the range's smallest and largest extremes in this experiment.
In this last plot, each value greater than 0 corresponds to a valid condition in line~\ref{checkCOrrectMin} of the \textit{ComputeSubMP} algorithm. 
This indicates that we found the smallest value of a distance profile, while pruning computations over the entire subsequence length range. 
As the subsequence length increases, VALMOD's pruning becomes less effective for the EMG (observe that there are no, or very few values above zero in the distances profiles for subsequence length \textit{4196}). 
On the other hand, we observe the presence of values above zero in the other datasets. This confirms that motifs in those cases are found, while pruning the search space.  
 
%{\bf ??? we need one sentence here: why are we going to talk about tlb? ???}
In order to further evaluate the pruning capability of VALMOD, we report the measurements for the Tightness of the Lower Bound (TLB)~\citep{Shieh,DBLP:conf/kdd/ZoumpatianosLPG15} performed during the previous experiment (Figure~\ref{Scalability1}). The TLB is a measure of the lower bounding quality; given two data series $t_1$ and $t_2$, the TLB is computed as follows: $LB_{dist}(t_1,t_2)/EuclideanDistance(t_1,t_2)$. Note that TLB takes values between 0 and 1. 
A TLB value of 1 means that the lower bound distance is the same as the Euclidean distance; this corresponds to the optimal case. 

\begin{figure}[tb]
 	\centering
 	\includegraphics[trim={0cm 0cm 0cm 0cm},scale=0.38]{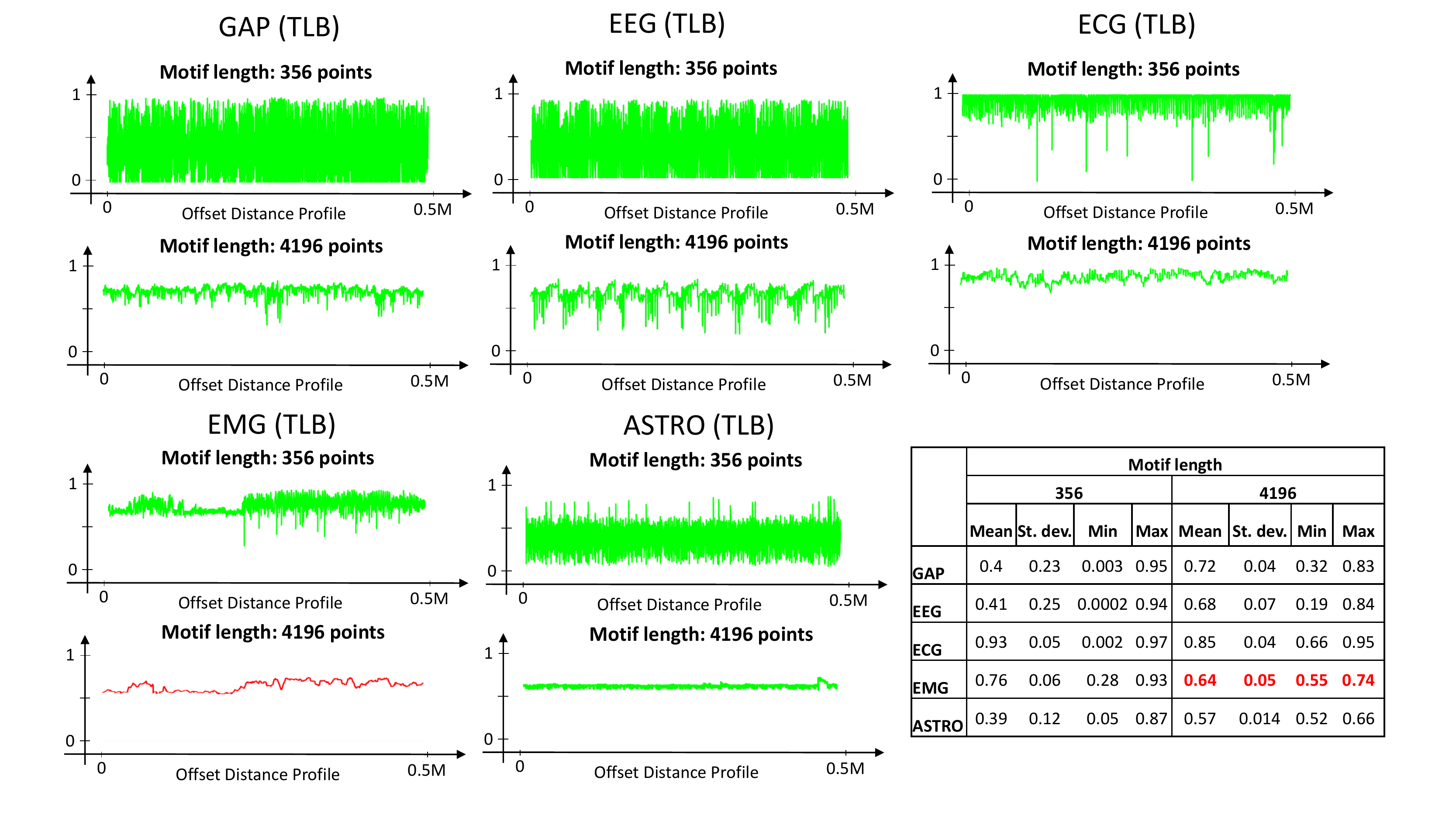}
 	\caption{Average of the tightness of the lower bound (TLB) for every Distance profile for all datasets. Subsequence lengths: 356/4196. (We report the results for the EMG dataset in red, which corresponds to VALMOD's worst case for lengths 4096-4196, as shown in Figure~\ref{Scalability1}.)}
 	\label{TLBfig}
 	\vspace*{-0.5cm}
\end{figure}

In Figure~\ref{TLBfig}, we show the average TLB for each (partial) distance profile. 
In the EMG dataset, when using the larger subsequence length, we observe a sharp decrease of the lower bounding quality (small TLB values), explaining the behavior observed for the EMG dataset (refer to Figure~\ref{EB-LB}(bottom-left)).
We also note similar results for the ASTRO dataset. As we have noted for this last case, the performance is not negatively affected, since we dispose of several partial distance profiles that provide the correct minimum distances, and thus permit us to find the motifs, without recomputing all the distance profiles. 
In contrast, in the other datasets, we note a smaller negative impact on TLB for the case of subsequence length \textit{4196}.

%{\bf ??? in figs10-12 why is emg in red and all the rest in green? add a sentence in each one of the three captions to explain this ???}

\begin{figure}[tb]
	\centering
 	\includegraphics[trim={1cm 6cm 0cm 2cm},scale=0.48]{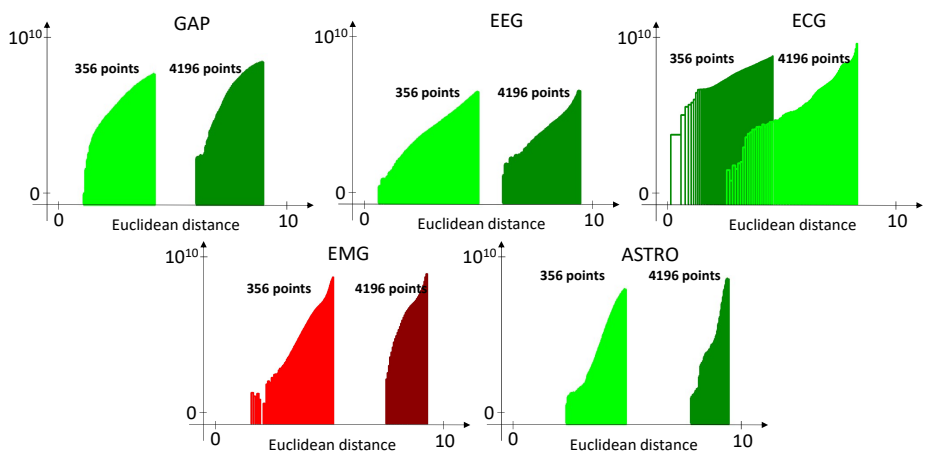}
 	\caption{Distribution of Euclidean distance of pairwise subsequences in all the datasets. Subsequence lengths: 356/4196. (We report the results for the  EMG dataset in red, which corresponds to VALMOD's worst case for lengths 4096-4196, as shown in Figure~\ref{Scalability1}.)}
 	\label{distroPairwise}
 	%\vspace*{-0.5cm}
 \end{figure}

In Figure~\ref{distroPairwise}, we also show the distance distribution of the pairwise subsequences, using the same datasets and subsequences lengths. 
Here, we plot the distances without length normalization, since the algorithm uses it to rank the motifs in the trailing part. 
For the EMG and ASTRO datasets, in the case of length \textit{4196}, the distance distribution includes many small and large values, which does not suggest the presence of motifs, but affects VALMOD negatively.  
Observe that in the other datasets, the values are more uniformly distributed over all the subsequence lengths. 
This denotes the presence of subsequence pairs that are substantially closer than the rest, which typically identifies the occurrence of motifs. 
In this case, VALMOD is able to prune more distance profile computations, leading to better performance.

%Note that in the discussion above, we have used the results of the ECG and EMG datasets only, which represent the extreme cases for VALMOD, exhibiting the overall best and worst performance, respectively.
%The rest of the datasets produce results that lie between these two extremes, and are omitted for brevity.

%We conclude here that the only issue with VALMOD is due to data dependent characteristics.
%This occurred in only one out of five runs on various datasets, and it only occurred on a single subsequence length, which did not allow us to gain any valuable insights as shown in Figure \ref{distroPairwise} (left) where we have high distances among the subsequences.

\begin{figure}[tb]
	\centering
	\includegraphics[trim={2.5cm 0cm 2cm 0cm},scale=0.45]{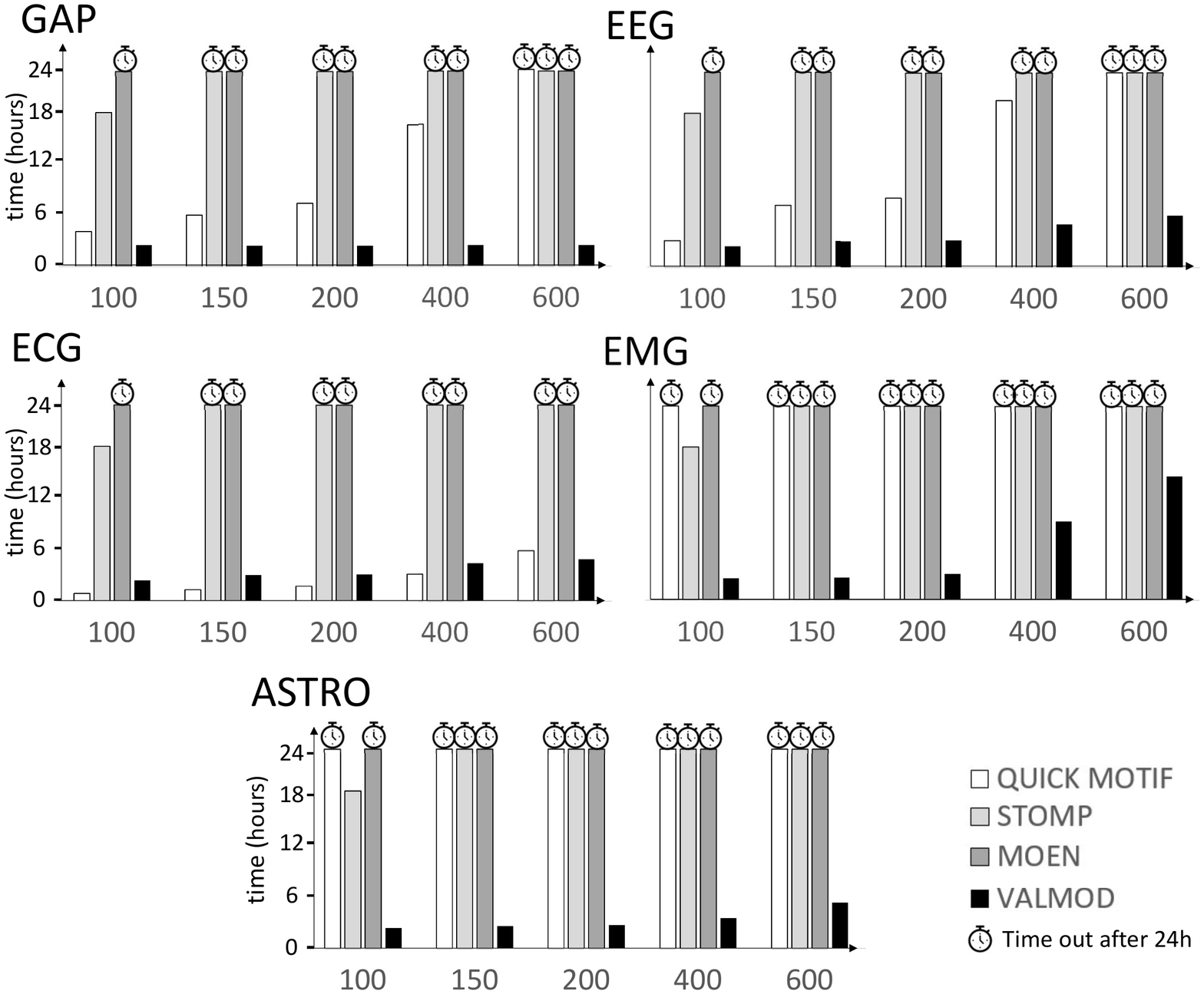}
	\caption{Scalability with increasing motif range.}
	\label{Scalability2}
	\vspace*{-0.5cm}
\end{figure}

\vspace{0.1cm}
\noindent{\bf Scalability Over Motif Range.} 
In Figure~\ref{Scalability2}, we depict the performance results as the motif range increases. 
VALMOD gracefully scales on this dimension, whereas the other approaches can seldom complete the task.
Not only does our technique address the intrinsic problem of STOMP and QUICK MOTIF, which independently process each subsequence length, but it also exhibits a substantial improvement over MOEN, the existing state-of-the-art approach for the discovery of variable length motifs.

\begin{figure}[tb]
	\centering
	\includegraphics[trim={2.5cm 0cm 2cm -1cm},scale=0.45]{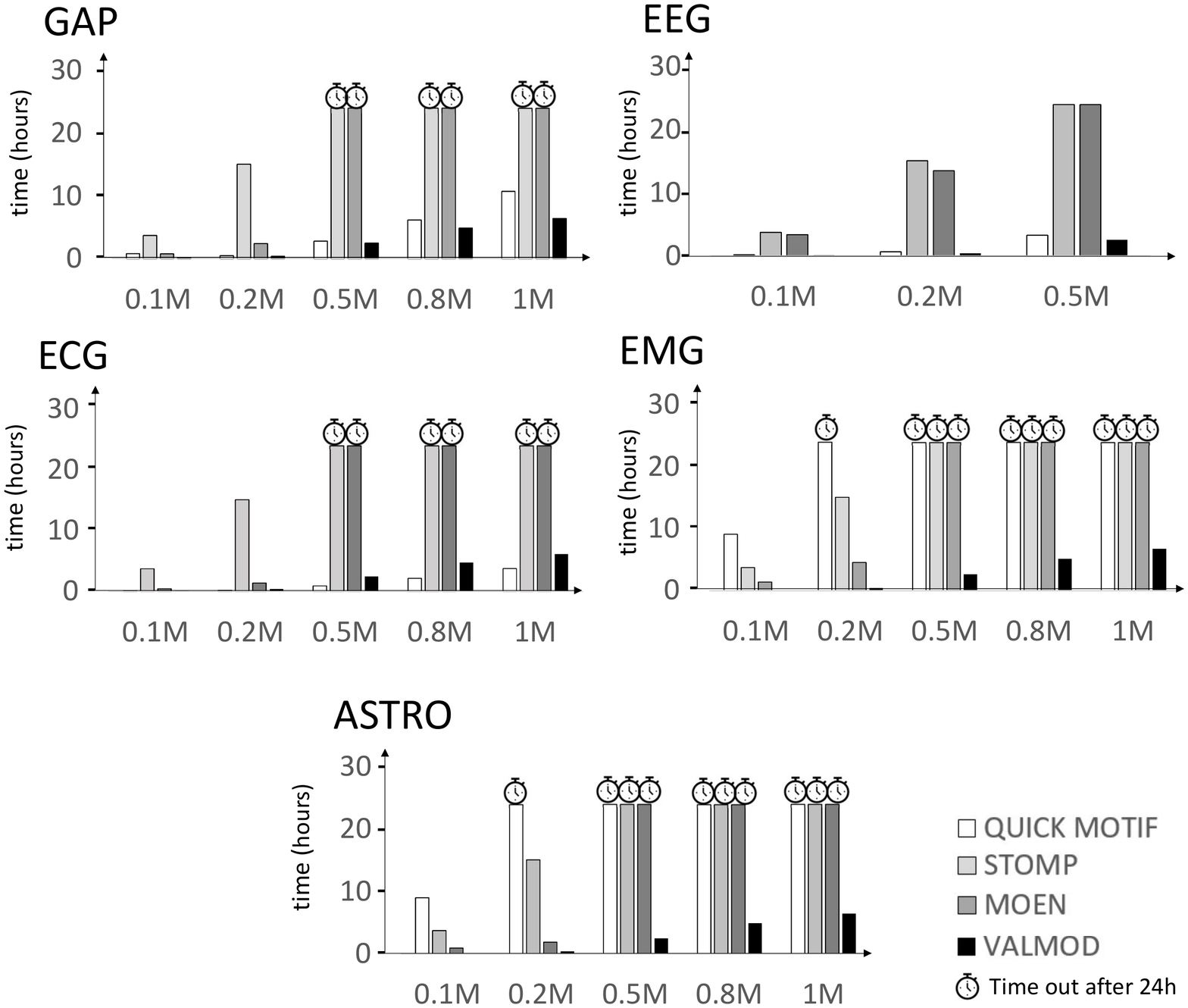}
	\caption{Scalability with increasing data series size.}
	\label{Scalability3}
	\vspace*{-0.3cm}
\end{figure}

\vspace{0.1cm}
\noindent{\bf Scalability Over Data Series Length.} 
In Figure~\ref{Scalability3}, we experiment with different data series sizes. 
For the EEG dataset we only report three measurements, since this collection contains no more than 0.5M points. 
We observe that QUICK MOTIF exhibits high sensitivity, not only to the various data sizes, but also to the different datasets (as in the previous case, where we varied the subsequence length). 
It is also interesting to note that QUICK MOTIF is slightly faster than VALMOD on the ECG dataset, which contains regular and similar heartbeat patterns, and is a relatively easy dataset for motif discovery. 
Nevertheless, QUICK MOTIF, as well STOMP and MOEN, fail to terminate within a reasonable amount of time for the majority of our experiments.
On the other hand, VALMOD does not exhibit any abrupt changes in its performance, scaling gracefully with the size of the dataset, across all datasets and sizes. 

\begin{figure}[tb]
	\centering
	\includegraphics[trim={2cm 14cm 4cm -0.5cm},scale=0.60]{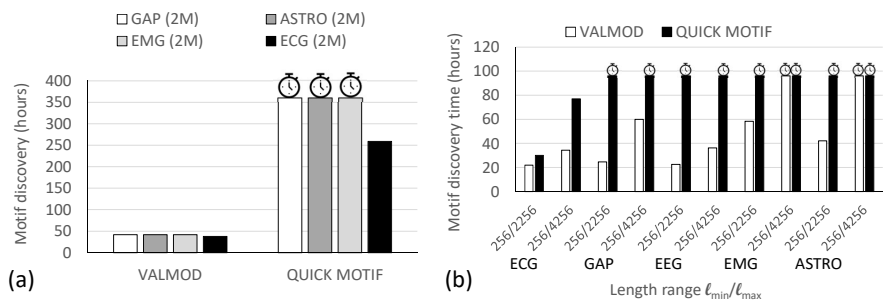}
	\caption{Scalability of VALMOD and QUICKMOTIF using large datasets (2M of points) and large length ranges.}
	\label{ExtremeExperiments}
	\vspace*{-0.3cm}
\end{figure}

\noindent{\bf Large Datasets and Length ranges.}
Here we report two further experiments that we have conducted on larger snippets of the datasets - namely, \textit{2} million points - and over a larger range of motif lengths.
To that extent, we want to test the scalability of our approach, considering two \textit{extreme} cases.  
We compare VALMOD to QUICKMOTIF, since the latter is the sole approach that can scale to data series lengths beyond half a million points, and to motif length ranges larger than \textit{100}.

In Figure~\ref{ExtremeExperiments}.(a), we report the motif discovery time on four datasets that contain \textit{2} million points. We pick the default length boundaries, namely $\ell_{min}=1024$ and $\ell_{max}=1124$, discovering motifs of each length in between them.
The results show that VALMOD gracefully scales, and is always one order of magnitude faster than QUICKMOTIF, which does not reach the timeout only in the case of the ECG datasets.
%This means that the index-based approach of QUICKMOTIF, which does not implement an ad-hoc variable length motif discovery, can not scale as VALMOD, which shows more robustness over the dataset size growth.

The same observations hold for the results of the experiments that vary the motif length range.
Figure~\ref{ExtremeExperiments}.(b), shows the results for length ranges \textit{2000} and \textit{4000}, on all five datasets in our study (at their default sizes).
Once again, QUICKMOTIF reaches the timeout state in all datasets, except for ECG, where for the larger length ranges is two times slower than VALMOD. On the other hand, VALMOD scales well and remains the method of choice (with the exception of the largest length ranges for the EMG and ASTRO datasets, where it reaches the timeout). 

The above results demonstrate the superiority of VALMOD, but also show its limits, which open possibilities for future work. 
%Nevertheless, we are not aware of any variable length motif discovery techniques that have been able to propose a solution, suitable for such large motif lengths range and dataset sizes.

\begin{figure}[tb]
	\centering
	\includegraphics[trim={0cm 3cm 15cm 2cm},scale=0.80]{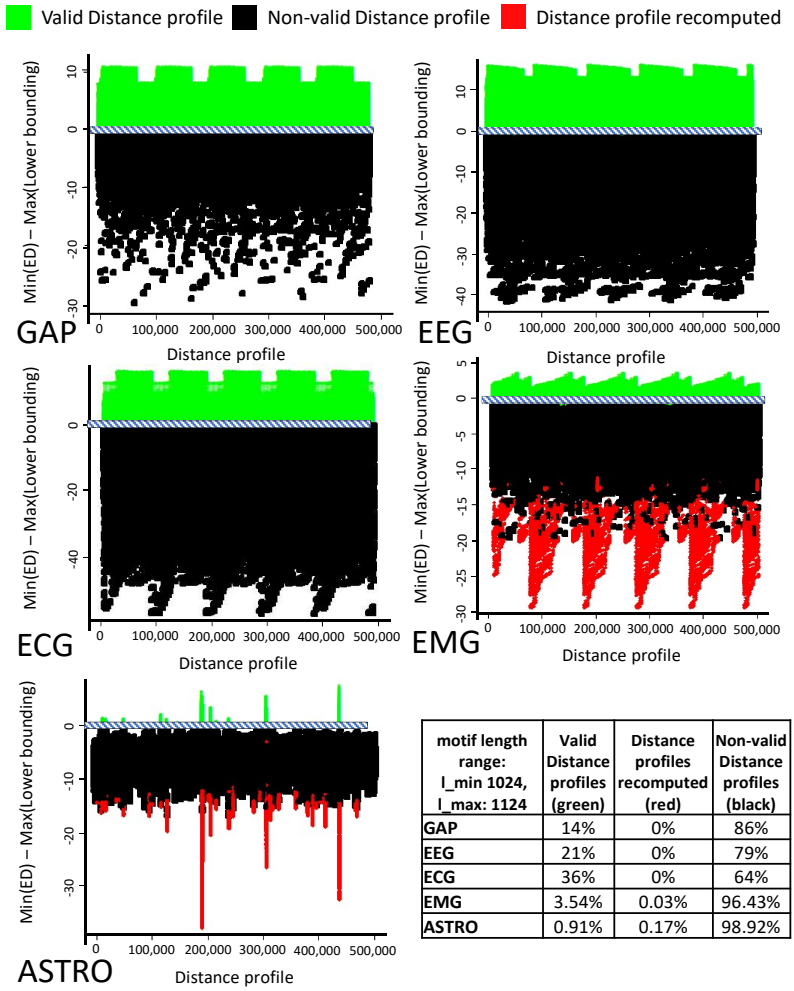}
	\caption{Partial distance profile repartition (\textit{valid}, \textit{non-valid}, \textit{recomputed}), in the motif discovery task on the five considered datasets. Default parameters are used in this experiment. }
	\label{DistanceProfileValidNValid}
	\vspace*{-0.3cm}
\end{figure}

\begin{figure}[tb]
	\centering
	\includegraphics[trim={0cm 13cm 12cm 3cm},scale=0.70]{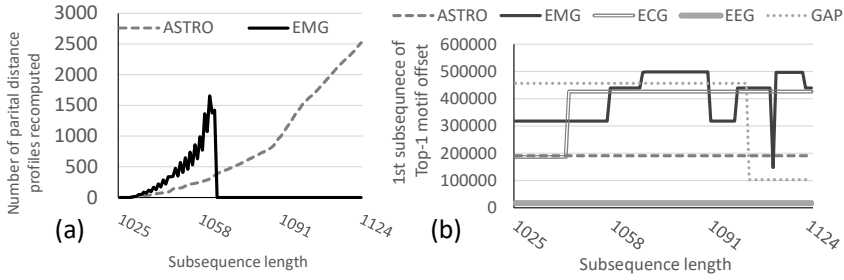}
	\caption{\textit{(a)} Distribution of \textit{recomputed} distance profiles for each subsequence length considered in the EMG and ASTRO datasets. \textit{(b)} Offset of the first subsequence in the discovered motif for all the length in the EMG And ASTRO datasets.}
	\label{DistanceProfileEMGASTRO}
	\vspace*{-0.3cm}
\end{figure}

\noindent{\bf Overall Pruning Power.}
In order to show the global effect of VALMOD's pruning power, we conduct an experiment recording the number of distance profile computations performed by procedure \textit{ComputeSubMP}, which extracts motifs of length greater than $\ell_{min}$, pruning the unpromising calculations.
We recall that this algorithm computes for each subsequence $T_{i,\ell}$ with $\ell>\ell_{min}$ a subset of distances (Euclidean and lower bounding), called partial distance profiles. If the smallest Euclidean distance computed is also smaller than the larger lower bounding distance, we know it is the true distance of the nearest neighbor of $T_{i,\ell}$. In this case, we call the partial distance profile \textit{valid}. Otherwise, we do not know the true nearest neighbor distance, and we call the partial distance profile \textit{non-valid}.
In order to identify the correct motifs, the algorithm only needs to recompute the entire \textit{non-valid} distance profiles that might contain distances shorter than those already found in the \textit{valid} distance profiles. 

In Figure~\ref{DistanceProfileValidNValid}, we depict the difference between the minimum Euclidean distance and the maximum lower bounding distance of each distance profile computed in the subsequence length range (\textit{1025/1124}). In the plots, the values above zero refer to the \textit{valid} ones (green points), whereas values under zero are either \textit{non-valid} (black points) or \textit{recomputed} (red/triangular points).
We observe that in the first three datasets, namely EEG, ECG and GAP, there are no distance profiles that are recomputed, meaning that the motifs are always found in the \textit{valid} (partial) distance profiles in the shortest time possible (\textit{best case}). 
Concerning the EMG and ASTRO datasets, several re-computations take place (red/triangle points). 
As we can see from the table in the bottom part of Figure~\ref{DistanceProfileValidNValid} though, the computed distance profiles are not more than the \textit{0.20\%} of the total. 
This means that the algorithm successfully prunes a high percentage of the computations, thanks also to the effectiveness of the proposed lower bounding measure.   
       
At this point, we can further analyze the reasons behind the pruning capability of our approach. To that extent, in Figure~\ref{DistanceProfileEMGASTRO}.\text{(a)} we plot the number of distance profiles that VALMOD recomputes at each subsequence length for the EMG and ASTRO datasets. 
These two datasets both contain noisy data, which influence re-computations. However, they differ according to the length for which these re-computations take place. 

Figure~\ref{DistanceProfileEMGASTRO}.\text{(b)} shows the position of the $Top-1$ motif along the subsequence length.
Note that the $Top-1$ motif is always placed around the same offset region in the ASTRO dataset, suggesting the presence of a few similar data segments, which is also verified by the high number of \textit{non-valid} distance profiles we observe in Figure~\ref{DistanceProfileValidNValid}(ASTRO).
On the other hand, in the EMG dataset, the motif location changes several times, denoting the presence of different segments, which contain motifs of different lengths. % {\bf ??? last sentence not very clear, what does similar/dissimilar mean? ???}
This is also confirmed by the more prevalent presence of \textit{valid} distance profile in the EMG dataset. 
In this last case, the re-computation number drops to zero as soon as the motif positions start to change, i.e., at length \textit{1058}, maintaining the same trend until the end.

\begin{figure}[tb]
	\centering
	\includegraphics[trim={2.5cm 0cm 3cm 0cm},scale=0.43]{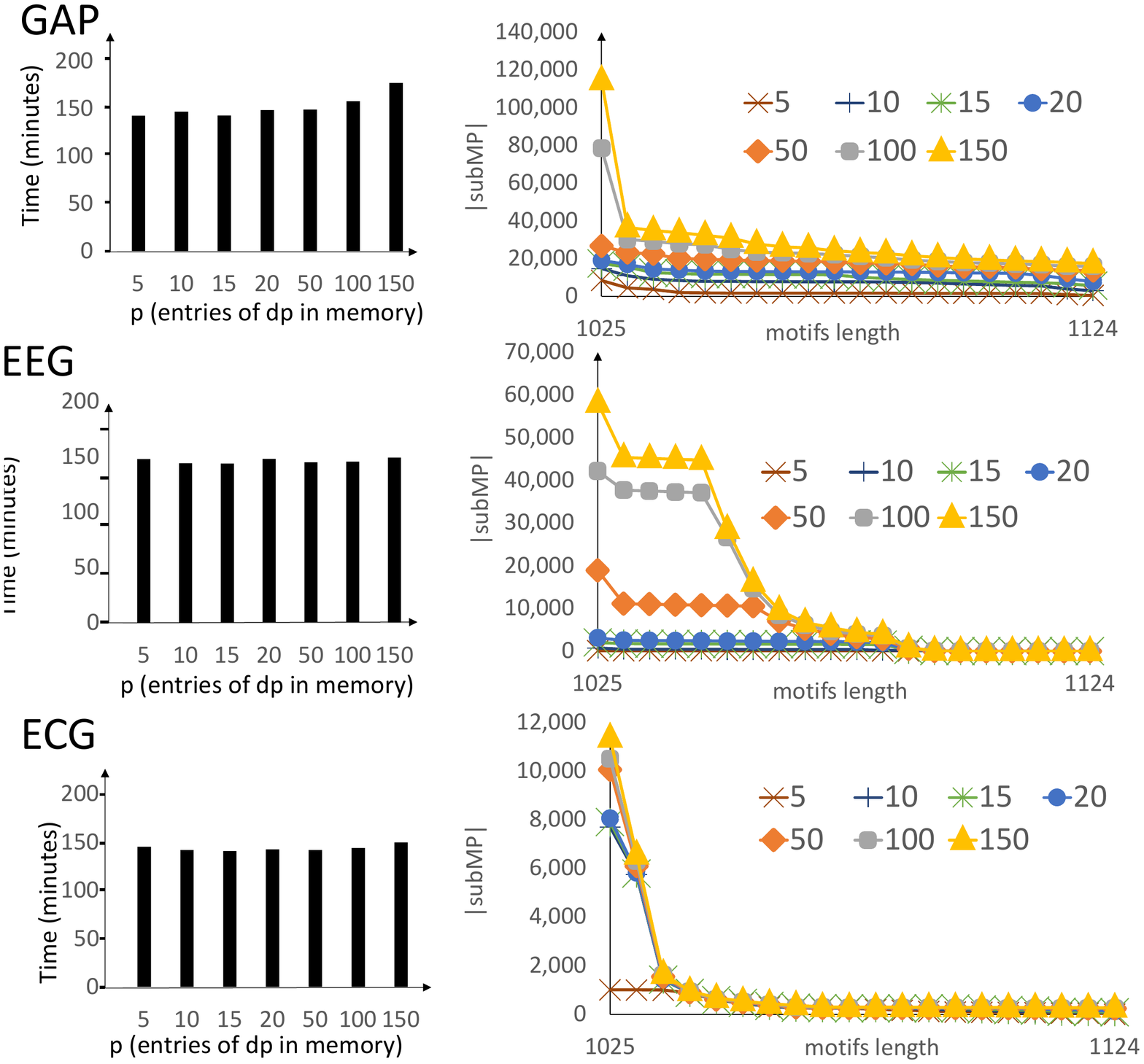}
	\includegraphics[trim={2.5cm 6cm 3cm 0cm},scale=0.43]{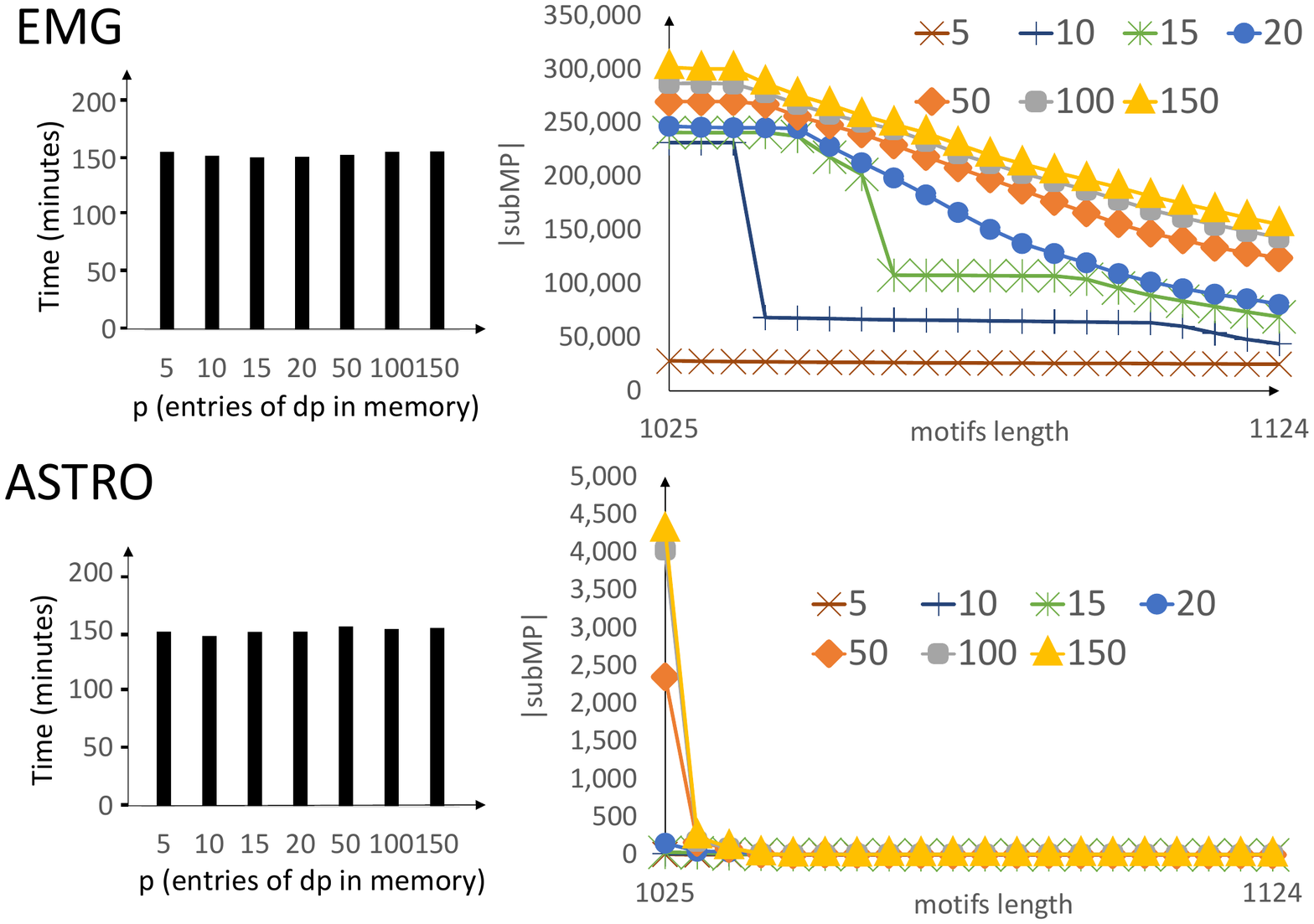}
	\caption{Scalability with increasing parameter p.}
	\label{Scalability4}
	\vspace*{-0.3cm}
\end{figure}

\noindent{\bf Effect of Changing Parameter $\mathbf{p}$.} 
In Figure~\ref{Scalability4}, we study the effect of parameter $p$ on VALMOD's performance. 
The $p$ value determines how many distance profile entries we compute and keep in the memory. 
Increasing $p$ leads to increased memory consumption, but could also translate to an overall speed-up, since having more distances may guarantee a larger margin between the greater lower bounding distance and the minimum true Euclidean distance in a distance profile. 
As we can see on the left side of the plot, increasing $p$ does not provide any significant advantage in terms of time complexity. 
Moreover, the plots on the right-hand side of the figure demonstrate that the size of the Matrix profile subset ($subMP$), computed by the \textit{computeSubMP} procedure, decreases in the same manner at each iteration (i.e., as we increase the length of the subsequences that the algorithm considers), regardless of the value of $p$. 
%along the subsequence lengths range. 
%Nevertheless, we observe that when $p$ is in the range $[5,20]$, the size of $subMP$ is considerably smaller (Figure~\ref{Scalability4}(right)). 
%Therefore, we choose \textit{50} {\bf ??? 50 or 20? ???} as the default value of $p$. 

It is important to note that irrespective of its size, $subMP$ \emph{always} contains the smallest distances of the matrix profile, namely the distances of the motif pair. 
Having a larger $subMP$ does not represent an advantage w.r.t. motif discovery, but rather an opportunity to view and analyze the subsequence pairs, whose distances are close to the motif.

%\subsection{Motif Evolution}
 %\begin{figure}[tb]
% 	\includegraphics[trim={3cm 8cm 0cm 0cm},scale=0.35]{MotifEVO/VALMAP_EVO}
% 	\caption{ }
% \end{figure}

\subsection{Motif Sets}

\begin{figure}[tb]
	\centering
	\includegraphics[trim={0.5cm 10cm 12cm 2.5cm},scale=0.65]{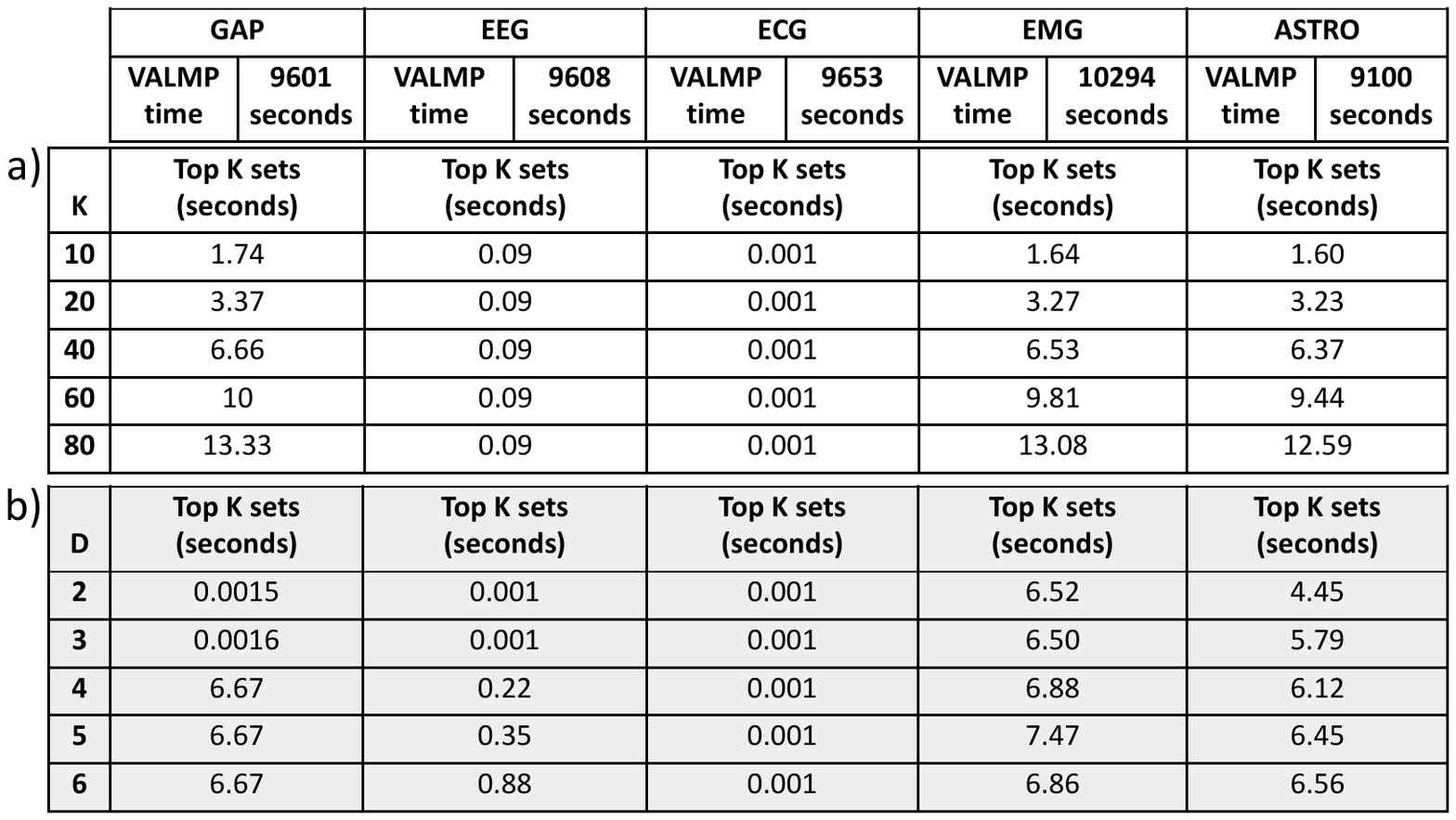}
	\caption{Time performance of variable length motif sets discovery. (a) Varying K (default D=4). (b) Varying radius factor D (default K=40).}
	\label{MotifSetsPerformance}
\end{figure}

We now conduct an experiment to show the time performance of identifying the variable length motif sets. 
We use the default values of Table~\ref{tableParameters}, varying $K$ and the radius factor $D$ for each dataset.
In Figure~\ref{MotifSetsPerformance} we report the results; we also show the time to compute $VALMP$ (the output of VALMOD). 
We note that once we build the pairs ranking of $VALMP$ ($heapBestKPairs$~in Algorithm~\ref{algo1_3}), we can run the procedure that computes the motif sets (Algorithm~\ref{computeVLMotifSets}).
The results show that this operation is 3-6 orders of magnitude faster than the computation of $VALMP$.
The advantage in time performance is pronounced for the $ECG$ and $EEG$ datasets, thanks to the pruning we perform with the partial distance profiles. 

The fast performance of the proposed approach also allows for a fast exploratory analysis over the radius factor, which would otherwise (i.e., with previous approaches) be extremely time-consuming to set for each dataset. 

\subsection{Discord Discovery }
{
In this last part, we conduct the experimental evaluation concerning discord discovery. 
In the following experiments, we use the same datasets as before.

We identify two state-of-the-art competitors to compare to our approach, the Motif And Discord (MAD) framework.
The first one, DAD (Disk Aware Discord Discovery)~\citep{DBLP:conf/icdm/YankovKR07}, implements an algorithm suitable for enumerating the \emph{fixed-length} $m^{th}$ discords of a data series collection stored on a disk.
We adapted this algorithm, as suggested by the authors, in order to extract discords from data series loaded in main memory. 
The second approach, GrammarViz~\citep{DBLP:conf/edbt/Senin0WOGBCF15}, is the most recent technique, which discovers \topkfirstd discords. It operates by means of grammar rules compression, which further operate on a summarized data series representation, in order to find the rare segments of the data (discords) in a reduced search space.
To the best of our knowledge, there exist no techniques capable of finding the \topkmd ranked variable-length discords as MAD, using a single execution of an algorithm.

\noindent{\bf $\mathbf{M^{th}}$ Discord Discovery.} 
In Figures~\ref{DAD_MAD}(a)-(b), we present the performance comparison between MAD and DAD for finding the $m^{th}$ discords, when we vary $m$, for all datasets.
(All other parameters are set to their default values, as listed in Table~\ref{tableParameters}.)
 
Since DAD discovers fixed-length $m^{th}$ discords, we report its execution time \emph{only} for the first length in the range, namely $\ell_{min}$.
We observe that MAD, which enumerates the $m^{th}$ discords of \textit{100} lengths ($\ell_{min}=1024$, $\ell_{max}=1124$) is still one order of magnitude faster than these DAD performance numbers, for all datasets, when $m$ is larger or equal to \textit{5}.
Moreover, the performance trend of MAD remains stable over all datasets, whereas DAD has different execution times.
%, which are clearly data dependent, as we note in the EMG and ASTRO dataset.
We observe that the computational time of DAD depends on the subsequence length, since it computes Euclidean distances in their entirety (only applying early abandoning based on the best so far distance). 
How effective this early abandoning mechanism is, depends on the characteristics of the data.
On the other hand, our algorithm computes all distances for the first subsequence length in constant time, and then prunes entire distance computations for the larger lengths. 
       
\begin{figure}[tb]
	\centering
	\includegraphics[trim={1cm 9cm 0cm 0cm},scale=0.60]{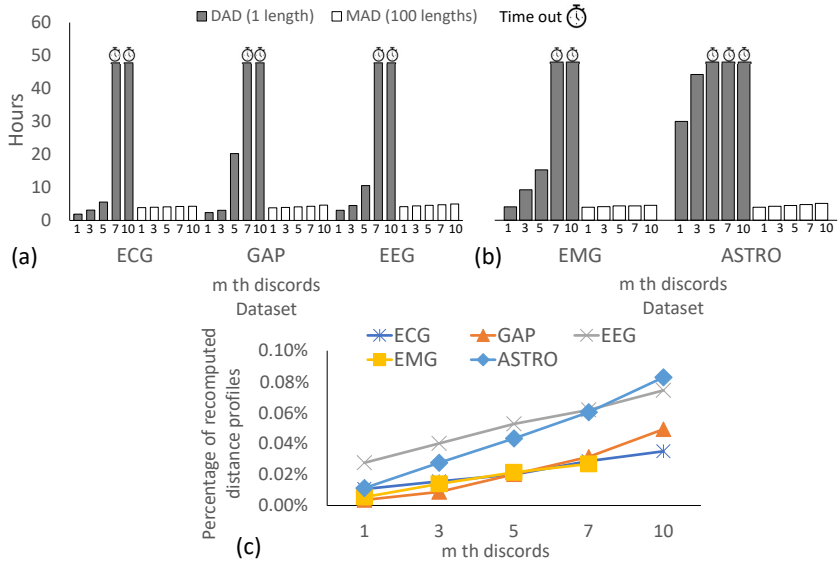}
	\caption{\textit{(a),(b)} DAD (one length) and MAD (\textit{100} lengths) \topkmd discords discovery time. (c) Percentage (on the total distance profiles) of \textit{non-valid} partial distance profiles recomputed by Algorithm~\ref{computeKMDiscordAlgo}}
	\label{DAD_MAD}
\end{figure}

In Figure~\ref{DAD_MAD}(c), we report the percentage of non-valid distance profiles that are recomputed, over the total number of distance profiles considered during the entire task of variable-length discord discovery.
We note that the number of re-computations is limited to no more than \textit{0.10\%}, in the worst case. 
This demonstrates the high computation pruning rate achieved by our algorithm, justifying the considerable speed-up achieved.

\noindent{\bf \topkfirstd Discord Discovery.} 
In Figure~\ref{Gviz_MAD}, we depict the performance comparison between GrammarViz and MAD. 
We do not report results for DAD, since it always reaches the imposed time-out, even for the variable length \topkfirstd discord discovery task.
Therefore, we consider \topkfirstd discords discovery, as previously introduced.
(We maintain the same parameter settings in this experiment.)
      
\begin{figure}[tb]
	\centering
	\includegraphics[trim={1cm 10cm 6cm 0cm},scale=0.60]{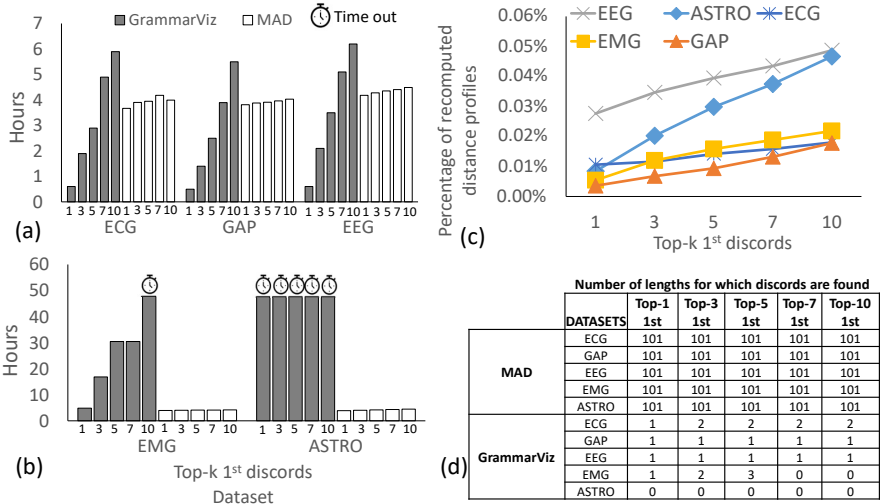}
	\caption{\textit{(a),(b)} GrammarViz and MAD (\textit{100} lengths) \topkfirstd discords discovery time. \textit{(c)} Percentage (on the total distance profiles) of \textit{non-valid} partial distance profiles recomputed by Algorithm~\ref{computeKMDiscordAlgo}.}
	\label{Gviz_MAD}
\end{figure}

First, we note that GrammarViz outperforms MAD in the first three datasets, for $k$ smaller or equal to \textit{5}, as depicted in Figure~\ref{Gviz_MAD}(a). 
Nevertheless, the experiment shows that MAD scales better over the number of discovered \topkfirstd discords, as its execution time increases only by a small constant factor.
A different trend is observed for GrammarViz, whose performance significantly deteriorates as $k$ increases from 1 to 6.

Moreover, this technique is highly sensitive to the dataset characteristics, as we observe in Figure~\ref{Gviz_MAD}(b), where the two noisy datasets, i.e., EMG and ASTRO, are considered.
This is a direct consequence of the data summarization sensitivity to the data characteristics, which then influences the ability to prune distance computations. 

In Figure~\ref{Gviz_MAD}(c), we report the percentage of non-valid distance profiles that MAD needed to recompute. 
In this case, too, this percentage is very low. 

To conclude, since GrammarViz is a variable length approach that selects the most promising discord lengths according to the distribution of the data summarization (by picking the lengths of the series, whose discrete versions represent a rare occurrence), we report in Figure~\ref{Gviz_MAD}(d) the number of lengths, for which discords are found. 
%{\bf ??? which fig? ???}
%{\bf ??? explain better why this is important, what these results mean for the algos ???}
We observe that our framework always enumerates and ranks discords of all lengths in the specified input range, based on the exact Euclidean distances of the subsequences. 
%This always allows to perform a complete exploratory analysis over the discords length. 
On the other hand, GrammarViz selects the most promising length based on the discrete version of the data, and only identifies the exact \topkfirstd discords for \textit{3} (out of \textit{100}) different lengths in the best case.
%{\bf ??? MAD is exact, and grammarviz is approximate, right? we never mention this here! ???}

\noindent{\bf \topkmd Discord Discovery.} 
Figure~\ref{MADtkm} depicts the execution time for the \topkmd discord discovery task, and the percentage of recomputed distance profiles for MAD, when varying $k$ and $m$. 
We observe that the pruning power remains high: the percentage of distance profile re-computations averages around $0.05$\%.

\begin{figure}[tb]
	\centering
	\includegraphics[trim={2cm 16cm 6cm 0cm},scale=0.60]{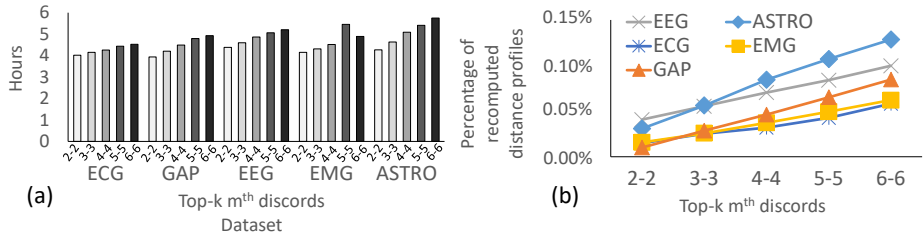}
	\caption{\textit{(a)} MAD (\textit{100} lengths)\topkmd discords discovery time on the five datasets. (b) Percentage (on the total distance profiles) of \textit{non-valid} partial distance profiles recomputed by Algorithm~\ref{computeKMDiscordAlgo}}
	\label{MADtkm}
\end{figure}

\noindent{\bf Utility of Variable-Length Discord Discovery.}
%\commentRed{?? message to convey ... if we use a range we can find the anomaly as discord... otherwise, with fixed length we would miss it.}
We applied MAD on a real use case, a data series containing the average number of taxi passengers for each half hour over 75 days at the end of 2014 in New York City~\citep{DBLP:journals/pvldb/RongB17}, depicted in Figure~\ref{taxiDiscords}(a). 
We know that this dataset contains an anomaly that occurred during the daylight savings time end, which took place the $2^{nd}$ of November 2014 at $2$am. At that time, the clock was set back at $1$am.    
Since the recording was not adjusted, two samples (corresponding to a \textit{1} hour recording) are summed up with the two subsequent ones. 
%This error, is directly detectable as the highest peak occurring in the data that we depict in Figure~\ref{taxiDiscords}(a). 

\begin{figure}[tb]
	\centering
	\includegraphics[trim={2cm 12.5cm 10cm 3cm},scale=0.60]{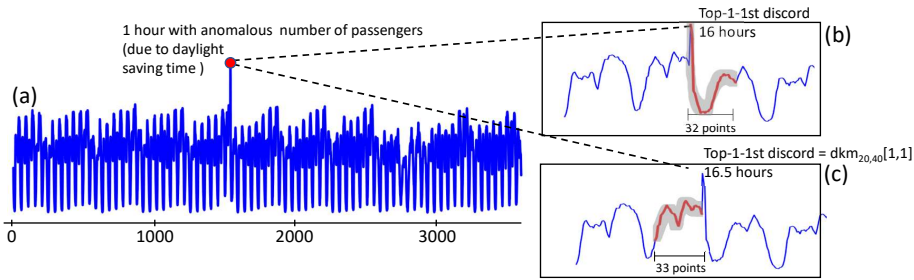}
	\caption{\textit{(a)} Data series reporting the number of taxi passengers over 75 days at the end of 2014 in New York City. \textit{(b)} $Top-1$ $1^{st}$ discord of length \textit{32}, which contains the abnormal peak generated by the double recording problem of daylight savings time. \textit{(c)} $Top-1$ $1^{st}$ discord of length \textit{33}, which represents an anomalous trend for the number of taxi passengers 
	%along 16.5 hours: it coincides to a trips re-organization of the passengers right before 
	due to 
	the daylight savings time.}
	\label{taxiDiscords}
\end{figure}

%We now discuss the insights produced by \emph{variable-length} \topkfirstd discords. 
We ran the variable-length discord discovery task using the length range $\ell_{min}=20$ and $\ell_{max}=48$, in order to cover subsequences that correspond to recordings between $10- and 24$ hours. 
Our algorithm correctly identifies the anomaly for subsequence length \emph{32}, shown in Figure~\ref{taxiDiscords}(b). 
Changing the window size does not allow the detection of the anomaly. 
For example, enlarging the window by \textit{just} 1 point, the \topkfirstd discord corresponds to a pattern \emph{before} the abnormality (refer to Figure~\ref{taxiDiscords}(c)).
%, which represents a higher passengers presence over \textit{16} hours, from the $1^{st}$ of November at $8.30$am to the $2^{nd}$ of November at $12.30$am. Since the New York Marathon~\citep{nycMarathon} took place the $2^{nd}$ of November 2014, we can associate this passengers increase to the athletes arrival in the city. 

These results showcase the importance of efficient variable-length discord discovery. 
It permits us to discover rare, or abnormal events with different durations, which can be easily missed in the fixed length discord discovery setting, where the analyst is constrained to examine a single length (or time permitting, a few fixed lengths).

%This experiment showcases the importance of variable length discord discovery. In fact, the double recording (abnormal peak) becomes part only of the $Top-1$ $1^{st}$ discord subsequence of length \textit{32}.
%Such fine-grained insight can easily be missed by setting a different length in the fixed length discord discovery setting.     
%(For example, enlarging the window by \textit{just} 1 point, the \topkfirstd discord corresponds to a pattern \emph{before} the abnormality.) 
%This discord of length \emph{32} has the highest length-normalized distance to its nearest neighbor, 
%%making it the $Top-1$ $1^{st}$ discord that appears in the ranking $dkm_{\ell_{min},\ell_{max}}$, 
%and is very dissimilar to the pattern representing the usual number of passengers of a single day. 
%This discord is connected to trips re-organization before the daylight saving time.
%%, mainly due to the public transportation re-scheduling.
%
%Figures~\ref{taxiDiscords}(b) and (c) show the \topkfirstd for lengths \textit{32} and \textit{33}, respectively. 
%We note that at length \textit{32}, the discord subsequence starts exactly with the two points containing the double recordings.
%Apart from this anomalous behavior, the remaining part of the pattern corresponds to the normal cycle of passengers we observe in the data.

\subsection{Exploratory Analysis: Motif and Discord Length Selection}

In this part, we present the results of an experiment we conducted to test the capability of MAD to suggest the most promising length/s for motifs and discords.
%As presented throughout the paper, this is a challenge that firstly involves scalability issues, and also the problem with interpreting the obtained results.

Given a data series, the user may have no clear idea about the motif/discord length. % or even a possible range of lengths. 
Therefore, we present use cases that examine the ability of MAD to perform a wide length-range search, providing the most promising results at the \emph{correct} length.

We used MAD for finding motifs and discords in the length range: $\ell_{min} = 256$ and $\ell_{max} = 4096$. We conducted this experiment in the first $500K$ points of the datasets listed in Table~\ref{datainfo}.
The considered motif/discord length range covers the user studies that have been presented so far in the literature (where knowledge of the exact length was always assumed). 
%We furthermore noted that going beyond the maximum length bound ($4096$), the motif subsequences tend to have very low similarity.
%In general, for a complex time series (i.e., lots of peaks and valleys) the curse of dimensionality ensures that long motifs are not very meaningful. That is to say, as the subsequence length gets longer, the motif distance approaches the discord distance.   

\noindent{\bf Scalability.} The MAD framework completed the motif/discord discovery task within \textit{2 days} (on average), enumerating the motifs and the $Top-1$ discords of each length in the given range. 
Concerning the competitors, we estimated that STOMP, which is the state-of-the-art solution for fixed length motif/discord discovery would take $320$ days for the same experiment (a little bit more than two hours for each of the lengths we tested).
QUICK MOTIF, which has data dependent time performance, takes up to more than \textit{6} days (projection) for all datasets but ECG (which completes in \textit{38} hours).
We note that the variable-length motif discovery competitor (MOEN) never terminates before \textit{24} hours when searching motifs of $600$ different lengths, while in this experiment, the length range is composed of $3841$ different lengths.      
Considering discord discovery, we observed that GrammarViz does not enumerate all the discords in the given length-range, since it selects the length according to the data summarizations. 
Thus, we are obliged to run this technique independently for each length, which would take at least $320$ hours in the best case (projection based on results of Figure~\ref{Gviz_MAD}).

\begin{figure}[tb]
	\centering
	\includegraphics[trim={0cm 10cm 15cm 0cm},scale=0.50]{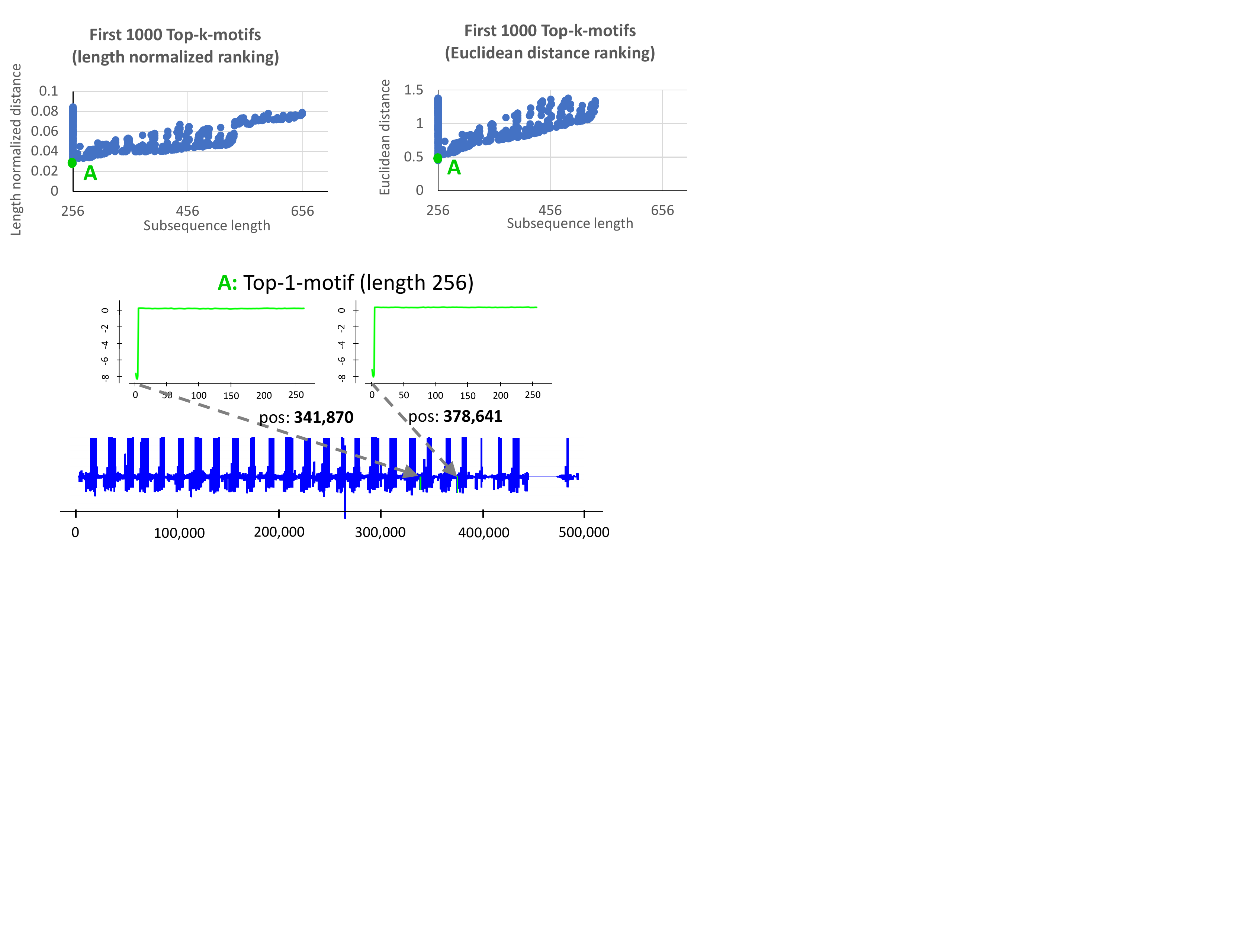}
	\caption{Top-1 motif (of length $256$) in the EEG data set. The subsequences pairs composing this motif have the smallest distance in both the Euclidean distance and length normalized ranking.}
	\label{EEGMotif_small}
\end{figure}

\begin{figure}[tb]
	\centering
	\includegraphics[trim={0cm 11.5cm 0cm 0cm},scale=0.46]{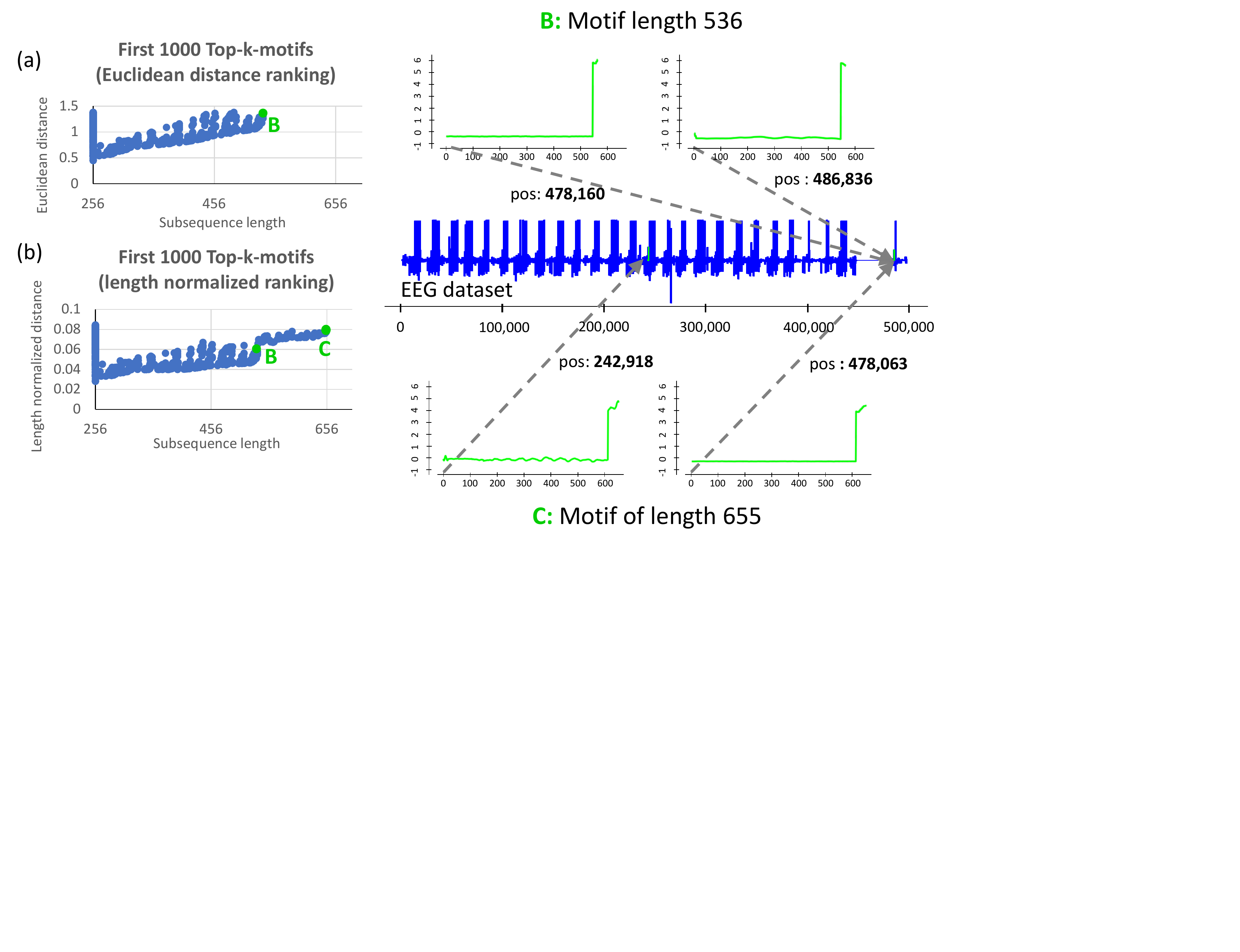}
	\caption{\textit{(a)} Top-1000 motifs according the length normalized distance (top), and the Euclidean Distance (bottom). \textit{(b)} Motif pair of the largest length ($656$) in the length normalized ranking (top) and motif pair of the largest length ($536$) in the Euclidean distance ranking (red/bottom).}
	\label{EEGMotif}
\end{figure}

\noindent{\bf Select the most promising length in Motif Discovery.} Once the search is completed, the MAD framework enumerates the motifs and discords ranking them in a second step, according to the proposed distance normalization strategy.
In Figure~\ref{EEGMotif_small}, we show the results of motif discovery for the EEG dataset.

%\commentRed{This data collection is the results of periodic EEG activity recorder during NREM sleep.
%Following the annotations of the domain experts, these data are characterized by the presence of cyclic sequences of cerebral activation and deactivation~\citep{CAP:dataset}.}
The objective of this experiment is to evaluate the proposed length-normalized correction strategy. 
In this regard, we compare the motifs sorted by using length-normalization, and by Euclidean distances. 

On the top part of Figure~\ref{EEGMotif_small}, we report the distance/length values of the $Top-1000$ motifs ranked by the length-normalized measure (left), which comprise a subset of the results we store in the VALMP structure (Algorithm~\ref{algo1}).
In the right part of the figure, we report the $Top-1000$ motifs ordered by their Euclidean distances.
 
We observe that the Top-$1$ motif, i.e., the subsequence pair with the smallest distance (marked by the letter A) is the same in both rankings.
We report this motif in the bottom part of Figure~\ref{EEGMotif_small}, which is composed of two quasi-identical patterns in the EEG data series.  

We now evaluate motifs of larger lengths in the same dataset, which may reveal other interesting and similar patterns at different resolutions (lengths). 
In Figure~\ref{EEGMotif}(a), we report again the distance/length values of the $Top-1000$ motifs ranked by their Euclidean distance, which reveal that the longest motif, marked as B, has length $536$.
We observe that this subsequence pair substantially differs from the $Top-1$ motif of Figure~\ref{EEGMotif_small}. 

Subsequently, in Figure~\ref{EEGMotif}(b), we report the longest motif (marked as C) of length $655$ that we found in the $Top-1000$ motif ranking, based on length-normalized distances.
We note that $6\%$ of the length-normalized motifs are longer than those in the $Top-1000$ of the Euclidean ranking.
The example of motif C, which is a longer version of B, shows that this pattern appears much earlier in the sequence than B.
%{\bf ??? it would be nice if we could explain what the motifs we present are ???}
%Discovering motif C gives us two new pieces of information. 
%First, we can consider it as a longer version of the motif B, since it has a very similar shape, though different absolute values towards the last part of the sequences. 
%Second, we note that the first subsequence of the motif C takes place earlier, thus in a very different data series position ($242,918$) respect to the subsequences of motif B.    
If we considered just the $Top-1000$ motifs ranked by their Euclidean distance, we would have missed this insight (motif C appears in the Euclidean distance ranking only in the $Top-4000$ motifs).

%Clearly, this new information can be found in the Euclidean ranking, but at a later stage.
%In fact, as we see from of Figure~\ref{EEGMotif}(c), motif C is highly penalized in the Euclidean ranking, since we need to consider the $Top-4000$ motifs before to find it.
%This fact is due to the highest number of short motifs ($256$ points), that have lower Euclidean distance than the larger motifs, as depicted in the plot of Figure~\ref{EEGMotif}(c).
%Using the length-normalized distances allows us to consider earlier several more motif occurrences of different lengths.    

\noindent{\bf Unfolding Top-k motifs.}
When considering the $Top-k$ motif ranking, we could manually inspect all the subsequence pairs.
%This clearly has a linear time complexity in terms of the data series size.
However, this is a cumbersome (and unnecessary) task for a user that would like to focus directly on the most interesting motifs.
In the previous experiment on EEG data, we examined the number of motifs we need to consider.
We now examine the value of $k$ that allows us to find interesting patterns within the $Top-k$ motif ranking.

\begin{figure}[tb]
	\centering
	\includegraphics[trim={0cm 10cm 0cm 0cm},scale=0.7]{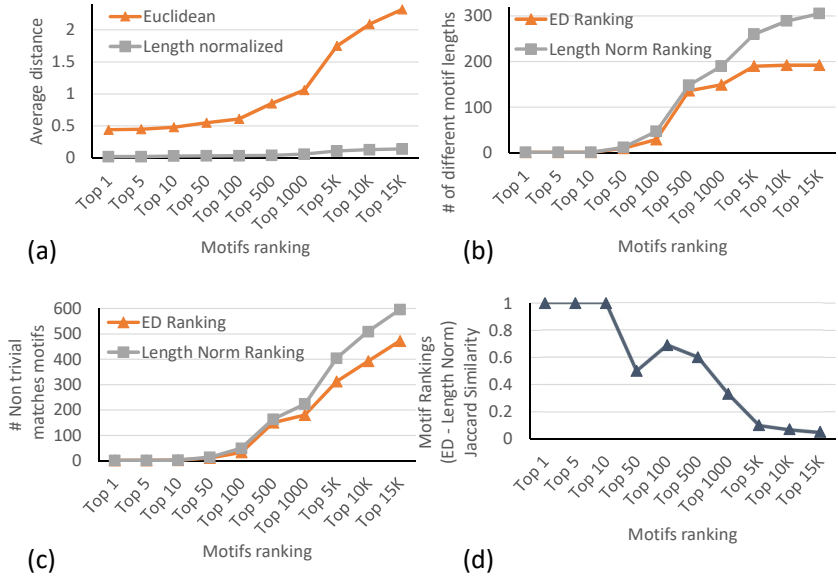}
	\caption{Euclidean and lenght-normalized $Top-k$ motifs properties. (a) Average distance (b) Number of motif lengths. (c) Number of non-	trivial match motifs. (d) Jaccard similarity of the ranking using the Euclidean and lenth-normalized Euclidean distances.}
	\label{TopkUnfolding}
\end{figure}
  
In Figure~\ref{TopkUnfolding}(a) we report the average distance for the $Top-k$ motif rankings that we built considering Euclidean and length-normalized distances, varying $k$.  
We note that the average distance exhibits a steep increase in the Euclidean distance rankings, starting from $k=500$. 
This is due to the presence of motifs of larger lengths, as depicted in Figure~\ref{TopkUnfolding}(b), since these pairs of longer subsequences have also a larger distance.
In this specific case, the user may choose to discard motifs beyond the $Top-500$, thus, disregarding several motifs of different lengths.
In contrast, we note that length-normalized distance is not heavily affected by longer motifs (Figure~\ref{TopkUnfolding}(a)).%, as the number of different motif lengths becomes up to $25\%$ higher as $k$ is larger than $500$. 
This will urge users to continue the exploration beyond the $Top-500$, and consider motifs of several different lengths that (as discussed earlier) represent different kinds of insights.

Another important factor to account in $Top-k$ motif analysis is the redundancy in the reported motifs.
In that respect, we can eliminate the motifs composed by subsequences that are trivial matches of motif subsequences that appear earlier in the ranking. 
In Figure~\ref{TopkUnfolding}(c), we plot the motifs that we retain (i.e., the motifs that are not trivial matches) from the Euclidean and length-normalized $Top-k$ rankings. 
We notice that as $k$ increases these retained motifs represent only a small subset of the motifs in the original rankings (up to $4\%$), which renders their examination easier. 
Furthermore, we observe that the Length-normalized $Top-k$ rankings contain up to $130$ more non-trivial match motifs than the Euclidean rankings, which translates to more useful results.

To conclude, we depict in Figure~\ref{TopkUnfolding}(d) the Jaccard similarity between the two ranking types (i.e., length-normalized and Euclidean) as we vary $k$.
While computing the intersection and the union of the two rankings, we discard the motifs that are trivial matches. 
As $k$ increases, % beyond $100$, 
and consequently the motif length increases as well (refer to Figure~\ref{TopkUnfolding}(b)), we observe that set similarity decreases.
This means that the new motifs of different lengths are not trivial matches %, and are not longer (or shorter) version 
of motifs found in higher ranking positions, but they represent new, useful results.

\begin{figure}[tb]
	\centering
	\includegraphics[trim={0cm 2cm 0cm 0cm},scale=0.38]{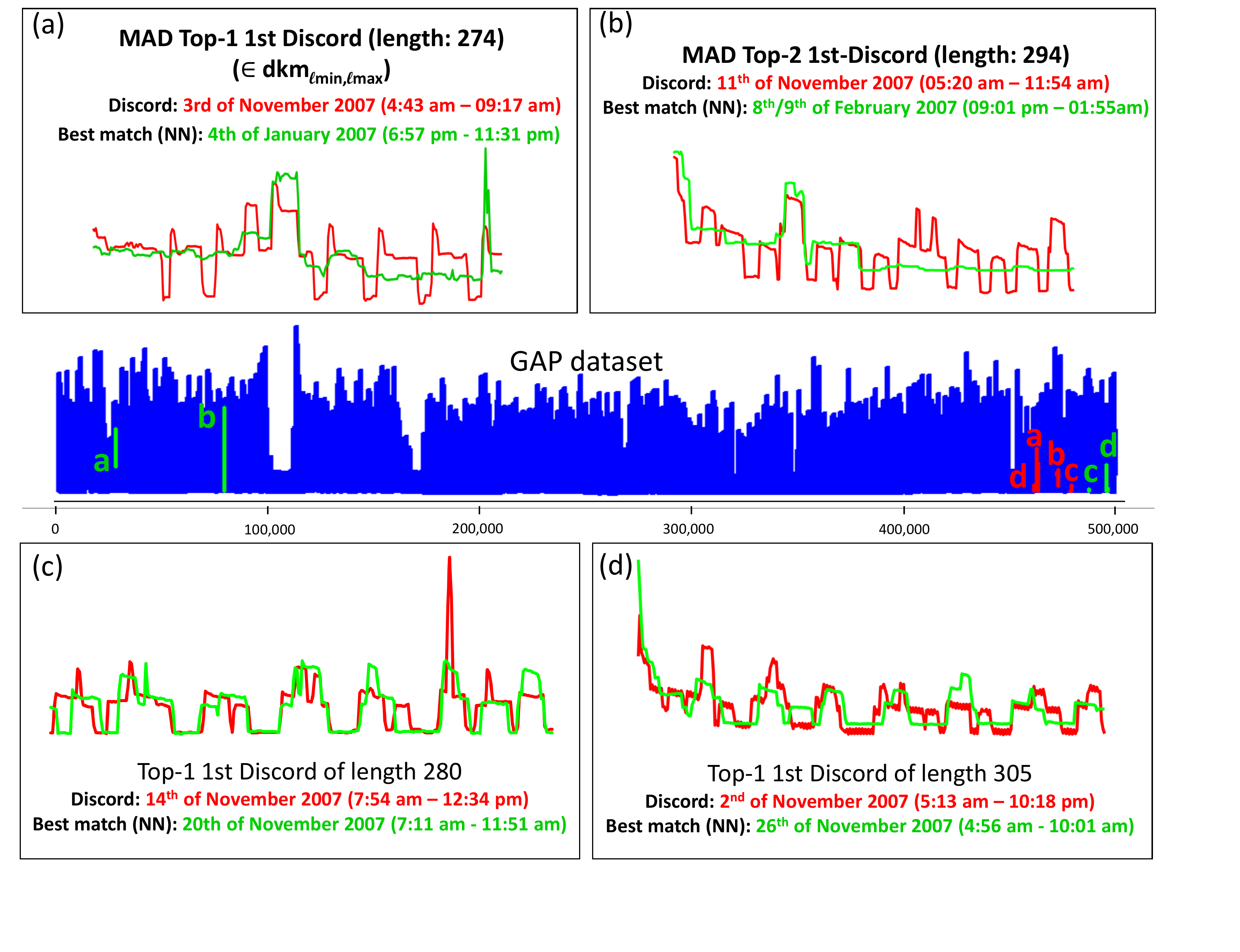}
	\caption{Four discords of different length in the GAP dataset. Each discord (red subsequence) is coupled with its nearest neighbor (green subsequence). \textit{(a)} The discord, with the highest length-normalized distance to its nearest neighbor has length $274$. \textit{(b)} Discord with the second highest length-normalized distance. \textit{(c)},\textit{(d)} discords with a smaller length-normalized distance to their nearest neighbor.}
	\label{GAPDiscord}
\end{figure}

\noindent{\bf Select the most promising length in Discord Discovery.}
In this part, we show the results of discord discovery performed in the GAP dataset.
We recall that in this case, the discord ranking performed according to their length normalized distances aims to favor smaller discords, which have a high point to point distance.

In Figure~\ref{GAPDiscord}, we report some of the discords we found in the length range $\ell_{min}=256$ and $\ell_{max}=4096$.
The discord with the highest length-normalized distance, \emph{best} $Top$-$1$ $1^{st}$ discord, is the one depicted in the top-left part of the figure, and has length $274$.
We plot it in red (dark), whereas its nearest neighbor appears in green (light).   
%Note, that this latter is the \emph{best} $Top$-$1$ $1^{st}$ discord of length $274$, which is stored in the $dkm_{\ell_{min},\ell_{max}}$ ranking in Algorithm~\ref{computeDkmAlllength}.
We note that this discord drastically differs from its nearest neighbor: it represents a fluctuating cycle of global power activity, while its nearest neighbor exhibits the expected behavior of two major peaks, in the morning and around noon.
In Figure~\ref{GAPDiscord}(b) we report the $Top$-$2$ $1^{st}$ discord in the length range $256$-$4096$ identified by MAD, which corresponds to the subsequence in that length range with the second highest length-normalized distance to its nearest neighbor. 
Once again, we observe a high degree of dissimilarity between the pattern of this discord and its nearest neighbor.
On the contrary, Figures~\ref{GAPDiscord}(c) and (d)
report the $Top$-$1$ $1^{st}$ discords for two specific lengths (i.e., 280 and 305, respectively). 
These discords correspond to patterns that are not significantly different from their nearest neighbors. 
Therefore, they represent discords that are less interesting than the ones reported by MAD in Figures~\ref{GAPDiscord}(a) and (b), which examines a large range of lengths.

%since there are seven cycles, which do not take place in the closest neighbor, denoting an abnormal occurrence in the data taking place an early Saturday morning, between \textit{4:43}am and \textit{9:17}am.
%%In general, we can expect a low and steady global active power evolution during the first period of this subsequence.
%In the best match of this discord, which spans a similar period of time (from \textit{6:57} to \textit{11:51}) we note that the global power activity has two peaks in the morning and close to noon, as we might expect.
%On the other hand, in the discord subsequence we note a presence of a fluctuating cycle, which takes place along the whole time range.

%In Figure~\ref{GAPDiscord}, we also report the discords of length $256$ and $4096$, the smallest and largest lengths examined, respectively.
%%Despite the fact that both discords have high similarity to they nearest neighbors, we can see that they are misaligned, and thus detected as discords. 
%Considering that the discord subsequences have almost the same time window positions of their nearest neighbors, we believe that these two occurrences represent a repeated patterns rather than abnormalities.
%{\bf ??? it is not true that the nearest neighbors of discords with lengths 256 and 4096 are in neighboring time windows: their NN are very far away! I do not see the argument here... ???}

This experiment demonstrates that MAD and the proposed discord ranking allows us to prioritize and select the correct discord length.

\section{Related Work}
\label{sec:related}

While research on data series similarity measures and data series query-by-content date back to the early 1990s~\citep{DBLP:conf/sofsem/Palpanas16}, \textit{data series motifs} and \textit{data series discords} were both introduced just fifteen and twelve years ago, respectively~\citep{DBLP:conf/kdd/ChiuKL03,DBLP:conf/adma/FuLKL06}.
Following their definition, there was an explosion of interest in their use for diverse applications. 
Analogies between \textit{data series motifs} and sequence \textit{motifs} exist (in DNA), and have been exploited. 
For example, discriminative motifs in bioinformatics~\citep{DBLP:conf/recomb/Sinha02} inspired discriminative data series motifs (i.e., data series shapelets)~\citep{citrusProduction}. 
Likewise, the work of Grabocka et al.~\citep{DBLP:journals/tkdd/GrabockaSS16} on generating idealized motifs is similar to the idea of consensus sequence (or canonical sequence) in molecular biology.
The literature on the general data series motif search is vast; we refer the reader to recent studies~\citep{ZhuZSYFMBK16,YehZUBDDSMK16} and their references.

The QUICK MOTIF~\citep{DBLP:conf/icde/LiUYG15} and STOMP~\citep{ZhuZSYFMBK16} algorithms represent the state of the art for fixed-length motif pair discovery. 
QUICK MOTIF first builds a summarized representation of the data using Piecewise Aggregate Approximation (PAA), and arranges these summaries in Minimum Bounding Rectangles (MBRs) in a Hilbert R-Tree index.
The algorithm then prunes the search space based on the MBRs.
On the other hand, STOMP is based on the computation of the matrix profile, 
%with a time cost of $O(n^2)$, 
in order to discover the best matches for each subsequence. 
The smallest of these matches is the motif pair. 
We observe that both of the above approaches solve a restricted version of our problem: they discover motif sets of cardinality two (i.e., motif pairs) of a fixed, predefined length. 
On the contrary, VALMOD removes these limitations and proposes a general and efficient solution.
Its main contributions are the novel algorithm for examining candidates of various lengths and corresponding lower bounding distance: these techniques help to reuse the computations performed so far, and lead to effective pruning of the vast search space.

We note that there are only three studies that deal with issues of variable length motifs, and attempt to address them~\citep{MinnenIES07,Gao0R16,YingchareonthawornchaiSRR13,DBLP:journals/datamine/GaoL18}. 
While these studies are pioneers in demonstrating the \textit{utility} of variable length motifs, they cannot serve as practical solutions to the task at hand for two reasons: 
(i) they are all approximate, while we need to produce exact results; and (ii) they require setting many parameters (most of which are unintuitive).
Approximate algorithms can be very useful in many contexts, if the amount of error can be bounded, or at least known. 
However, this is not the case for the algorithms in question. 
%While they have shown reasonable results on small, smooth datasets, it is not clear how they would fare in much longer, and much noisier datasets such as the seismological dataset we consider in Section~\ref{sec:seismology}. 
Certain cases, such as when analyzing seismological data, the threat of litigation, or even criminal proceedings~\citep{Bertolaso}, would make any analyst reluctant to use an approximate algorithm.

The other work that explicitly considers variable length motifs is MOEN~\citep{DBLP:journals/kais/MueenC15}. 
Its operation is based on the distance computation of subsequences of increasing length, and a corresponding pruning strategy based on upper and lower bounds of the distance computed for the smaller length subsequences.
Unlike the algorithms discussed above, MOEN is exact and requires few parameters. 
%However, as we shall see in our experimental section, MOEN is not competitive in time-performance with our proposed approach.
% (in fairness, our proposed approach leverages the Matrix Profile, of which the author of~\citep{DBLP:journals/kais/MueenC15} was an inventor).
However, it has been tuned for producing only a single motif pair for each length in the range, and as our evaluation showed, it is not competitive in terms of time-performance with our approach.
This is due to its relatively loose lower bound and sub-optimal search space pruning strategy, which force the algorithm to perform more work than necessary.

Exact discord discovery is a problem that has attracted lots of attention. 
The approaches that have been proposed in the literature can be divided in the following two different categories.
First, the \textit{index-based} solutions, i.e., Haar wavelets~\citep{DBLP:conf/adma/FuLKL06,DBLP:conf/sdm/BuLFKPM07} and SAX~\citep{Keogh2005,Keogh2007,DBLP:conf/edbt/Senin0WOGBCF15}, where series are first discretized and then inserted in an index structure that supports fast similarity search. 
Second, the \textit{sequential scan} solutions~\citep{YehZUBDDSMK16,Liu2009,DBLP:conf/adma/FuLKL06,Parameter-Free_Discord,DBLP:conf/icdm/YankovKR07,ZhuZSYFMBK16}, which consider the direct subsequence pairwise distance computations, and the corresponding search space optimization.

Indexing techniques are based on the discretization of the real valued data series, with several user defined parameters required for this operation. 
In general, selecting and tuning these parameters is not trivial, and the choices made may influence the behavior of the discord discovery algorithm, since it is strictly dependent on the quality of the data representation. 
In this regard, the most recent work in this category, GrammarViz~\citep{DBLP:conf/edbt/Senin0WOGBCF15}, proposes a method of \topkfirstd discord search based on grammar compression of data series represented by discrete SAX coefficients.
These representations are then inserted in a hierarchical structure, which permits us to prune unpromising candidate subsequences.
%{\bf ??? grammarviz is indexed-based? does it use an index structure? ???}
The intuition is that rare patterns are assigned to representations that have high \textit{Kolmogorov complexity}. This means that a rare SAX string is not compressible, due to the lack of repeated terms.
%In this work the search is performed, considering a hierarchical structure, which allows to firstly visit the most rare sequences, performing thus the pruning of the sequences that are well repeated in the dataset, having higher probability not to correspond to any discords.

The state of the art for the sequential scan methods is represented by STOMP, since computing the matrix profile permits to discover, in the same fashion as motifs, the \topkfirstd discords. 
Surprisingly, just one work exists that addresses the problem of $m^{th}$ discord discovery~\citep{DBLP:conf/icdm/YankovKR07}. 
The authors of this work, proposed the Disk Aware discords Discovery algorithm (DAD), which is based on a smart sequential scan performed on disk resident data.
This algorithm is divided into two parts. 
The first is discord candidate selection, where it identifies the sequences, whose nearest neighbor distance is less than a predefined range. 
%A technique to estimate an optimal range is also proposed in the paper. 
%{\bf ??? what does that mean? explain with 1-2 sentences ???}
The second part, which is called refinement, is applied in order to find the exact discords among the candidates. 
%To prune the Euclidean distance computations an early abandoning technique is used.
Despite the good performance that this algorithm exhibits in finding the first discord, when $m$ is greater than one, %and thus we are interested in finding multiple discords, 
it becomes hard to estimate an effective range.
%, which in general becomes large. This, will not permit to scale, 
In turn, this leads to scalability problems, due to the explosion of the number of distances to compute.

%{\bf ??? where are all the rest discord papers (the ones we reference in norma)? ???}

In summary, while there exists a large and growing body of work 
%that both uses motif discovery in diverse domains, or offers enhancements/generalizations of basic motif search, 
on the motif and discord discovery problems, 
this work offers the first scalable, parameter-light, \textit{exact} variable-length algorithm in the literature for solving both these problems. 

\section{Conclusions}
\label{sec:conclusions}

Motif and discord discovery are important problems in data series processing across several domains, and key operations necessary for several analysis tasks.
Even though much effort has been dedicated to these problems, no solution had been proposed for discovering motifs and discords of different lengths.

In this work, we propose the first framework for variable-length motif and discord discovery. 
We describe a new distance normalization method, as well as a novel distance lower bounding technique, both of which are necessary for the solution to our problem. 
We experimentally evaluated our algorithm by using five real datasets from diverse domains. 
The results demonstrate the efficiency and scalability of our approach (up to 20x faster than the state of the art), as well as its usefulness.

In terms of future work, we would like to further improve the scalability of VALMOD.
%, which would lead to faster performance even for larger motif and discord length ranges.
We also plan to extend VALMOD in order to efficiently compute a complete matrix profile for each length in the input range.
This would enable us to support more diverse applications, such as discovery of \textit{shapelets}~\citep{YeK09}.
%(on top of variable length motif discovery).

%\begin{acknowledgements}
%\end{acknowledgements}

\bibliographystyle{spbasic}      % basic style, author-year citations
\bibliography{GeneralBIB}   % name your BibTeX data base

\end{document}